\documentclass[11pt,a4wide]{article}
\pdfoutput=1

\usepackage{jheppub}

\usepackage[english]{babel}
\usepackage[babel]{csquotes}
\usepackage{amsfonts}
\usepackage{array, tabularx}
\usepackage{mathtools}
\usepackage{amsthm}
\usepackage{url}
\usepackage{multicol, multirow}
\usepackage{emptypage}
\usepackage{verbatim}
\usepackage{dsfont}
\usepackage{hyperref}
\usepackage{float}
\usepackage{amsmath}
\usepackage{amssymb,tikz}
\usepackage{verbatim}
\usepackage{xspace}
\usepackage{youngtab}
\usepackage{tikz-cd}

\allowdisplaybreaks[2]

\usepackage[font=small,format=hang,labelfont={sf,bf}]{caption}

\usepackage{tensor}

\usepackage{multirow}

\newcommand{\tableheading}[1]{$\ell=#1$ &$\mathcal O$ & $\Delta$ & $\gamma$ ($n=3$) & $\gamma$ ($n=4$) & $\gamma$ ($n$ gen.)}
\newcommand{\tableheadingF}[1]{$\ell=#1$ &$\mathcal O$ & $\Delta$ & $\gamma$ ($n=4$) & $\gamma$ ($n$ gen.) }

\usepackage{youngtab}

\allowdisplaybreaks

\definecolor{darkblue}{rgb}{0.1,0.1,0.7}
\hypersetup{colorlinks,
           linkcolor={darkblue},
           citecolor={darkblue},
           urlcolor={darkblue}
}

\newcommand{\XXXbar}{{\overline{XXX}}}
\newcommand{\XXbar}{{\overline{XX}}}
\newcommand{\TotS}{Z_4}
\newcommand{\eps}{{\varepsilon}}
\renewcommand{\O}{{\mathcal O}}

\newcommand{\lsp}{\hspace{1pt}}

\renewcommand{\geq}{\geqslant}
\renewcommand{\leq}{\leqslant}

\title{Anomalous Dimensions in Hypercubic Theories}

\author{Alexander Bednyakov,$^{a}$}
\author{Johan Henriksson,$^{b}$}
\author{Stefanos R.\ Kousvos$^{b}$}
\affiliation{
$^a$ Bogoliubov Laboratory of Theoretical Physics,
Joint Institute for Nuclear Research, 141980 Dubna, Russia\\
$^b$ Department of Physics, University of Pisa and INFN, section of Pisa. Largo Pontecorvo 3, I-56127 Pisa, Italy\\}

\emailAdd{bednya@theor.jinr.ru} \emailAdd{johan.henriksson@df.unipi.it} \emailAdd{stefanos.kousvos@df.unipi.it}

\abstract{We perform a comprehensive perturbative study of the operator spectrum in multi-scalar theories with hypercubic global symmetry. This includes working out symmetry representations and their corresponding tensor structures. 
These structures are then used to compute the anomalous dimensions of scalar operators with up to four fields and arbitrary representations to six-loop order. 
Moreover, we determine one-loop anomalous dimensions for a large number of low-lying operators in the spectrum which include more powers of the fundamental field and/or insertions of derivatives.
As an aside we show how projectors used in the conformal bootstrap can be conveniently reused in computations of anomalous dimensions. The results of our study are of use to the conformal bootstrap. 
They also illuminate features of conformal perturbation theory and the large $n$ expansion.
Our results may be of interest for various crossover phenomena in statistical field theory.
In total, we compute the scaling dimension of more than 300 operators, of which 16 are computed to six-loops. Our analysis is exhaustive with respect to group theory up to rank 4 for any number of flavours $n$, and also exhaustive with respect to which representations exist for $n \leq 4$.}

\begin{document}

\maketitle

\section{Introduction}

Scalar field theories in $d=4-\varepsilon$ dimensions provide the simplest non-trivial examples of fully interacting conformal field theories one may study. They also describe numerous phase transitions in statistical field theory \cite{Pelissetto:2000ek}. Given the modern resurgence of the conformal bootstrap \cite{Rattazzi:2008pe}, see also \cite{Poland:2018epd} for a review, there has been a renewed interest in the space of CFTs in general, as well as the space of multi-scalar CFTs in $d=4-\varepsilon$ in particular \cite{Osborn:2017ucf,Rychkov:2018vya,Osborn:2020cnf,Hogervorst:2020gtc,Codello:2020lta,Zinati:2019gct}. With the exception of the $O(n)$ vector models, however, most studies have focused mostly on the existence, or not, of fixed points. Beyond the knowledge of the space of CFTs, which could be viewed as a qualitative understanding, we would also like to achieve a quantitative understanding of such CFTs. In other words, instead of focusing on the space of CFTs itself, we focus on one single family of CFTs and attempt to compute as much of its spectrum as possible. This family of CFTs is the hypercubic family with $C_n := S_n \ltimes (\mathbb Z_2)^n$ global symmetry.\footnote{Conveniently thought of as the theory of $n$ coupled Ising models which may be permuted with each other by the $S_n$ factor.} This theory has indeed already been subjected to some precision RG studies, such as \cite{Adzhemyan:2019gvv}, however in our work we will study all scalar operators with up to four fields to six-loop order. Moreover, we will make a one-loop study that generalises the work \cite{Antipin:2019vdg} to include spinning operators and scalar operators with derivatives.

The hypercubic CFT has the UV Lagrangian\footnote{Note that $\lambda_2 = u/3 = g_1/3$ and $\lambda_1 = v = g_2$ in the notation of Refs.~\cite{Aharony:1973zz} and \cite{Adzhemyan:2019gvv}, respectively.}
\begin{equation}
\mathcal{L}=\frac{1}{2}\partial_\mu \phi_a \partial_\mu \phi_a -\frac{\lambda_1 \delta_{abcd}+\lambda_2 (\delta_{ab}\delta_{cd}+\delta_{ac}\delta_{bd}+\delta_{ad}\delta_{bc})}{4!}\phi_a \phi_b \phi_c \phi_d\,,
\label{hypercubiclagrangian}
\end{equation}
defined in $d<4$ spacetime dimensions. 
Let us motivate the choice of considering the hypercubic symmetry. 
Beyond being physically relevant for phase transitions, it provides (one of) the simplest deformations of the $O(n)$-symmetric multi-scalar theory. Phrased slightly differently, it is the simplest critical theory which is not the $O(n)$ model. While indeed non-trivial, the $O(n)$ vector model enjoys a number of simplifications compared to the most generic multi-scalar theory that can be written down. Firstly, from the perturbative point of view, it is the only single coupling quartic deformation of the free theory. This fact is related to it being the most symmetric scalar field theory, i.e. it has the smallest number of invariant tensors, all of which may be built using powers of the two-index Kronecker delta $\delta_{ij}$, and it leads to considerable simplifications in the conformal bootstrap. 
For example, if one wishes to study any theory that is not the $O(n)$ vector model, one often needs to make various assumptions to first exclude the $O(n)$ model from parameter space (given that it is the maximally symmetric theory), in order to then be able to study the theory of interest.\footnote{Notably, in the $C_3$ case we will see that this problem is exacerbated by the fact that the $C_3$ and $O(3)$ fixed points are incredibly close, in the RG sense of the word.} Another notable simplification in the case of the $O(n)$ model, is the continuous nature of its symmetry which provides conserved currents,  and simplifies its large $n$ expansion. 
Whereas in the $C_n$ case, beyond the obvious omission of conserved currents, the large $n$ expansion is also considerably harder to carry out, see \cite{Binder:2021vep}. Hence, if one were to achieve satisfactory understanding of the $C_n$ family, we believe that one would be considerably closer to understanding the most general multi-scalar CFT possible.

In the present work we make use of the recent six loop results for generic multi-scalar theories that appeared in \cite{Bednyakov:2021ojn} in conjunction with the group theoretic results of \cite{Kousvos:2021rar} (see also \cite{Antipin:2019vdg} and \cite{BenAliZinati:2021rqc}) in order to compute the most precise estimates to date for a large number of operators in various representations of the $C_n$ global symmetry. In addition to this, we will compute the scaling dimensions of more than 300 operators at one loop. 

The text has been organised as follows, and each section is, for the most part, self contained.
In section~\ref{sec:introGroupTheory} we give a brief introduction to the hypercubic group $C_n$, highlighting its relation with the familiar case of $O(n)$, of which it is a subgroup.
In section~\ref{chapter2} we demonstrate how to calculate scaling dimensions by deforming the Lagrangian with the operator whose dimension we want to read off. Consequently, we detail a method for writing down representations of the global symmetry and explicit operators transforming in them. We do this using global symmetry projectors which make direct contact to numerical conformal bootstrap intuition. Then, in section~\ref{chapter3}, we proceed to outline a method with which we may calculate the scaling dimension of any\footnote{I.e. in any representation of the local and global symmetry.} operator at one loop. For a discrete group (i.e. fixed integer $n$), this method can be completely automated, provided that the conjugacy classes and the character table are given. 
In the non-automated implementation we demonstrate an alternative (but equivalent) method to the use of projectors for fixing the tensor structures, valid for any value of $n$ in $C_n$. With this formal machinery established we present all our results in section~\ref{chapter5}. The reader primarily interested in the results may proceed directly to section~\ref{chapter5} after a quick glance at section~\ref{sec:introGroupTheory}. 
To the best of our knowledge, various technical details are explicitly presented for the first time in our work. Let us lastly point out that throughout this work upper and lower indices will be taken to be indistinguishable and equivalent.

\section{Gentle introduction to representations of the hypercubic group}
\label{sec:introGroupTheory}

In this section, we give a brief overview of the representations of the hypercubic group $C_n$. More details are left for later in the manuscript. We contrast with the perhaps more familiar case of $O(n)$, of which it is a subgroup. This in practice means that it has more invariant tensors. Typically, when breaking the symmetry from $O(n)$ to $C_n$, the irreps of the former split into multiple irreps of the latter.

The dimension of the hypercubic group is
\begin{equation}
\dim C_n=2^nn!\,,
\end{equation}
which is easily accounted for ``physically'' by noting that the  $2^n$ factor corresponds to ``spin-flips'' of $n$ spins, and $n!$ corresponds to the possible permutations of the $n$ spins.
Aspects of the representation theory for the hypercubic group are nicely described in \cite{Antipin:2019vdg}. The group can also be treated as a special case of replica groups $S_n \ltimes G^n$ when $G=\mathbb Z_2$, see \cite{Kousvos:2021rar}.

For fixed integer $n$, the hypercubic group is finite. Therefore, in principle, the representation theory can be dealt with using finite-group theory. In the case $n=3$, it becomes the familiar octahedral group $O_h$ with $48$ elements. For $n=4$, it is a finite group of $384$ elements; we collect some specific information about this case in appendix~\ref{app:n4reps}.

\paragraph{Branching from $\boldsymbol{O(n)}$}
In general the hypercubic group is a discrete subgroup of $O(n)$. We therefore expect branching rules
\begin{equation}
R_{O(n)}\longrightarrow \bigoplus_i R_i\,.
\end{equation}
The first few irreps of the hypercubic group can be easily understood from this decomposition.

The singlet $S$ and vector $V={\tiny \yng(1)}$ of $O(n)$ simply become the singlet $S$ and vector $V$ of $C_n$:
\begin{equation}
S_{O(n)}\longrightarrow S, \qquad \dim S=1, \qquad V_{O(n)}\longrightarrow V,\quad \dim V=n\,.
\end{equation}
The first operators in these irreps are $S=\phi^i\phi^i$, and $\phi^i$.

The rank-2 symmetric tensor $T={\tiny \yng(2)}$ of $O(n)$ decomposes into a traceless diagonal part $X$, and an off-diagonal part $Z$,
\begin{equation}
T\longrightarrow X\oplus Z, \qquad \dim X=n-1,\quad \dim Z=\frac{n(n-1)}2.
\end{equation}
The sum $\dim X+\dim Z=\frac{(n-1)(n+2)}2 = \dim T$, as expected. The first operators in these irreps are constructed out of two fields, with representatives $X_{ii}=\phi_i\phi_i-\frac1n\sum_k\phi_k\phi_k$ and $Z_{ij}=\phi_i\phi_j$, $i\neq j$.\footnote{The associated crossover exponents of \cite{Aharony1976,Aharony:2022ajv} are related to the dimensions of $X$ and $Z$ by $\phi_{\mathrm{axis}}=\frac{d-\Delta_X}{d-\Delta_S}$ and $\phi_{\mathrm{diag}}=\frac{d-\Delta_Z}{d-\Delta_S}$.}

Likewise, the rank-$m$ traceless symmetric tensor $T_m={\tiny \yng(3)\cdots\yng(1)}$ of $O(n)$ decomposes into a sum of irreps,
\begin{equation}
\label{eq:TmDecomp}
T_m\longrightarrow Z_m\oplus \ldots, \qquad \dim Z_m=\left(\begin{smallmatrix}n\\m\end{smallmatrix}\right).
\end{equation}
Here $Z_2=Z$.
The first operator in the $Z_m$ irrep is always a scalar $\phi_{i_1}\phi_{i_2}\cdots\phi_{i_m}$, with all indices $i_j$ unequal.
Note that the construction of the $Z_m$ irrep, and thus the general form of the decomposition \eqref{eq:TmDecomp} requires $n\geqslant m$.\footnote{Since, for example, one cannot demand that there is a tensor with four different indices, if one only has access to three fields.} Otherwise finite-$n$ effects occur.

The rank-$2$ antisymmetric tensor $A={\tiny \yng(1,1)}$ of $O(N)$ remains irreducible, and becomes an antisymmetric irrep which we call $B$:
\begin{equation}
A\longrightarrow B, \qquad \dim B=\frac{n(n-1)}2.
\end{equation}
The same holds for the rank-$m$ antisymmetric irrep of $O(n)$, $A_m\longrightarrow B_m$, with $\dim B_m=\left(\begin{smallmatrix}n\\m\end{smallmatrix}\right)$. Contrary to the case of $O(n)$, the irreps $B_m$ actually contain scalar primary operators with no inserted derivatives. Such operators can be written down in the form $\phi_{[i_1}\phi_{i_2}^3\phi_{i_3}^5\cdots\phi_{i_m]}^{2m-1}$, $m\geqslant 3$.\footnote{Note that there are also operators with derivatives which are generally lower in the spectrum. The operator $\phi_{[i}\phi_{j]}^3$ in the $B_2=B$ irrep is a primary operator in the free theory, but becomes proportional to the equation-of-motion operator in the interacting theory, see the discussion in footnote \ref{Redundant}. The smallest scalar $B$ operator then has six fields.}

Table~\ref{tab:irrepsLowLying} contains a complete list of low-lying irreps, specifically those that exist at $n=3$ (upper half) and $n=4$ (entire table). More branching rules are given in appendix~\ref{app:branchingRules}.

\paragraph{Tensor products}
The tensor product of two vectors takes the form
\begin{equation}
V\otimes V=(S\oplus X\oplus Z)_{\mathrm{sym}}\oplus (B)_{\mathrm{antisym}}\,.
\label{OPEPhiPhi}
\end{equation}
In conformal bootstrap applications, this decomposition is relevant for the $\phi$ four-point system.\footnote{This system has been considered in previous works. In the numerical study \cite{Stergiou:2018gjj}, the representations $(S,X,Z,B)$ were denoted $(S,V,Y,A)$, in  \cite{Kousvos:2018rhl,Kousvos:2019hgc} they were denoted $(S,X,Y,A)$, whereas in \cite{Rong:2017cow} which studied $S_4 \times \mathbb Z_2 \sim S_3 \ltimes \mathbb Z_2^3$ they were called $(S,T^\prime,n,A)$. In the analytic study \cite{Dey:2016mcs}, they were denoted $(S,V,T,A)$. 
In the scalar spectrum study \cite{Antipin:2019vdg}, representations were only referred to by their Young bi-tableau form (see below). Moreover, in \cite{Kousvos:2018rhl}, the irreps $(XV,\XXbar)$ were denoted $(A',\bar S)$.} In its conformal block decomposition, the symmetric irreps contain operators of even spin, and the antisymmetric irrep contains operators of odd spin. 
This is determined by whether the irreps on the right hand side of \eqref{OPEPhiPhi} are symmetric or antisymmetric under swapping the $\phi$'s on the left hand side between themselves. However, there are of course primary operators of all spins in all irreps. 
These can in turn be probed by larger bootstrap systems. They appear in OPEs between non-identical operators, where the product of operators can be neither symmetric nor antisymmetric, since they are distinct objects. For the reader's convenience we give more tensor products in appendix~\ref{app:tensorProducts}.

\begin{table}
\centering
\caption{Some low-lying irreps. The list is complete for irreps up to $n\leqslant 4$. The $n=3$ symbols refer to the standard names for the irreps in the cubic group $C_3=O_h$, see table~\ref{tab:characterTable}.}\label{tab:irrepsLowLying}
{\small
\renewcommand{\arraystretch}{1.65}
\begin{tabular}{|ccc|cc|c|c|c|c|}
\hline 
Symbol & YT &  $n=3$ symbol & dimension & rank
\\
\hline
$S$  & ([$n$],\raisebox{-0pt}{$\bullet$})& $A_{1g}$ & $1$&  $0$ 
\\
$V$  & ([$n-1$],\raisebox{-0pt}{\tiny \yng(1)})& $T_{1u}$ & $n$&  $1$ 
\\
$X$  & ({\renewcommand{\arraystretch}{0.6}$\begin{matrix}[n-1]\\\!\!\!\!\!\!\!\!\!\!\!\!\text{\tiny \yng(1)}\end{matrix}$},\raisebox{-0pt}{$\bullet$})& $E_g$ & $n-1$&  $2$ 
\\
$Z$  & ([$n-2$],\raisebox{-0pt}{\tiny \yng(2)})& $T_{2g}$ & $\frac{n(n-1)}2$&  $2$ 
\\
$B$  & ([$n-2$],\raisebox{-3pt}{\tiny \yng(1,1)})& $T_{1g}$ & $\frac{n(n-1)}2$&$2$ 
\\
$XV$  &  ({\renewcommand{\arraystretch}{0.6}$\begin{matrix}[n-2]\\\!\!\!\!\!\!\!\!\!\!\!\!\text{\tiny \yng(1)}\end{matrix}$},\raisebox{-0pt}{\tiny \yng(1)})& $T_{2u}$ & $n(n-2)$& $3$ 
\\
$VB$  & ([$n-3$],\raisebox{-3pt}{\tiny \yng(2,1)})& $E_u$ & $\frac{n(n-1)(n-2)}3$& $3$ 
\\
$Z_3$  &([$n-3$],\raisebox{-3pt}{\tiny \yng(3)})& $A_{2u}$ & $\frac{n(n-1)(n-2)}6$&  $3$ 
\\
$B_3$  & ([$n-3$],\raisebox{-6pt}{\tiny \yng(1,1,1)})& $A_{1u}$ &$\frac{n(n-1)(n-2)}6$  &$3$ 
\\
$\XXbar$  & ( {\renewcommand{\arraystretch}{0.6}$\begin{matrix}[n-2]\\\!\!\!\!\!\!\!\!\!\!\!\!\text{\tiny \yng(1,1)}\end{matrix}$},\raisebox{-0pt}{$\bullet$})& $A_{2g}$ & $\frac{(n-1)(n-2)}2$ &$4$
\\
\hline
$Z_4$ &  ([$n-4$],\raisebox{0pt}{\tiny \yng(4)})  && $\frac{ n(n-1)(n-2)(n-3)}{24}$ & $4$  
\\
$XX$  & ( {\renewcommand{\arraystretch}{0.6}$\begin{matrix}[n-2]\\\!\!\!\!\!\!\!\!\text{\tiny \yng(2)}\end{matrix}$},$\bullet$) && $\frac{n(n-3)}2$ & $4$ 
\\
$XZ$  &( {\renewcommand{\arraystretch}{0.6}$\begin{matrix}[n-3]\\\!\!\!\!\!\!\!\!\!\!\!\!\text{\tiny \yng(1)}\end{matrix}$},\raisebox{0pt}{\tiny \yng(2)}) && $\frac{n(n-1)(n-3)}2$ & $4$ 
\\
$XB$  &( {\renewcommand{\arraystretch}{0.6}$\begin{matrix}[n-3]\\\!\!\!\!\!\!\!\!\!\!\!\!\text{\tiny \yng(1)}\end{matrix}$},\raisebox{-3pt}{\tiny \yng(1,1)}) && $\frac{n(n-1)(n-3)}2$ & $4$ 
\\
$BZ$  &  ([$n-4$],\raisebox{-3pt}{\tiny \yng(3,1)})  && $\frac{ n(n-1)(n-2)(n-3)}{8}$ & $4$ 
\\
$BB$  &  ([$n-4$],\raisebox{-3pt}{\tiny \yng(2,2)})  && $\frac{ n(n-1)(n-2)(n-3)}{12}$ & $4$ 
\\
$VB_3$ &  ([$n-4$],\raisebox{-6pt}{\tiny \yng(2,1,1)}) & & $\frac{ n(n-1)(n-2)(n-3)}{8}$ & $4$ 
\\
$B_4$ &  ([$n-4$],\raisebox{-9pt}{\tiny \yng(1,1,1,1)})  && $\frac{ n(n-1)(n-2)(n-3)}{24}$ & $4$ 
\\
$\XXbar V$  &( {\renewcommand{\arraystretch}{0.6}$\begin{matrix}[n-3]\\\!\!\!\!\!\!\!\!\!\!\!\!\text{\tiny \yng(1,1)}\end{matrix}$},\raisebox{-3pt}{\tiny \yng(1)}) && $\frac{n(n-2)(n-3)}2$ & $5$
\\
$\XXXbar$  &( {\renewcommand{\arraystretch}{0.6}$\begin{matrix}[n-3]\\\!\!\!\!\!\!\!\!\!\!\!\!\text{\tiny \yng(1,1,1)}\end{matrix}$},$\bullet$) && $\frac{(n-1)(n-2)(n-3)}6$ & $6$ 
\\\hline
\end{tabular}
}
\end{table}

\paragraph{Construction of ``composite irreps''}
Beyond rank 2, it is convenient to construct the irreps from building blocks made out of lower irreps, thinking of them as ``composite irreps''. We would like to stress that despite this formulation, all irreps are indecomposable.\footnote{We will later explicitly see that we can build composite irreps by taking products of lower irreps and enforcing an additional constraint. It is this additional constraint that ensures that our irreps are distinct from a generic tensor product, which would in general be reducible.} As an example, at rank 3, we can construct four new irreps:
\begin{equation}
Z_3,\qquad B_3, \qquad XV, \qquad VB.
\end{equation}
Here $Z_3$ and $B_3$ are the completely symmetric/antisymmetric combination of three vectors, under the constraint that all indices are unequal.  The irreps $XV$ and $VB$ have components that look like a vector $V$ juxtaposed with an $X$ and $B$ operator respectively, again demanding that the indices in the different building blocks take different values. This way of constructing representative elements in irreps will be used in the construction of tensor structures below.\footnote{We note that in general, this notation does not give the complete characterization of the irrep, only a way to construct representative elements. The Young tableaux reprepresentation below gives a precise characterization of the irreps.} When combining $X$ factors, we use a bar to denote antisymmetric combinations, \emph{viz.} $\overline{XX}$.

\paragraph{Young tableaux for irreps}

As described in detail in \cite{Antipin:2019vdg}, the irreducible representations of the hypercubic group $C_n$ can be uniquely labeled by a pair of Young tableaux with a total number of $n$ boxes:
\begin{equation}
\label{eq:Young-tableaux-notation}
R =(Y_{\mathrm{left}}(n-p), Y_{\mathrm{right}}(p)).
\end{equation}
In this notation, the right Young tableau $Y_{\mathrm{right}}$ represents $p$ objects that are odd under the $\mathbb Z_2$ centre (odd powers of fields) transforming in an irreducible representation of the $p$-element symmetric (permutation) group $S_p$. The left Young tableau $Y_{\mathrm{left}}$ represents another $n-p$ objects that are even under $\mathbb Z_2$ (even powers of fields) transforming in an irrep of $S_{n-p}$. In this notation, the singlet and the vector representations are given by
\begin{equation}
S=([n],\bullet), \qquad
V= ([n-1],\text{\tiny \yng(1)}),
\end{equation}
where the notation $[k]$ means a row with $k$ boxes.
Using the Young tableaux notation \eqref{eq:Young-tableaux-notation}, it is possible to exhaustively list the irreps that exist for a given value of $n$. For instance, a quick combinatorial exercise shows that Table~\ref{tab:irrepsLowLying} is complete up to $n\leqslant4$. 
The Young tableaux notation makes the following properties clear:
\begin{itemize}
\item The dimension, following from a dimension formula given in \cite{Antipin:2019vdg}:
\begin{equation}
\dim(Y_{\mathrm{left}}(n-p), Y_{\mathrm{right}}(p))=\begin{pmatrix}n\\p
\end{pmatrix}\dim( Y_{\mathrm{left}})\dim(Y_{\mathrm{right}}),
\end{equation}
where $\dim( Y_{\mathrm{left}})$ and $\dim(Y_{\mathrm{right}})$ are computed using the standard hook rule formula for $S_{n-p}$ and $S_p$ respectively.
\item The rank of the irrep, which is given by the number of boxes in $Y_{\mathrm{right}}$, plus twice the number of boxes in rows $2,\,3,\,\ldots$ in $Y_{\mathrm{left}}$.
\item The symmetry properties between the ``$X$ building blocks'' in the irrep, which are given by rows $2,\,3,\,\ldots$ in $Y_{\mathrm{left}}$.
\end{itemize}
This concludes our brief overview of the representation theory of the hypercubic group. Details on the remaining irreps will be found throughout the draft.

\section{Perturbative setup for six-loop results}
\label{chapter2}

We initiate our study of the spectrum from the mostly familiar Lagrangian picture. This will allow us to extract the scaling dimensions of all possible operators built with up to four powers of the field and no derivatives.\footnote{In principle we can extract any operator using this methodology, but general group-independent explicit expressions in the literature are only known for the aforementioned set of operators.} Our methodology is rather straightforward, we will deform the free theory with all possible such operators. Then, we will calculate the beta function of each operator transforming in each irreducible representation of the group. Naturally, all quadratic and cubic (in powers of the field) couplings and their corresponding beta functions will be zero at the fixed point. All non-singlet quartic couplings will also be zero. However, the derivatives of the beta functions, which can be related to the scaling dimensions of the operators in the CFT, will not be zero. Note that in the following section we will consider all possible products of the $S$, $V$, $Z$ and $X$ representations, which mirror OPEs in CFT. These products will be sufficient to generate all the tensor structures needed to extract the scaling dimensions for our desired set of operators. As a by-product, we will also give more details on the construction of many of the irreps in table~\ref{tab:irrepsLowLying}.

\subsection{Six-loop beta functions}

Let us start by considering the following Lagrangian density \cite{Bednyakov:2021ojn}
\begin{equation}
\mathcal{L} =\frac{1}{2}\partial_\mu \phi_a \partial_\mu \phi_a - t_a \phi_a - \frac{m_{ab}^2}{2}\phi_a \phi_b - \frac{h_{abc}}{3!}\phi_a \phi_b \phi_c  -\frac{\lambda_{abcd}}{4!}\phi_a \phi_b \phi_c \phi_d \,,
\label{lagr}
\end{equation}
which contains the most general multi-scalar interactions with up to four powers of the field. Note that in principle one can also include the cosmological constant, however we will not use it in the present work, see \cite{Bednyakov:2021ojn}. The beta functions for the theory in \eqref{lagr} are known to six loop order. As we will see this suffices to extract the scaling dimensions to order $\varepsilon^6$. The methodology proceeds as follows: in $d=4-\varepsilon$ dimensions for a given global symmetry there exists a controlled fixed point for which the singlet quartic coupling(s) $\lambda^S_{abcd} = \sum\limits_{\mathrm{singlets}} g^S_i T^{S_i}_{abcd}$ are non-zero, and all other couplings are zero. The crucial observation is that the derivative $\frac{\partial \beta (g)}{\partial g}|_{F.P.} =\Delta -d $ (we let $g$ denote any coupling) evaluated at the fixed point is non-zero, and determines the scaling dimension as a power series in $\varepsilon$.

Let us be a bit more concrete. Assume that we would like to determine the dimension of some quartic non-singlet operator called $R = T^R_{abcd}\phi_a\phi_b\phi_c \phi_d$. We must thus deform the free theory with $\lambda_{abcd}=\lambda^R_{abcd}+\lambda^S_{abcd}$, where $\lambda^R_{abcd} = g^R T^R_{abcd}$. Note that the second term, $\lambda^S_{abcd}$, is needed in order to reach a non-trivial fixed point. The beta function is then 
\begin{equation}
	\frac{d \lambda_{abcd}}{d t} \equiv  \beta_{abcd}  = \beta^R_{abcd} + \beta^S_{abcd}, \qquad t = \log \mu. 
\label{beta1}
\end{equation}
Using a tensor that satisfies $P^R_{abcd} \lambda^{R}_{abcd} = g^R$ and  $P^R_{abcd} \lambda^{R^\prime}_{abcd} =0$ for any $R^\prime \neq R$, henceforth called a projector, we obtain
\begin{equation}
	\beta (g^R) = P^R_{abcd}\beta_{abcd}\,,
\label{beta2}
\end{equation}
where we have defined $\beta(g)=\frac{d g}{dt}$.
This is then sufficient to obtain the scaling dimension of the operator multiplying $g^R$ in the Lagrangian. For completeness, let us note that there may be more than one distinct operator in the same representation and with the same number of fields. This is indeed the case in the hypercubic theories we study in this work. In this case one must include all such operators in the computation. For example suppose that we had two $R$ operators and three $S$ operators. We would then have 
\begin{equation}
\lambda_{abcd}=\lambda^{R_1}_{abcd}+\lambda^{R_2}_{abcd}+\lambda^{S_1}_{abcd}+\lambda^{S_2}_{abcd}+\lambda^{S_3}_{abcd}\,.
\label{coupling1}
\end{equation}

In this case we would proceed with the same steps as above, the only difference being that the scaling dimensions in the $R$ representation would be obtained from the eigenvalues of the matrix 
\begin{equation}
\label{mixingmatrix}
\begin{pmatrix}
\frac{\partial \beta (g^{R_1})}{\partial g^{R_1}} & \frac{\partial \beta (g^{R_1})}{\partial g^{R_2}} \\ \frac{\partial \beta (g^{R_2})}{\partial g^{R_1}} & \frac{\partial \beta (g^{R_2})}{\partial g^{R_2}}
\end{pmatrix},
\end{equation}
evaluated, as usual, at the fixed point values of the couplings. 

In practice, one can use the data files included with \cite{Bednyakov:2021ojn}, in which the general six-loop $\beta_{abcd}$ is expressed in terms of products of $\lambda_{abcd}$ with all but four external indices contracted. For the hypercubic theory, we substitute into $\beta_{abcd}$ the following expression\footnote{Since only $\lambda_1$ and $\lambda_2$ can be non-zero at the fixed point, it is sufficent to keep only terms linear in $g^R$ within $\beta_{abcd}$. Remember that the couplings $g^R$ correspond to non-singlet deformations of the lagrangian \eqref{hypercubiclagrangian}.} 
\begin{equation}
	\lambda_{abcd} \to \lambda_1 \delta_{abcd}+\lambda_2 (\delta_{ab}\delta_{cd}+\delta_{ac}\delta_{bd}+\delta_{ad}\delta_{bc}) + \sum\limits_i g^{R_i} T^{R^i}_{abcd}. 
	\label{eq:lambda_subst}
\end{equation}
We contract the resulting $\beta_{abcd}$ with the required projector \eqref{beta2} to extract a particular $\beta(g^{R})$ 
entering \eqref{mixingmatrix}.
Since all $T^R_{abcd}$ (and $P^R_{abcd}$) in the hypercubic theory can be written in terms of fully symmetric Kronecker symbols $\delta^{(p)}_{a_1,a_2,...,a_p}$ (see below), index contraction can be efficiently carried out by repeated use of the rules 
\begin{equation}
	 \delta^{(p+1)}_{c a_1 a_2 \cdots a_p} \delta^{(q+1)}_{cb_{1} b_2 \cdots b_q} = \delta^{(p+q)}_{a_1 a_2 \cdots a_p b_1 b_2\cdots b_q}, \qquad
	 \delta^{(p+2)}_{c a_1 a_2 \cdots a_p c} = \delta^{(p)}_{a_1 a_2 \cdots a_n}, \qquad \delta^{(2)}_{cc} = \delta_{cc} = n.
		\label{eq:delta_rules}
\end{equation}
As a result, $\beta(g^R)$ becomes a scalar function of $\lambda_{1}, \lambda_{2}$, and $g^{R_i}$. 

Before proceeding, let us note that projectors are ubiquitous in the conformal bootstrap, where they are used to determine the sum rules which constrain the theory. An intuitive way to think of them, is that they are the tools used to ``inject'' information about the global symmetry into what would otherwise be a calculation for a generic theory. The same will be true in our present work, where they will "input" the global symmetry into the group independent results of \cite{Bednyakov:2021ojn}. 

\subsection{Tensor structures for OPE and crossing}

We will be interested in constructing all possible representations, as well as their corresponding tensor structures, that can appear in the product of two, three, or four powers of the order parameter field $\phi_a$, which transforms in the defining representation of the hypercubic group $C_n := S_n \ltimes (\mathbb Z_2)^n$. This specific analysis will be sufficient to construct all representations that contain operators with $\leq 4$ fields and no derivatives. These are precisely the operators for which the beta functions are known at six loops. The group theory for a generic group that may be parametrised as $S_n \ltimes G^n $, with $G$ arbitrary, has been analysed in \cite{Kousvos:2021rar}. Hypercubic theories are then a special case, with $G=\mathbb Z_2$.

The hypercubic group may also be seen as the symmetry of $n$ coupled Ising models, which may be permuted with each other. In this picture, each $\mathbb Z_2$ in $S_n \ltimes (\mathbb Z_2)^n$ corresponds to an Ising model, and $n$ labels the number of them. This point of view can be found in \cite{Komargodski:2016auf}; it is also the basis for the (controlled) large $n$ perturbation theory description \cite{Binder:2021vep}.

To start we will use the fact that tensor products are associative, using this we will be able to systematically calculate all possible representations that can appear in an arbitrary product of fields. It can be checked that our results nicely overlap with \cite{Antipin:2019vdg}. Note that in our case we will present the results in a fashion that allows an immediate generalisation to theories with a general $S_n \ltimes G^n$ symmetry. Our analysis is done in a way that also makes contact to the numerical bootstrap straightforward, where tensor products of fields mirror the operator product expansion (OPE).

\subsubsection{Products of two fields}\label{twofields}
We consider the tensor product of two fields, $\phi_a$ and $\phi_b$, which we decompose onto irreducible representations (irreps) of the hypercubic group. From e.g. \cite{Kousvos:2021rar} we have (we will not distinguish between upper and lower indices)
\begin{equation}
\phi_a \times \phi_b \sim \delta_{ab}S + X_{ab}+Z_{ab}+B_{ab}\,,
\label{2ptOPE}
\end{equation}
where each term on the RHS corresponds to a different irrep. The key observation now is that we can recover the form of a specific operator on the RHS of \eqref{2ptOPE} by contracting the left hand side with a projector\footnote{We call the the tensor structures we will use projectors, but some times we might not actually normalise them to have a unit product, i.e. $P^2 = P$. They will however always be orthogonal to each other.} of the global symmetry. More explicitly
\begin{equation}
R_{ab} = P_{abcd}^R \phi^c \phi^d,
\label{projcontaction}
\end{equation}
where repeated indices imply summation and the label $R$ represents a specific irrep we are interested in. The projectors are found by computing the tensor structure that corresponds to a 4-pt function $\langle \phi^a \phi^b \phi^c \phi^d \rangle $ when decomposed onto a specific irrep. A number of these have been computed due to their utility in the numerical conformal bootstrap. Let us quote the explicit form of the projectors that are relevant in the $\phi_a \otimes \phi_b $ product:
\begin{equation}
\begin{split}
& P^S_{abcd}=\frac{1}{n}\delta_{ab}\delta_{cd}, \\
& P^X_{abcd}=\delta_{abcd}-\frac{1}{n}\delta_{ab}\delta_{cd},\\
& P^Z_{abcd}=\frac{1}{2}(\delta_{ac}\delta_{bd}+\delta_{ad}\delta_{bc}-2\delta_{abcd}),\\
& P^B_{abcd}=\frac{1}{2}(\delta_{ac}\delta_{bd}-\delta_{ad}\delta_{bc}),
\end{split}
\label{4indexproj}
\end{equation}
where $\delta_{abcd}$ is just the 4-index generalization of $\delta_{ab}$, i.e. it is non-zero only if all indices are equal. Contracting these expressions onto a product of two order parameter fields we obtain our representation of interest. For clarity we present a representative element for each irrep:
\begin{equation}
\begin{split}
& S_{11} = P_{11cd}^S \phi^c \phi^d = (\phi_1^2 + \phi_2^2 + \ldots  + \phi_n^2), \\
& X_{11} = P_{11cd}^X \phi^c \phi^d = \phi_1^2 - \frac{1}{n} (\phi_1^2 + \phi_2^2 + \ldots  + \phi_n^2) ,\\
& Z_{12} = P_{12cd}^Z \phi^c \phi^d =\frac{1}{2}( \phi_1 \phi_2+\phi_2 \phi_1) = \phi_1 \phi_2, \\
& B_{12} = P_{12cd}^B \phi^c \phi^d =\frac{1}{2}( \phi_1 \phi_2- \phi_2 \phi_1) = 0.
\end{split}
\label{quadraticoperators}
\end{equation}
Notably the last operator can only be written down by inserting derivatives, hence it vanishes. The (group theoretical) dimensions of these representations are 
\begin{equation}
\begin{split}
&  \dim S = 1,\\
&  \dim X = n-1,\\
&  \dim Z = \frac{n(n-1)}{2},\\
&  \dim B = \frac{n(n-1)}{2},\\
\end{split}
\label{2ptdimensions}
\end{equation}
and, indeed 
\begin{equation}
\dim S + \dim X  + \dim Z + \dim B  = n^2.
\end{equation}
Say, for example, that we would like to calculate $\Delta_X$ i.e. the scaling dimension of the quadratic $X$ operator. To do this, we would deform the lagrangian of the theory with the term 
\begin{equation}
\delta \mathcal{L} = g P^X_{11ab} \phi^a \phi^b,
\end{equation}
and compute its beta function. Note that $gP^X_{11ab}$ is equal to $(m^2_{ab})^X$ in the notation of \eqref{lagr}. Of course, at the fixed point we should have $g=0$. But the derivative of the beta function with respect to the coupling should be non-zero and equal to $\Delta_X-d$, where $d$ is the dimensionality, in our case $d=4-\varepsilon$. More explicitly 
\begin{equation}
	\frac{\partial \beta(g)}{\partial g}\Big|_{F.P.} = \Delta_X -d.
\end{equation} 
Notably, the beta functions for these types of deformations can be recovered from the general six loop beta functions of \cite{Bednyakov:2021ojn}, by replacing $m^2_{ab} \rightarrow g P^X_{11ab}$. The above analysis is sufficient to recover the scaling dimensions of all spin-$0$ bilinear fields with no derivatives inserted.

\subsubsection{Products of three fields}
Having fully classified the product of two fields, let us proceed to the product of three fields. Using the  associativity of the tensor product we have 
\begin{equation}
\phi^a \phi^b \phi^c = \phi^a \times (\phi^b \times \phi^c) \sim \phi^a \times (\delta_{bc} S + X_{bc} + Z_{bc} + B_{bc}),
\label{threefields}
\end{equation}
where we have also plugged in the decomposition of two fields from the previous section. We must now analyse the RHS of \eqref{threefields} term by term. To compute operators in various representations we use the relation 
\begin{equation}
R_{abc}=(P_6)^R_{abcdef}\phi^d \phi^e \phi^f ,
\label{threefieldscontraction}
\end{equation}
where 
\begin{equation}
(P_6)^R_{abcdef}=P^R_{abcdgh}P^{R_2}_{ghef}.
\label{threepointprojectors}
\end{equation}
The structures with four indices on the RHS of \eqref{threepointprojectors} are precisely the ones computed in the previous section. In other words, $R_2$ takes as values either $S$ or $X$ or $Z$. Note that we omitted $B$ since it is identically zero. This can be further simplified noting the identity
\begin{equation}
P^{R_2}_{abcd}R_{cd}=R_{ab},
\end{equation}
which follows from the fact that by definition a projector projects an irrep back to itself. Thus we have
\begin{equation}
P^R_{abcdef}\phi^d R^{ef} =P^R_{abcdef}P^{R_2}_{efgh} \phi^d  R^{gh},
\end{equation} 
from which we conclude 
\begin{equation}
(P_6)^R_{abcdef}=P^R_{abcdgh}P^{R_2}_{ghef}=P^R_{abcdef}.
\label{threepointprojectorssimplified}
\end{equation}

It now remains to compute the structures with six indices $P^R_{abcdgh}$. These correspond to the tensor structures of the 4-pt function $\langle \phi_a R_{bc} \phi_d R_{ef} \rangle$ when decomposed onto irreps.

\paragraph{The product $\phi^a \otimes S$}
Let us start by analysing the $\phi_a \otimes S$ product. This one is rather simple since only one representation can appear on its RHS, namely the defining representation $\phi_a$. We have $P^V_{abcdef}=\delta_{ad} \delta_{bc}\delta_{ef}$. This in turn gives us the operator

\begin{equation}
V_{111}=\phi_1 (\phi_1^2 + \phi_2^2 + \ldots  + \phi_n^2),
\label{vectorinphis}
\end{equation}
which is part of the defining (or ``vector'') representation.

\paragraph{The product $\phi^a \otimes X_{bc}$}
The product $\phi^a \otimes X_{bc}$  can be decomposed onto two irreducible representations, the defining irrep $V$, and another one that we will call $XV$. For these we have
\begin{equation}
\begin{split}
& P^{V} _{abcdef}= \delta_{abcdef}-\frac{1}{n}(\delta_{bc}\delta_{adef}+\delta_{ef}\delta_{abcd})+\frac{1}{n^2}\delta_{ad}\delta_{bc}\delta_{ef}  ,\\
& P^{XV}_{abcdef}=-\delta_{abcdef}+\frac{n-1}{n}\delta_{ad}\delta_{bcef}+\frac{1}{n}(\delta_{bc}\delta_{adef}+\delta_{ef}\delta_{abcd})-\frac{1}{n}\delta_{ad}\delta_{bc}\delta_{ef},
\end{split}
\label{6ptphiX}
\end{equation}
where the irreps satisfy
\begin{equation}
\begin{split}
& \dim V = n,\\
& \dim XV = n(n-2),
\end{split}
\end{equation}
and indeed
\begin{equation}
\dim V + \dim XV = n(n-1).
\end{equation}
Using \eqref{threefieldscontraction} we obtain
\begin{equation}
\begin{split}
& V_{111} = \left(1-\frac{1}{n}\right)\phi_1 X_{11} \sim \phi_1 \left(\phi_1^2 -\frac{1}{n} (\phi_1^2 + \phi_2^2 +\ldots + \phi_n^2)\right),\\
& XV_{122} = \frac{n-1}{n}\phi_1 X_{22} + \frac{1}{n}\phi_1 X_{11}.
\end{split}
\end{equation}
Note that the expression for $V_{111}$ seems to differ from the expression for $V_{111}$ in \eqref{vectorinphis}, this is because a given operator can appear with more than one independent structures. A different structure for this operator will also be found in the $\phi_a \otimes Z_{bc}$ product.

\paragraph{The product $\phi_a \otimes Z_{bc}$}
In the $\phi_a \otimes Z_{bc}$ product, the irreps $V$ and $XV$ appear again, and in addition we have two new irreps which we call $VB$ and $Z_3$. Since these are rather lengthy we write the expressions for their tensor structures in Appendix~\ref{ApppendixPhiZ}. However, we may briefly explain the qualitative structure of the product. Firstly, we need to separate two different cases, since they will not mix under action of the group. The first case corresponds to when the index $a$ is equal to one of the two indices $b$ or $c$. The second case corresponds to all three indices being different. In the first case we have

\begin{equation}
\phi_a Z_{bc} \sim \left(\frac{\delta_{ab}}{n-1}\phi_r Z_{rc}+\left(\phi_a Z_{bc}-\frac{\delta_{ab}}{n-1}\phi_r Z_{rc}\right)\right)+(b \leftrightarrow c);
\label{phizaeqb}
\end{equation}
the specific $\frac{1}{n-1}$ factor, instead of $\frac{1}{n}$, is due to the fact that necessarily $b \neq c$ as observed in section~\ref{twofields}. The first term on the RHS of \eqref{phizaeqb} corresponds to the defining ($V$) representation, the second corresponds to the $XV$ representation. We proceed to the second case now, the decomposition is schematically
\begin{equation}
\phi_a Z_{bc} \sim (\phi_a Z_{bc}+\phi_b Z_{ac}+\phi_c Z_{ba})+(\phi_a Z_{bc}-\phi_b Z_{ac})+(\phi_a Z_{bc}-\phi_c Z_{ba}).
\label{phizaneqb}
\end{equation}
The first term on the RHS of \eqref{phizaneqb} now corresponds to a totally symmetric irrep which we call $Z_3$, whereas the last two terms correspond to what we call $VB$. The $VB$ operators are quite interesting in the sense that using three fields only, there are no operators at spin $0$. 

The aforementioned irreps have the following dimensions 
\begin{equation}
\begin{split}
& \dim V = n,\\
& \dim XV = n(n-2),\\
& \dim VB = \frac{n(n-1)(n-2)}{3},\\
& \dim Z_3 = \frac{n(n-1)(n-2)}{6},\\
\end{split}
\label{3ptdimensions}
\end{equation}
which add up to $\frac{n^2(n-1)}{2}$ as expected.

Let us now comment on the fact that the defining representation $V$ seems to appear with more than one structures. More concretely, we have seen it appear as $\phi_1^3 -\frac{\phi_1}{n}(\phi_1^2 +\ldots )$ and $\phi_1(\phi_1^2 + \ldots )$. In the $\phi_a \otimes Z_{bc}$ product, it now also appears as $\phi_1 (\phi_2^2+\ldots  + \phi_n^2)$, however we see that this is a linear combination of the previous two forms. We are thus left with two linearly independent combinations, which is indeed what is expected. The two combinations should lead to a beta function in matrix form that will have two eigenvectors. The one eigenvector will be proportional to the equation of motion and have a scaling dimension equal to $d-\Delta_{\phi}$. This operator is a redundant operator.\footnote{Note that $\phi^3$ on its own is not a scaling eigenoperator, so it cannot be classified as a primary or a descendant. The actual scaling eigenoperator is proportional to the equation of motion, and is thus redundant and needs to be removed from the spectrum of the CFT. The dimension of this operator can of course be computed with our methods, and satisfies the relation $\Delta_{\mathrm{redundant}} = d- \Delta_{\phi}$, as required for the equation of motion operator. We thank Andreas Stergiou for discussions on this point.  A detailed analysis on the implications of such operators will appear elsewhere. \label{Redundant}} The other operator is not redundant and has the second lowest scaling dimension in the defining representation after $\phi_a$ itself.

To read off the scaling dimensions of ``$\phi^3$ type'' operators we may again use the results of \cite{Bednyakov:2021ojn}, this time making the replacement $h_{abc} \rightarrow g(P_6)^R_{111abc}$.\footnote{We pick the index values "$111$" for presentation purposes, it suffices that we pick a set of index values for which the operator of choice is not identically zero.}

We have now concluded the analysis of all possible operators that can be built with three powers of the order parameter field and no derivatives.

\subsubsection{Products of four fields}
Given that the products of four fields quickly become rather complicated we present them in Appendix~\ref{fourfields}.

\subsection{Summary of irreps}
Below we summarise all irreps that can appear by taking  all possible products between irreps of up to rank two (modulo the antisymmetric irrep). We also give their corresponding dimensions, and denote which of the products of $S$, $V$, $X$ and $Z$ they can appear in:
\begin{align}
  \dim  S =1 &: \lsp \lsp \lsp \lsp  V  \otimes  V  , \lsp S \otimes S ,\lsp X \otimes X , \lsp Z \otimes Z ,\nonumber\\
  \dim  X =n-1 &: \lsp \lsp \lsp \lsp  V  \otimes  V  , \lsp S \otimes X ,\lsp X \otimes X , \lsp Z \otimes Z , \nonumber\\
    \dim  Z =\frac{n(n-1)}{2} &: \lsp \lsp \lsp \lsp  V  \otimes  V  , \lsp S \otimes Z ,\lsp X \otimes Z , \lsp Z \otimes Z , \nonumber\\
  \dim  B =\frac{n(n-1)}{2} &: \lsp \lsp \lsp \lsp  V  \otimes  V  , \lsp Z \otimes Z , \lsp Z \otimes X , \nonumber\\
   \dim  V  =n &: \lsp \lsp \lsp \lsp  V  \otimes S , \lsp  V  \otimes X , \lsp  V  \otimes Z  ,\nonumber\\
  \dim XV =n(n-2) &: \lsp \lsp \lsp \lsp  V  \otimes X , \lsp  V  \otimes Z  , \nonumber\\
 \dim VB =\frac{n(n-1)(n-2)}{3} &: \lsp \lsp \lsp \lsp   V  \otimes Z , \nonumber\\
  \dim Z_3 =\frac{n(n-1)(n-2)}{6} &: \lsp \lsp \lsp \lsp   V  \otimes Z , \nonumber\\
 \dim XX =\frac{n(n-3)}{2} &: \lsp \lsp \lsp \lsp  X \otimes X , \lsp Z \otimes Z , \nonumber\\
 \dim \XXbar =\frac{(n-1)(n-2)}{2} &: \lsp \lsp \lsp \lsp  X \otimes X  ,\nonumber\\
 \dim XZ =\frac{n(n-1)(n-3)}{2} &: \lsp \lsp \lsp \lsp  Z \otimes X , \lsp Z \otimes Z,  \nonumber\\
 \dim XB =\frac{n(n-1)(n-3)}{2} &: \lsp \lsp \lsp \lsp   Z \otimes Z  ,\nonumber\\
 \dim Z_4 =\frac{n(n-1)(n-2)(n-3)}{24} &: \lsp \lsp \lsp \lsp   Z \otimes Z ,\nonumber\\
  \dim BB =\frac{n(n-1)(n-2)(n-3)}{12} &: \lsp \lsp \lsp \lsp   Z \otimes Z ,  \nonumber\\
 \dim BZ =\frac{n(n-1)(n-2)(n-3)}{8} &: \lsp \lsp \lsp \lsp   Z \otimes Z  .
\end{align}
Out of the above, we will be able to extract the scaling dimensions of $S$, $X$, $Z$, $B$, $V$, $XV$, $Z_3$, $XX$, $XZ$ and $Z_4$ to six loops from the Lagrangian as outlined in the previous section. Our results agree with those of \cite{Antipin:2019vdg}, as well as the ones in section~5.1.1 of \cite{Osborn:2017ucf}, but extend to many more orders in perturbation theory. For the rest of the operators, we outline techniques to compute their scaling dimensions at one loop and compute a large number of them in the following sections.

\section{Perturbative setup for generic one-loop results}
\label{chapter3}

We now turn to more general calculations, in which we compute anomalous dimensions of operators built with generic numbers of fields $\phi$, as well as generic insertions of derivatives. Additionally, we show how the group theory information can be encoded in auxiliary vectors, an alternative but completely equivalent formulation to projectors. We also point out how the two are related. For earlier incarnations of the method we will be using, see \cite{Kehrein:1992fn,Kehrein:1994ff}, as well as \cite{Liendo:2017wsn,Antipin:2019vdg}.

In practice, we will be using the implementation of \cite{Hogervorst:2015akt,Hogervorst:2015tka}, whose starting point is conformal perturbation theory. It uses the fact that the leading-order correction to the scaling dimension of a primary operator $\mathcal O$ is proportional to the OPE coefficients of said operator with the operators driving the flow (i.e. the "$\phi^4$" singlet operators, when in $d=4-\varepsilon$), 
\begin{equation}
\label{eq:gammaOfPrimary}
\gamma_{\mathcal O}=K \tensor C{^\O_{\O_{\mathrm{int}}\O}} +O(\eps^2).
\end{equation}
Here we consider quartic interactions of the form
\begin{equation}
\mathcal O_{\mathrm{int}}=\frac1{4!}\left(\lambda_1\delta_{abcd}+\lambda_2 (\delta_{ab}\delta_{cd}+\delta_{ac}\delta_{bd}+\delta_{ad}\delta_{bc})\right)\phi^a\phi^b\phi^c\phi^d
\end{equation} and we work in conventions where $K=\frac{\pi^2}{12}$ (consistent with \cite{Hogervorst:2015akt}).

To leading order in small $\varepsilon$, the couplings take the values
\begin{align}
\text{Hypercubic}:&& \frac{\lambda_1}{16\pi^2}&=\frac{n-4}{3n}\varepsilon ,&   \frac{\lambda_2}{16\pi^2}&=\frac\varepsilon{3n}
\\
O(N):&&  \frac{\lambda_1}{16\pi^2}&=0,& \frac{\lambda_2}{16\pi^2}&=\frac\varepsilon{N+8}, 
\\
\text{Decoupled Ising}:&&  \frac{\lambda_1}{16\pi^2}&=\frac\varepsilon3,&   \frac{\lambda_2}{16\pi^2}&=0.
\end{align}
From these couplings it is clear that the one-loop spectrum at $n=4$ agrees with that of $O(N)$ at $N=4$, as pointed out in \cite{Antipin:2019vdg}.

In \eqref{eq:gammaOfPrimary}, we used the fact that for a scaling operator, the correction is directly proportional to the OPE coefficient, which requires first constructing the scaling operator under consideration. In practice, it is convenient to use the OPE with the interaction term as a means to construct the scaling operators, which in turn are either primaries or descendants. This introduces a matrix (called the one-loop anomalous dimension operator) acting on a set of operators
\begin{equation}
\label{eq:setofoperatorsEx}
\hat \Gamma \O_i =K \sum_{\O_j\in \mathcal X_{p,R,L}}\tensor C{^{\O_i}_{\O_{\mathrm{int}}\O_j}}\O_j.
\end{equation}
Here the sum goes over a set $ \mathcal X_{p,R,L}$ of operators (not necessarily scaling operators) of the same global symmetry representation ($R$) and Lorentz ($L$) irreps, and the same number of fields $p$ as $\O_i$. The restriction to this subset is because for $\phi^4$ theories, we can restrict the one-loop computation to consider only operators with the same number of fields $\phi$ and gradients $\partial$ as the operator $\O_i$.

As an example, consider singlet ($R=S$) scalar ($L=\{\ell=0\}$) operators with four fields. Hence
\begin{equation}
\mathcal X_{4,S,\ell=0}=\{\O_1=(\delta_{ij}\phi^j\phi^j)^2,\,\O_2=\delta_{ijkl}\phi^i\phi^j\phi^k\phi^l\}.
\end{equation}
In this case, it is straight-forward to compute the OPE coefficients via Wick contractions\footnote{Note that in these expressions, the operators are not canonically normalised and therefore the OPE coefficients with one raised index are not symmetric, which can indeed be seen from \eqref{eq:GactionExample}.}
\begin{align}
K\tensor C{^{\O_1}_{\O_{\mathrm{int}}\O_1}} & = \frac{2 (n-4)}{3 n}+\frac{16}{3 n}+\frac{2}{3},
&
K\tensor C{^{\O_1}_{\O_{\mathrm{int}}\O_2}} & = \frac{4 (n-4)}{3 n},
\nonumber
\\
K\tensor C{^{\O_2}_{\O_{\mathrm{int}}\O_1}} & = \frac{2}{n},
&
K\tensor C{^{\O_2}_{\O_{\mathrm{int}}\O_2}} & = \frac{2 (n-4)}{n}+\frac{4}{n},
\label{eq:GactionExample}
\end{align}
where we ommitted the overall factor $\varepsilon$. 
This gives the eigenvalues $\frac{4(n-1)}{3n}$ and $2$, and the corresponding eigenoperators (left eigenvectors) are
\begin{equation}
(\phi^4)_{\mathrm{rel.}}=-\O_1+2\O_2, \qquad (\phi^4)_{\mathrm{irr.}}=3\O_1+(n-4)\O_2.
\end{equation}
At this point, there are no descendant operators. 

When considering operators with derivatives, many of the eigenoperators will correspond to descendants. The subset of primaries among them can be found by considering the subspace of eigenoperators, inserted at $0$, that are annihilated by the generator $K^\mu$ of special conformal transformations. It is defined such that it annihilates the field $\phi(0)$, 
\begin{equation}
[K_\mu,\phi(0)]=0,
\end{equation}
and satisfies standard commutation relations with $P_\mu=i\partial_\mu$ and the other generators of the conformal algebra (see e.g. \cite{Poland:2018epd}).

For operators in non-trivial Lorentz irreps, we make use of the implementation of \cite{Hogervorst:2015akt} to compute the OPEs, see appendix~C of \cite{Hogervorst:2015tka}, as well as \cite{Henriksson:2022rnm}, for more details. The main complication in this work is to deal with the global symmetry irreps, to which we now turn.

\subsection{Implementation with tensor structures}

We will write operators inside $\mathcal X_{p,R,L}$ as 
\begin{equation}
\O=\mathbf R_{i_1\cdots i_p}\mathbf L_{\mu_1\cdots\mu_s} \partial^{\mu_1}\cdots \phi^{i_1}\partial^{\mu_a}\cdots \phi^{i_2}\cdots \partial^{\mu_b}\phi^{i_p}
\end{equation}
and consider all possible ways of distributing the derivatives on the $p$ fields. In this paper we limit ourselves to traceless symmetric spin-$\ell$ Lorentz irreps -- for which we use $\mathbf L_{\mu_1\cdots\mu_s}=l_{\mu_1}l_{\mu_2}\cdots l_{\mu_\ell}\eta_{\mu_{\ell+1}\mu_{\ell+2}}\cdots \eta_{\mu_{s-1}\mu_s}$, with $l^\mu l_\mu=0$ -- although it is also possible to consider more general Lorentz irreps.

For the global symmetry, one needs to construct tensor structures $\mathbf R_{i_1\cdots i_p}$ appropriate for the symmetry group under consideration. These can be taken to be proportional to projectors, but as we will see there are also more compact choices. Knowing the symmetry/antisymmetry properties of each irrep we can write down the tensor structures directly. 

For finite groups, as will be described a bit later on, there is an alternative method using the character table. Here instead we explain how to construct tensor structures for the hypercubic group, where we keep $n$ generic, or equivalently, large enough to avoid accidental relations at finite $n$.\footnote{For finite $n$, there are relations between the Kronecker symbols, see for example equation~2.22 in \cite{Kousvos:2018rhl}.}

First we introduce a set of structures for the irrep $S$. They simply consist of rank-$p$ Kronecker symbols, 
\begin{equation}
\label{eq:singletstructure}
\mathbf S^{(p)}_{a_1a_2\cdots a_{p}}=\delta^{(p)}_{a_1a_2\cdots a_p},
\end{equation}
for $p$ even. 
A general structure for singlet operators is a product of such structures. For instance, in the example in \eqref{eq:setofoperatorsEx} above, we have $\phi^{a_1}\phi^{a_2}\phi^{a_3}\phi^{a_4}$ contracted with $\mathbf S^{(2)}_{a_1a_2}\mathbf S^{(2)}_{a_3a_4} $ to form $\O_1$ and with $\mathbf S^{(4)}_{a_1a_2a_3a_4}$ to form $\O_2$.

In general, given a structure, $\mathbf R_{i_1\cdots i_p}$ for an irrep $R$, one can always form larger structures of the same irrep by adjoining the singlet structures~\eqref{eq:singletstructure}. This is easy to understand, in the sense that the invariant tensors of the group are spanned by Kronecker deltas. Hence, multiplying an irrep/operator by more singlet structures we do not alter any of its symmetry properties. The difference with $O(n)$ is that one can now have Kronecker deltas with more than two indices. Operators built by adjoining more structures will in turn correspond to operators with more powers of $\phi$, i.e. with higher scaling dimensions.

As an example let us consider the $Z_{12}$ element of the $Z$ representation of \eqref{quadraticoperators} for two, four and six powers of the defining field. We have 
\begin{equation}
\begin{split}
 \mathcal X_{2,Z,0} &=\{\phi_1\phi_2 \} ,
\\
 \mathcal X_{4, Z,0} &=\{\phi_1\phi_2^3,\, \phi_1\phi_2(\delta_{ij}\phi_i \phi_j) \},
\\
 \mathcal X_{6,Z,0} &=\{\phi_1 \phi_2^5,\, \phi_1^3 \phi_2^3,\, \phi_1 \phi_2 (\delta_{ij}\delta_{kl}\phi_i \phi_j \phi_k \phi_l),\,\phi_1 \phi_2 (\delta_{ijkl}\phi_i \phi_j \phi_k \phi_l), \,\phi_1 \phi_2^3 (\delta_{ij}\phi_i \phi_j)  \},
\end{split}
\label{eq:operatorbasesZ}
\end{equation}
where it is understood that one should symmetrise uncontracted indices. Below we will see how to write these representatives with the use of tensors.

We will now describe how to construct the structures for the different irreps. By going through each irrep one-by-one, the corresponding tensor structure will illuminate our naming conventions for the irreps.
\begin{description}
\item[Structures for irrep $\boldsymbol V$] We introduce
\begin{equation}
\mathbf{V}^{(n)}_{a_1a_2\cdots a_{n}}=v^b\delta^{(n+1)}_{ba_1a_2\cdots a_n},
\end{equation}
where $v$ is an ancillary vector. Notice that if we put $v=(1,0,0,\ldots ,0)$ this is equivalent to simply plugging in the index value $b=1$. Hence, auxiliary vectors will allow us to avoid making reference to explicit index values. They will also allow us to symmetrize indices in a compact way. 
\end{description}

\paragraph{Structures for rank-2 irreps}

There are three irreps of rank 2, corresponding to the trace-free diagonal part ($X$), the symmetric part ($Z$) and the anti-symmetric part ($B$) of the matrix $\phi^i\phi^j$. 

\begin{description}
\item[Irrep $\boldsymbol X$] For this irrep we want to impose the condition that the tensor is traceless and that the transforming indices are equal. We introduce an ancillary vector $x_a$ satisfying $x^ax_a=0$ in order to enforce tracelessness, and define the structures
\begin{equation}
\label{eq:structureX}
\mathbf X^{(p)}_{a_1a_2\cdots a_p}=x^cx^d\delta^{(p+2)}_{cda_1\cdots a_p},
\end{equation}
for $p$ even. Notice that this is equivalent to contracting the projectors of the $X$ representation with the auxiliary vectors $x^a$. Similar statements will apply to the other irreps as well. Alternatively, it is also convenient to implement the properties for the $\mathbf X^{(p)}_{a_1a_2\cdots a_p}$ tensor directly without reference to the auxilliary vector $x^i$.\footnote{These properties follow from the definition \eqref{eq:structureX} and include ${\mathbf X^{(2)}_a}^a=0$, ${\mathbf X^{(p)}_{ab_1b_2\cdots b_{p-2}}}^a=\mathbf X^{(p-2)}_{b_1b_2\cdots b_{p-2}}$, ${\mathbf X^{(p)}_{a_1\cdots a_{p-1}}}^{b}\delta^{(q)}_{bc_1\cdots c_{q-1}}=\mathbf X^{(p+q-2)}_{a_1\cdots c_{q-1}}$.}
\item[Irrep $\boldsymbol Z$] For this irrep we want to impose that the indices are symmetric but necessarily take unequal values. We therefore introduce an ancillary vector $z_a$ that satisfies $z^az_a=0$ and $z^az^b\delta^{(p)}_{abc_1\ldots c_{p-2}}=0$. We then define the structures 
\begin{equation}
\mathbf Z^{(p,q)}_{a_1a_2\cdots a_{p}b_1b_2\cdots b_q}=z^bz^c\delta^{(p+1)}_{ca_1a_2\cdots a_p}\delta^{(q+1)}_{db_1b_2\cdots b_q},
\end{equation}
for $p$ and $q$ odd.
\item[Irrep $\boldsymbol B$] Here we introduce a rank-$2$ antisymmetric tensor $b_{cd}=-b_{dc}$, and define the structures
\begin{equation}
\mathbf B^{(p,q)}_{a_1a_2\cdots a_{p}b_1b_2\cdots b_q}=b^{cd}\delta^{(p+1)}_{ca_1a_2\cdots a_p}\delta^{(q+1)}_{db_1b_2\cdots b_q},
\end{equation}
for $p$ and $q$ odd.
\end{description}

\paragraph{Structures for rank-3 irreps}
The construction of the completely antisymmetric and completely symmetric irreps of arbitrary rank proceeds in the expected way. 
\begin{description}
\item[Irrep $\boldsymbol{B_3}$] We introduce
\begin{equation}
\mathbf{B3}^{(m,n,p)}_{a_1a_2\cdots a_mb_1b_2\cdots b_nc_1c_2\cdots c_p}=b^{ijk}\delta^{(m+1)}_{ia_1a_2\cdots a_m}\delta^{(n+1)}_{jb_1b_2\cdots b_n}\delta^{(p+1)}_{kc_1c_2\cdots c_p},
\end{equation}
where $b^{ijk}$ is completely antisymmetric.
\item[Irrep $\boldsymbol{Z_3}$] We introduce
\begin{equation}
\mathbf{Z3}^{(m,n,p)}_{a_1a_2\cdots a_mb_1b_2\cdots b_nc_1c_2\cdots c_p}=z^iz^jz^k\delta^{(m+1)}_{ia_1a_2\cdots a_m}\delta^{(n+1)}_{jb_1b_2\cdots b_n}\delta^{(p+1)}_{kc_1c_2\cdots c_p}.
\end{equation}
\end{description}

More interesting are the remaining rank-3 irreps, which can be seen as composites. Here we use the fact that we do not need to construct the entire irrep, just a representative, and therefore we can use auxiliary identities to  annihilate interfering terms (traces).
\begin{description}
\item[Irrep $\boldsymbol{VB}$] For the irrep $VB$, we introduce the structure
\begin{equation}
\mathbf{VB}^{(m,n,p)}_{a_1a_2\cdots a_mb_1b_2\cdots b_nc_1c_2\cdots c_p}=\mathbf{V}^{(m)}_{a_1a_2\cdots a_{m}}
\mathbf B^{(n,p)}_{b_1b_2\cdots b_{n}c_1c_2\cdots c_p}
\end{equation}
and demand that $v_ib^{ij}=0$.
\item[Irrep $\boldsymbol{XV}$] For the irrep $XV$ we introduce
\begin{equation}
\mathbf{XV}^{(m,n)}_{a_1a_2\cdots a_mb_1b_2\cdots b_n}=\mathbf X^{(m)}_{a_1a_2\cdots a_m} \mathbf{V}^{(n)}_{b_1b_2\cdots b_{n}}
\end{equation}
and demand that $v_ax^a=0$ and $v^ax^b\delta^{(p)}_{abc_1\ldots c_{p-2}}=0$.
\end{description}

\paragraph{Structures for rank 4 irreps} For rank 4 irreps the construction of the appropriate structures follows exactly the same logic. Given their more extended nature, they are given in appendix~\ref{app:moreStructuresDil}.

\paragraph{Example}
As a comprehensive example, let us consider the determination of scalar $Z$ operators up to $\Delta=6$. 

The operator bases \eqref{eq:operatorbasesZ} can now be written as
\begin{align}
 \mathcal X_{2,Z,0} &=\{\mathbf Z^{(1,1)}_{ab}\phi^a \phi^b \} ,
\\
 \mathcal X_{4, Z,0} &=\{\mathbf Z^{(1,3)}_{abcd},\,\mathbf Z^{(1,1)}_{ab}\delta_{cd} \} \phi^a\phi^b\phi^c\phi^d,
\\
 \mathcal X_{6,Z,0} &=\{\mathbf Z^{(1,5)}_{abcdef},\,\mathbf Z^{(3,3)}_{abcdef},\,\mathbf Z^{(1,3)}_{abcd}\delta_{ef},\,\mathbf Z^{(1,1)}_{ab}\delta_{cdef},\,\mathbf Z^{(1,1)}_{ab}\delta_{cd}\delta_{ef} \}\phi^a\phi^b\phi^c\phi^d\phi^e\phi^f.
\label{eq:operatorBasPost60}
\end{align}
In these bases, we find the matrices
\begin{align}
\Gamma\big|_{ \mathcal X_{2,Z,0}}&=\begin{pmatrix}
\frac{2}{3n}
\end{pmatrix}
,\qquad 
\Gamma\big|_{ \mathcal X_{4,Z,0}}=\begin{pmatrix}
 1 & \frac{1}{n} \\
 \frac{4 (n-4)}{3 n} & \frac{2 (n+6)}{3 n}
\end{pmatrix},
\\
\Gamma\big|_{ \mathcal X_{6,Z,0}}&=\begin{pmatrix}
 \frac{10 (n-1)}{3 n} & 0 & \frac{10}{3 n} & 0 & 0 \\
 0 & \frac{2 (n+1)}{n} & \frac{2}{n} & 0 & 0 \\
 \frac{2 (n-4)}{n} & \frac{2 (n-4)}{3 n} & \frac{5 n+22}{3 n} & 0 & \frac{1}{n} \\
 \frac{8 (n-4)}{3 n} & 0 & \frac{8}{3 n} & \frac{2 (n+1)}{n} & \frac{2}{n} \\
 0 & 0 & \frac{8 (n-4)}{3 n} & \frac{4 (n-4)}{3 n} & \frac{2 (2 n+17)}{3 n}
\end{pmatrix}.
\end{align}
The entry $\frac2{3n}$ of $\Gamma\big|_{ \mathcal X_{2,Z,0}}$ and eigenvalues $\frac{12+5n\pm \sqrt{n^2+24n-48}}{6n}$ of $\Gamma\big|_{ \mathcal X_{4,Z,0}}$ agree perfectly with previous determinations in \cite{Osborn:2017ucf,Antipin:2019vdg}. They also agree with the leading term in the six-loop result below. The eigenvalues of the last matrix take no simple form, and must be stored as roots of a degree-five polynomial.\footnote{The large-$n$ expansion of the roots of this polynomial is a bit peculiar. We find that two roots agree at the first two orders: $\gamma=2+\frac6n+\ldots$. Because of this, the expansion produces a non-integer power of $n$, and we find that the roots expand as $\gamma=2+\frac6n\pm\frac{\sqrt{96}}{n^{3/2}}+\frac{14}{n^2}+\ldots$}
At $n=3$, the five roots of the polynomial take the values $\{2.4672,\,2.76274,\,\frac{28}9,\,3.87613,\,4.56059\}$. The eigenvalue $\frac{28}9$ however does not correspond to a local operator $n=3$ due to finite-$n$ effects, and should therefore be removed from the $n=3$ spectrum.\footnote{We can understand this by noting that for $n=3$, the structure $\mathbf Z^{(1,5)}_{abcdef}$ becomes linearly dependent on the other structures in \eqref{eq:operatorBasPost60}, which can be understood from equation~2.22 of \cite{Kousvos:2018rhl}. In practice, we perform these type of checks by implementing an alternative computation for $n=3$ and $n=4$, described in section~\ref{sec:implFiniteGroup} below, and compare the lists.} 

Finally, let's consider the case of four fields and a pair of contracted derivatives. We work with the basis
\begin{align}
\nonumber
\mathcal X_{4,Z,\eta}=\{&\mathbf Z^{(1,3)}_{abcd} \partial_\mu\phi^a\partial^\mu\phi^b\phi^c\phi^d,\,\mathbf Z^{(1,3)}_{abcd} \phi^a\partial_\mu\phi^b\partial^\mu\phi^c\phi^d,\,\mathbf Z^{(1,1)}_{ab}\partial_\mu\phi^a\partial^\mu\phi^b\phi^c\phi^c,\\&\mathbf Z^{(1,1)}_{ab}\partial_\mu\phi^a\phi^b\partial^\mu\phi^c\phi^c,\,\mathbf Z^{(1,1)}_{ab}\phi^a \phi^b\partial_\mu\phi^c\partial^\mu\phi^c\},
\end{align}
where the notation $\eta$ reminds us that there is a pair of Lorentz indices contracted by $\eta_{\mu\nu}$. The matrix is
\begin{equation}
\Gamma\big|_{ \mathcal X_{4,Z,\eta}}=\begin{pmatrix}
\frac{3 n-2}{3 n} & \frac{2}{3 n} & \frac{1}{3 n} & \frac{2}{3 n} & 0 \\
 \frac{2}{3 n} & \frac{3 n-2}{3 n} & 0 & \frac{2}{3 n} & \frac{1}{3 n} \\
 \frac{4 (n-4)}{3 n} & 0 & \frac{2 (n+2)}{3 n} & \frac{8}{3 n} & 0 \\
 \frac{2 (n-4)}{3 n} & \frac{2 (n-4)}{3 n} & \frac{2}{3 n} & \frac{2 (n+4)}{3 n} & \frac{2}{3 n} \\
 0 & \frac{4 (n-4)}{3 n} & 0 & \frac{8}{3 n} & \frac{2 (n+2)}{3 n}
\end{pmatrix}.
\end{equation}
We find eigenvalues $1, 2\times \frac23, \frac{12+5n\pm \sqrt{n^2+24n-48}}{6n}$. For each eigenvalue, we then construct a basis in the corresponding (left) eigenspace, and act with the generator $K_\nu$ of special conformal transformations. We find that in the two-dimensional eigenspace $\frac23$, $K_\nu$ has a one-dimensional kernel, and we therefore conclude that there must be a primary operator of the form $\square\phi^4$ and anomalous dimension $\frac23$. The eigenspaces of the other eigenvalues have full rank after the action of $K_\nu$, and thus do not correspond to primaries.\footnote{The remaining eigenvalues are descendants of spin-1 operators $\partial\phi^4$ (dimensions $1$ and $\frac23$) and the scalars $\phi^4$ we saw above. Note that the spin-$1$ operator with anomalous dimension $1$ is removed from the spectrum since it becomes related to the broken spin-$2$ current $\mathcal J_{2,Z}$.}

In total, we have now completed the computation of the data found in table~\ref{tab:irrepZscalar}, which shows the nine scalar $Z$ operators with 4d dimension $\leqslant6$.

\subsection{Implementation for finite groups}
\label{sec:implFiniteGroup}

In the case of a finite group, there is an alternative approach to the use of tensor structures. In this approach, one considers a basis that can capture all possible representations, and only after the eigenvalues are found, one determines the decomposition into irreps of the corresponding eigenvectors. This is done using the character table, and the transformation properties of the field $\phi^i$ under the elements of the group.
In this section, we will explain the general theory, keeping the cubic group at $n=3$ as the main example. For the computations in the results section of this paper, we have performed the computation in the cases of $n=3$ and $n=4$. 

Let us present the character table and other useful information for the cubic group $C_3$, which in the literature is known under the conventional name $O_h$. In appendix~\ref{app:n4reps}, we repeat this for $n=4$.

The character table of $O_h$ is given in table~\ref{tab:characterTable}. Here we show our symbols for the irreps, alongside the standard literature symbols (recall the meaning of the subscripts $g$ ``even'', $u$, ``uneven'' referring to the transformation under the global $\mathbb Z_2$ subgroup which acts on all fields simultaneously).
\begin{table}
\centering
\caption{Character table for $C_3=O_h$. We give the names of the irreps in the notation used in this paper, and of the irreps and conjugacy classes in standard literature notation. Note that $J_{ij}$ is shorthand for $\phi_i \phi_j - \phi_j \phi_i $.}\label{tab:characterTable}
{\small
\renewcommand{\arraystretch}{1.25}
\begin{tabular}{|cc|cccccccccc|c|}
\hline 
\multicolumn{2}{|c|}{Irrep} & $E$ & $8C_3$ & $3C_2$& $6C_2$& $6C_4$  & $i$ &  $8S_3$ & $3\sigma_h$& $6\sigma_d$& $6S_4$  & Functions
\\
\hline
$S$  & 
$A_{1g}$ &  
 1 & 1 & 1 & 1 & 1 & 1 & 1 & 1 & 1 & 1 & $\phi_1^2+\phi_2^2+\phi_3^2$
\\
$\XXbar$  & 
$A_{2g}$  &
 1 & 1 & 1 & -1 & -1 & 1 & 1 & 1 & -1 & -1 
&
\\
$X$  & 
 $E_g$  &
 2 & -1 & 2 & 0 & 0 & 2 & -1 & 2 & 0 & 0 
& $(\phi_1^2-\phi_2^2,\phi_1^2-\phi_3^2)$
\\
$B$  & 
$T_{1g}$ & 
 3 & 0 & -1 & -1 & 1 & 3 & 0 & -1 & -1 & 1 
& $(J_{23},J_{31},J_{12})$
\\
$Z$  & 
$T_{2g}$ & 
 3 & 0 & -1 & 1 & -1 & 3 & 0 & -1 & 1 & -1 
& $(\phi_1\phi_3,\phi_2\phi_3,\phi_1\phi_2)$
\\
$B_3$  & 
$A_{1u}$ &
 1 & 1 & 1 & 1 & 1 & -1 & -1 & -1 & -1 & -1 
& $\epsilon^{ijk}\phi_i\phi_j\phi_k$
\\
$Z_3$  &  
$A_{2u}$ &
 1 & 1 & 1 & -1 & -1 & -1 & -1 & -1 & 1 & 1 
& $\phi_1\phi_2\phi_3$
\\
$VB$  & 
 $E_u$ & 
 2 & -1 & 2 & 0 & 0 & -2 & 1 & -2 & 0 & 0 
&
\\
$V$  & 
$T_{1u}$ & 
 3 & 0 & -1 & -1 & 1 & -3 & 0 & 1 & 1 & -1 
& $(\phi_1,\phi_2,\phi_3)$
\\
$XV$  & 
 $T_{2u}$ &
 3 & 0 & -1 & 1 & -1 & -3 & 0 & 1 & -1 & 1 
&
\\
\hline
\end{tabular}
}
\end{table}

In addition to the character table, we need a representative transformation in each conjugacy class. This information is usually available in databases for finite groups. For our $n=3$ cubic group we can take
\begin{align}\nonumber
 {   \scriptsize    \begin{pmatrix}1 & 0 & 0    \\    0 & 1 & 0    \\    0 & 0 & 1    \\   \end{pmatrix}    }   \in E,\  {   \scriptsize    \begin{pmatrix} 0 & \!\!-1 & 0    \\    0 & 0 & 1    \\    \!\!-1 & 0 & 0    \\   \end{pmatrix}    }  \in 8C_3,\   {   \scriptsize    \begin{pmatrix} \!\!-1 & 0 & 0    \\    0 & \!\!-1 & 0    \\    0 & 0 & 1    \\   \end{pmatrix}    }  \in 3C_2,\   {   \scriptsize    \begin{pmatrix} \!\!-1 & 0 & 0    \\    0 & 0 & \!\!-1    \\    0 & \!\!-1 & 0    \\   \end{pmatrix}    }  \in 6C_2,\   {   \scriptsize    \begin{pmatrix} 0 & 1 & 0    \\    \!\!-1 & 0 & 0    \\    0 & 0 & 1    \\   \end{pmatrix}    }   \in 6C_4,
\\
  {   \scriptsize    \begin{pmatrix} \!-1 & 0 & 0    \\    0 & \!\!-1 & 0    \\    0 & 0 & \!\!-1    \\   \end{pmatrix}    }\in i,\     {   \scriptsize    \begin{pmatrix} 0 & 0 & 1    \\    1 & 0 & 0    \\    0 & \!\!-1 & 0       \end{pmatrix}    }   \in 8S_3,\  {   \scriptsize    \begin{pmatrix} 1 & 0 & 0    \\    0 & 1 & 0    \\    0 & 0 & \!\!-1       \end{pmatrix}    }  \in 3\sigma_h,\  {   \scriptsize    \begin{pmatrix} 1 & 0 & 0    \\    0 & 0 & 1    \\    0 & 1 & 0       \end{pmatrix}    } \in 6\sigma_d,\    {   \scriptsize    \begin{pmatrix} 0 & 1 & 0    \\    \!\!-1 & 0 & 0    \\    0 & 0 & \!\!-1      \end{pmatrix}    }\in 6S_4,
\label{eq:CCrepresentatives}
\end{align}
all acting on $(\phi_1,\phi_2,\phi_3)^T$.

\paragraph{Example} As an example, consider the space of scalar operators with $p=3$ fields. In this space, we choose the basis
\begin{equation}
\label{eq:Threefields-spin0}
\mathcal X_{3,0}=\{\phi_1^3,\,\phi_1^2\phi_2,\,\phi_1^2\phi_3,\,\phi_1\phi^2_2,\,\phi_1\phi_2\phi_3,\,\phi_1\phi_3^2,\,\phi_2^3,\,\phi_2^2\phi_3,\,\phi_2\phi_3^2,\,\phi_3^3\}.
\end{equation}

The computation of the dilatation operator is performed exactly as above, except that we now explicitly perform all summations over repeated indices:
\begin{equation}
\mathcal O_{\mathrm{int}}=\frac1{4!}\left(\lambda_1\left(\phi_1^4+\phi_2^4+\phi_3^4\right)+3\lambda_2 \left(\phi_1^2+\phi_2^2+\phi_3^2\right)^2\right).
\end{equation}
In the basis \eqref{eq:Threefields-spin0}, we find the anomalous dimension matrix 
\begin{equation}
\Gamma\big|_{\mathcal X_{3,0}}=\frac19{\scriptsize\begin{pmatrix}6 & 0 & 0 & 3 & 0 & 3 & 0 & 0 & 0 & 0 \\
 0 & 6 & 0 & 0 & 0 & 0 & 1 & 0 & 1 & 0 \\
 0 & 0 & 6 & 0 & 0 & 0 & 0 & 1 & 0 & 1 \\
 1 & 0 & 0 & 6 & 0 & 1 & 0 & 0 & 0 & 0 \\
 0 & 0 & 0 & 0 & 6 & 0 & 0 & 0 & 0 & 0 \\
 1 & 0 & 0 & 1 & 0 & 6 & 0 & 0 & 0 & 0 \\
 0 & 3 & 0 & 0 & 0 & 0 & 6 & 0 & 3 & 0 \\
 0 & 0 & 1 & 0 & 0 & 0 & 0 & 6 & 0 & 1 \\
 0 & 1 & 0 & 0 & 0 & 0 & 1 & 0 & 6 & 0 \\
 0 & 0 & 3 & 0 & 0 & 0 & 0 & 3 & 0 & 6
\end{pmatrix}}.
\end{equation}
The eigenvalues, multiplied with their multiplicities on the left, are $\left\{3\times\frac49,\,3\times\frac59,\,1\times\frac23,\,3\times 1\right\}$.

Let us consider first the eigenspace of $\frac49$. By computing the corresponding eigenspace, we find that the basis can be taken as
\begin{equation}
\left\{\phi_1^2(\phi_1-\phi_2-\phi_3),\,
\phi_2^2(\phi_2-\phi_1-\phi_3),\,
\phi_3^2(\phi_3-\phi_1-\phi_2)\right\}.
\end{equation}
By inspection it transforms as a vector. However, using the information about the character table we can determine this fact systematically. First we determine how these states transform among each other under the action of the representatives \eqref{eq:CCrepresentatives} of the conjugacy classes. Taking the traces of the resulting matrices gives a set of characters $\vec\chi =( 3 , 0, -1 , -1 , 1 , -3 , 0 , 1 , 1 , -1 )$. Finally, using orthogonality relations for the characters we find that 
\begin{equation}
\langle \vec \chi_R,\vec \chi\rangle = \begin{cases}
1 & R=V ,
\\
0 & \text{else}.
\end{cases}
\end{equation}
Thus we have determined that there is an eigenoperator with the anomalous dimension $\frac49$ transforming in the vector representation. 

Proceeding in the same manner with the other states gives
\begin{align}
\frac59&:  & &\{\phi_1(\phi_2^2-\phi_3^2),\,\phi_2(\phi_3^2-\phi_1^2),\,\phi_3(\phi_1^2-\phi_2^2)\} & \langle \vec \chi_R,\vec \chi\rangle&=\delta_{R,XV},
\\
\frac 23&: & & \{\phi_1\phi_2\phi_3\} & \langle \vec \chi_R,\vec \chi\rangle&=\delta_{RZ_3},
\\
\label{eq:extraEigenvalue}
1&: &  & \{\phi_1(2\phi_1^2+3\phi_2^2+3\phi_3^2),\phi_2(2\phi_2^2+3\phi_1^2+3\phi_3^2),\phi_3(2\phi_3^2+3\phi_1^2+3\phi_2^2)\} & \langle \vec \chi_R,\vec \chi\rangle&=\delta_{RV},
\end{align}
so we conclude that the remaining $\phi^3$-type operators are an $XV$ operator with anomalous dimension $\frac59$, a $Z_3$ operator with anomalous dimension $\frac23$ and a vector ($V$) operator with anomalous dimension $1$. The $V$ operator with anomalous dimension $1$ is not a scaling eigen-operator on its own. Instead it is one component of the equation-of-motion operator, which also has a component along the $\square \phi$ direction, see the discussion in and around footnote \ref{Redundant}.

In the example presented above with no inserted derivatives, we did not need to perform the primary check. In the more general case, one performs the primary check in the following way. For each eigenvalue, compute the corresponding eigensystem. Then, act with $K_\nu$ and compute its kernel. Consequently, write a basis for the kernel of $K_\nu$, and compute the characters for the transformations of the basis elements, and ultimately the inner products of the characters with the rows of the character table. All these steps can be systematically implemented, for instance in Mathematica.
The correct identification of the number of primary operators can also be checked by a character decomposition, following \cite{Dolan:2005wy,Henning:2017fpj,Meneses:2018xpu,Henriksson:2022rnm}.\footnote{The characters and the plethystic exponential for conformal symmetry is explained in these references. For the characters of the global symmetry group, one simply uses $\mathrm{PLexp}(\chi(g))=\exp(\sum_k \frac1k\chi(g^k))$, where $g$ are the group elements. In practice, one needs only a representative group element in each conjugacy class.}

\section{Results}
\label{chapter5}

In this section we present the collective results of our study. In section~\ref{sec:resultsSixloop} we give the results of the six-loop computations applied to scalar operators with $\leqslant4 $ fields and no derivatives. In section~\ref{sec:results-oneloop}, we give the results from the one-loop dilatation operator, focusing on the ten irreps that exist in $n=3$. In appendix~\ref{app:ResultsNgtr4}, we give results for the remaining 10 irreps that exist for $n=4$. We stress that all our results have been computed for $n$ generic. 

All our results, together with literature results from \cite{Dey:2016mcs}, are available in a data file with the Arxiv submission of this paper. We give some more details on this data file in appendix~\ref{app:datafile}.

\subsection{Results from six-loop beta functions}
\label{sec:resultsSixloop}

Even though our results are included in the data file, in order to discuss some qualitative features of the operator spectrum, we will present tables with resummed anomalous dimensions. For demonstration purposes we choose to perform Padé$_{3,3}$ resummations. These were found to be surprisingly accurate in the specific case of the Ising CFT, when compared to non-perturbative estimates in \cite{Henriksson:2022gpa}. The reader is of course encouraged to use our results with their resummation of choice. We remind the reader of the hypercubic Lagrangian \eqref{hypercubiclagrangian}, which for convenience we repeat here:\footnote{Note that we dropped all terms that vanish at the fixed point for simplicity of demonstration.}
\begin{equation}
\mathcal{L}=\frac{1}{2}\partial_\mu \phi_a \partial_\mu \phi_a -\frac{\lambda_1 \delta_{abcd}+\lambda_2 (\delta_{ab}\delta_{cd}+\delta_{ac}\delta_{bd}+\delta_{ad}\delta_{bc})}{4!}\phi_a \phi_b \phi_c \phi_d.
\end{equation}
It has four fixed points. These are the free theory $\lambda_1 =\lambda_2 =0$, the theory of $n$ decoupled Ising models $\lambda_1 \neq 0 $ and $\lambda_2 = 0$, the $O(n)$ theory $\lambda_1 = 0 $ and $\lambda_2 \neq 0$ and the fully interacting hypercubic fixed point where $\lambda_1 \neq 0$ and $\lambda_2 \neq 0$. Hence, by calculating anomalous dimensions of operators first as a function of $\lambda_1$ and $\lambda_2$ (i.e. $\gamma_O = \gamma_O (\lambda_1, \lambda_2)$) we can calculate the scaling dimensions at the three different non-trivial fixed points by simply plugging in the corresponding values of the couplings at the end. 
This will be useful in comparing properties of the different fixed points. It is instructive to do this, since the hypercubic theory can be seen as a deformation of $n$ decoupled Ising models, or alternatively as a deformation of an $O(n)$ symmetric theory with a term that breaks the symmetry down to $C_n$. 
We are also aware of more accurate determimnations for some of the scaling dimensions in the Ising and $O(n)$ theories, such as \cite{Kos:2016ysd, Chester:2019ifh,Chester:2020iyt}, nevertheless we have chosen to present all results computed with the same method for comparison purposes between the different fixed points. 
See also the recent Monte Carlo study of \cite{Hasenbusch:2022zur}. 
Lastly, let us also point out that we chose to present results up to three significant digits, since with fewer digits a lot of the scaling dimensions would coincide, i.e. our choice of significant digits does {\bf not} imply any specific error bar. 

In tables \ref{TableQuadratic}, \ref{TableCubic} and \ref{TableQuartic}, we present the values of the scaling dimensions corresponding to operators built with one, two, three and four powers of the order parameter field. These are evaluated at $n=3$. The text in the $O(n)$ column refers to which representation the operators transform in at the $O(n)$ fixed point.\footnote{$S$ denotes the singlet, whereas $T_m$ denotes the $m$ index traceless symmetric irrep.} The label "$EOM$" is used to point out that an operator is redundant due to the equation of motion. In table \ref{TableQuarticVanishing} we evaluate at $n=4$ the operators that vanish for $n=3$. These vanish due to their group theoretic dimension being proportional to $n-3$. For example, one cannot construct an irrep that needs four indices which all take different values for $n<4$, hence $Z_4$ vanishes. 
In the ``Decoupled Ising'' column we also present what the operators correspond to at the decoupled Ising fixed point. Lastly, we give analytic expressions for the anomalous dimensions truncated at two loops for the readers convenience:
\begin{align}
\Delta_\phi&= 1-\frac{\varepsilon}{2}+\varepsilon^2\frac{n^2+n-2 }{108 n^2}+O(\varepsilon^3)
\,,\nonumber\\
\Delta_S&= 2-\varepsilon+\varepsilon \frac{2 (n-1)}{3 n}-\varepsilon^2\frac{(n-1) \left(19 n^2-326 n+424\right)}{162 n^3}+O(\varepsilon^3)
\,,\nonumber\\
\Delta_X&= 2-\frac{2 \varepsilon (n+1)}{3 n}+\varepsilon^2\frac{19 n^3+131 n^2-538 n+424 }{162 n^3}+O(\varepsilon^3)
\,,\nonumber\\
\Delta_Z&= 2-\varepsilon+\frac{2}{3 n}\varepsilon  +\varepsilon^2\frac{3 n^3-127 n^2+530 n-424 }{162 n^3}+O(\varepsilon^3)
\,,\nonumber\\
\Delta_{(V)_1}^{EOM}&= 3-\frac{\varepsilon}{2}-\varepsilon^2\frac{n^2+n-2 }{108 n^2}+O(\varepsilon^3)
\,,\nonumber\\
\Delta_{(V)_2}&=3-\frac{3\varepsilon}2+\varepsilon\frac{2 (n-1)}{3 n}+\varepsilon^2\frac{-35 n^3+693 n^2-1506 n+848 }{324 n^3}+O(\varepsilon^3)
\,,\nonumber\\
\Delta_{XV}&= 3+\varepsilon \left(\frac{2}{3 n}-\frac{7}{6}\right) +\varepsilon^2\frac{41 n^3-399 n^2+1206 n-848 }{324 n^3}+O(\varepsilon^3)
\,,\nonumber\\
\Delta_{Z_3}&=3-\frac23\varepsilon+\frac{2}{n}\varepsilon +\varepsilon^2\frac{3 n^3-257 n^2+922 n-848 }{108 n^3}+O(\varepsilon^3)
\,,\nonumber\\
\Delta_{S_{(1)}}&= 4-\varepsilon^2\frac{(n-1) (17 n^2-4 n+212)  }{27 n^2 (n+2) }+O(\varepsilon^3)
\,,\nonumber\\
\Delta_{S_{(2)}}&= 4-2\varepsilon+\varepsilon\frac{4 (n-1)}{3 n}-\varepsilon^2\frac{(n-1) (19 n^3-72 n^2-660 n+848)  }{81 n^3 (n+2) }+O(\varepsilon^3)
\,,\nonumber\\
\Delta_{X_{(1)}}&= 4-\varepsilon+\varepsilon^2\frac{107 n^2-640 n+848 }{108 n^2-81 n^3}+O(\varepsilon^3)
\,,\nonumber\\
\Delta_{X_{(2)}}&=4-\frac{4 \varepsilon}{3 n}+\varepsilon^2\frac{-153 n^4+1329 n^3-4618 n^2+5984 n-1696 }{81 n^3 (3 n-4)}+O(\varepsilon^3)
\,,\nonumber\\
\Delta_{Z_{(1)}}&= 4-\varepsilon\frac{ 7 n-12+\sqrt{n (n+24)-48}}{6 n}-\frac{\varepsilon^2 (n (n (13 n+964)-2612)+2544)}{324 n^3} \nonumber\\
&\quad
+\frac{\varepsilon^2 (n (10464-n (n (19 n-1172)+4780))-10176)}{324 n^3 \sqrt{n (n+24)-48}}+O(\varepsilon^3)
,\nonumber\\
\Delta_{Z_{(2)}}&= 4+\varepsilon\frac{12-7 n+\sqrt{n (n+24)-48} }{6 n}-\frac{\varepsilon^2 (n (n (13 n+964)-2612)+2544)}{324 n^3} \nonumber\\
&\quad +\frac{\varepsilon^2 (n (n (n (19 n-1172)+4780)-10464)+10176)}{324 n^3 \sqrt{n (n+24)-48}}+O(\varepsilon^3),
\nonumber\\
\Delta_{B}^{EOM}&=4-\varepsilon
\,,\nonumber\\
\Delta_{Z_4}&= 4-2\varepsilon+ \frac{4}{n}\varepsilon+\varepsilon^2\frac{n^3-129 n^2+390 n-424 }{27 n^3}+O(\varepsilon^3)
\,,\nonumber\\
\Delta_{XX}&= 4-\frac{4 \varepsilon (n-1)}{3 n}+\varepsilon^2\frac{19 n^3-273 n^2+678 n-424 }{81 n^3}+O(\varepsilon^3)
\,,\nonumber\\
\Delta_{XZ}&= 4+\varepsilon \left(\frac{8}{3 n}-\frac{5}{3}\right) +\varepsilon^2\frac{11 n^3-330 n^2+924 n-848 }{81 n^3}+O(\varepsilon^3).
\label{twoloops2}
\end{align}

\begin{table}
\caption{Defining field and quadratic operators at $n=3$.}
\label{TableQuadratic}
\centering
\begin{tabular}{|c|c|lr|lr|}
\hline
 Irrep & $C_3$ & $O(3)$ & & Decoupled Ising &
\\\hline
 $V$ & 0.5184  &0.5184 & $V$   & 0.5174  & $\phi$
\\\hline
 $S$ & 1.564  & 1.562 &  $S$&1.409  & $\phi^2$ 
\\
 $X$ & 1.204  & 1.211 &  $T_2$&1.409  & $\phi^2$  
\\
 $Z$ & 1.207  & 1.211 &  $T_2$&1.035  & $(\phi)(\phi)$
\\\hline
\end{tabular}
\end{table}

\begin{table}
\caption{Cubic (in powers of the field) operators at $n=3$.}
\label{TableCubic}
\centering
\begin{tabular}{|c|c|lr|lr|}
\hline
 Irrep & $C_3$ & $O(3)$ & & Decoupled Ising &
\\\hline
 $(V)_1$ & 2.481  & 2.481 &  $EOM$ & 2.482  &  $EOM$  
\\
 $(V)_2$ & 2.054  & 2.043  & $T_3$ & 1.924  &  $\phi (\phi^2)$  
\\
$XV$        &2.043  & 2.043 &$T_3$   &1.924  &$\phi (\phi^2)$  
\\
$Z_3$ &  2.066 & 2.043 &$T_3$  & 1.552 & $(\phi) (\phi) (\phi)$
\\\hline
\end{tabular}
\end{table}

\begin{table}
\caption{Quartic operators at $n=3$.}
\label{TableQuartic}
\centering
\begin{tabular}{|c|c|lr|lr|}
\hline
 Irrep & $C_3$ & $O(3)$ & & Decoupled Ising &
\\\hline
 $(S)_1$ & 3.784  & 3.782  & $S$  & 3.805 & $\phi^4$  
\\
 $(S)_2$ & 3.011  & 2.992  & $T_4$ & 2.819 & $(\phi^2)^2$ 
\\
$(X)_1$  & 3.003 & 2.992 & $T_4$ & 2.819  &  $(\phi^2)^2$
\\
$(X)_2$ &  3.554 & 3.550 & $T_2$ & 3.805 & $\phi^4$
\\
$(Z)_1$  & 2.988 & 2.992 & $T_4$ & 2.438 & $(\phi)(\phi)(\phi^2)$
\\
$(Z)_2$ & 3.544  & 3.550 & $T_2$ & 3.035  & $\partial^2 (\phi)(\phi)$
\\
$B$   & 3.000  & 2.992 & $T_4$ & 3.000 & $EOM$
\\\hline
\end{tabular}
\end{table}

\begin{table}
\caption{Quartic operators that vanish at $n=3$, evaluated at $n=4$.}
\label{TableQuarticVanishing}
\centering
\begin{tabular}{|c|c|lr|lr|}
\hline
 Irrep & $C_4$ & $O(4)$ & & Decoupled Ising &
\\\hline
 $Z_4$ & 2.570  & 2.890  & $T_4$  & 2.070 & $(\phi)(\phi)(\phi)(\phi)$  
\\
 $XX$ & 2.873  & 2.890  & $T_4$ & 2.819 & $(\phi^2)^2$ 
\\
$XZ$  & 2.725 & 2.890 & $T_4$ & 2.438  &  $(\phi)(\phi)(\phi^2)$
\\\hline
\end{tabular}
\end{table}

\subsection{Results of one-loop computations}
\label{sec:results-oneloop}

In this section, we present the results from the computation of one-loop anomalous dimensions using the dilatation operator. Here we present results for the irreps that exist for $n=3$, additional irreps can be found in appendix~\ref{app:ResultsNgtr4}. The same results are also available in computer-readable format in an ancillary data file, as detailed in appendix~\ref{app:datafile}.

We have only implemented operators in spin-$\ell$ Lorentz representations (including of course the case $\ell=0$), meaning that mixed-symmetry tensors or parity-odd operators have not been considered.\footnote{The spin-$\ell$ operators are the more readily accessible ones in e.g. the conformal bootstrap, since they are the only ones exchanged in the four-point functions of scalars. To study parity-odd operators in $d=3$ one typically needs to use external spinning operators, which is a considerably more difficult task. Note however that some operators in Lorentz mixed-symmetry-tensor irreps are low-lying in the spectrum, the lowest of which is a $B_3$ operator of the form $\phi_{[i}\partial^{[\mu}\phi_j\partial^{\nu]}\phi_{k]}$.}
In the tables, we have only kept primaries, and not operators that become related to the equations-of-motion operator, like the one with eigenvalue $1$ in \eqref{eq:extraEigenvalue}.

In general, our implementations have been limited to the following
\begin{itemize}
\item Singlet ($S$): operators with $\Delta\leqslant 8$ (for spin 3 and 5: $\Delta\leqslant 9$). Table~\ref{tab:singletsscalars} for scalars, table~\ref{tab:singletspinning13} for spin $1$--$3$ and table~\ref{tab:singletspinning46} for spin $4$--$6$.

\begin{table}
\centering
\caption{Scalar singlet ($S$) operators. If the ``$n$ gen.'' term contains no ellipses ($\ldots$), it is exact.}\label{tab:singletsscalars}
{\small
\renewcommand{\arraystretch}{1.25}
\begin{tabular}{|c|c|l|lll|}
\hline
\tableheading0
\\\hline
$1$ & $ \phi^2 $ & $ 2 - \varepsilon + \varepsilon\gamma $ & $\frac{4}{9} $ & $ \frac{1}{2} $ & $ \frac{2 (n-1)}{3 n} $
 \\ 
$2$ & $ \phi^4 $ & $ 4 - 2\varepsilon + \varepsilon\gamma $ & $\frac{8}{9} $ & $ 1 $ & $ \frac{4 (n-1)}{3 n} $
 \\ 
$3$ & $ \phi^4 $ & $ 4 - 2\varepsilon + \varepsilon\gamma $ & $2 $ & $ 2 $ & $ 2 $
 \\ 
$4$ & $ \phi^6 $ & $ 6 - 3\varepsilon + \varepsilon\gamma $ & $2.739344 $ & $ \frac{5}{2} $ & $ 2-\frac{2}{n}+\ldots $
 \\ 
$5$ & $ \square \phi^4 $ & $ 6 - 2\varepsilon + \varepsilon\gamma $ & $\frac{8}{9} $ & $ 1 $ & $ \frac{4 (n-1)}{3 n} $
 \\ 
$6$ & $ \phi^6 $ & $ 6 - 3\varepsilon + \varepsilon\gamma $ & $\frac{10}{3} $ & $ \frac{7}{2} $ & $ \frac{8}{3}+\frac{202}{21 n}+\ldots $
 \\ 
$7$ & $ \phi^6 $ & $ 6 - 3\varepsilon + \varepsilon\gamma $ & $4.705100 $ & $ \frac{9}{2} $ & $ 5-\frac{30}{7 n}+\ldots $
 \\ 
$8$ & $ \square^2 \phi^4 $ & $ 8 - 2\varepsilon + \varepsilon\gamma $ & $0.2418791 $ & $ 0.2934557 $ & $ 0+\frac{40}{27 n}+\ldots $
 \\ 
$9$ & $ \phi^8 $ & $ 8 - 4\varepsilon + \varepsilon\gamma $ & $4.567545 $ & $ \frac{14}{3} $ & $ \frac{8}{3}-\frac{8}{3 n}+\ldots $
 \\ 
$10$ & $ \square \phi^6 $ & $ 8 - 3\varepsilon + \varepsilon\gamma $ & $1.875210 $ & $ \frac{11}{6} $ & $ \frac{7n-\sqrt{n^2+12 n-28}}{3 n}$
 \\ 
$11$ & $ \square^2 \phi^4 $ & $ 8 - 2\varepsilon + \varepsilon\gamma $ & $0.5618756 $ & $ \frac{5}{9} $ & $ \frac{10}{9}-\frac{40}{9 n}+\ldots $
 \\ 
$12$ & $ \square^2 \phi^4 $ & $ 8 - 2\varepsilon + \varepsilon\gamma $ & $1.196245 $ & $ 1.262100 $ & $ \frac{4}{3}+\frac{44}{27 n}+\ldots $
 \\ 
$13$ & $ \phi^8 $ & $ 8 - 4\varepsilon + \varepsilon\gamma $ & --- & $ \frac{14}{3} $ & $ \frac{10}{3}+\frac{404}{21 n}+\ldots $
 \\ 
$14$ & $ \square \phi^6 $ & $ 8 - 3\varepsilon + \varepsilon\gamma $ & $2.791456 $ & $ \frac{17}{6} $ & $ \frac{7n+\sqrt{n^2+12 n-28}}{3 n}$
 \\ 
$15$ & $ \phi^8 $ & $ 8 - 4\varepsilon + \varepsilon\gamma $ & $6.157130 $ & $ 6 $ & $ 4+\frac{32}{3 n}+\ldots $
 \\ 
$16$ & $ \phi^8 $ & $ 8 - 4\varepsilon + \varepsilon\gamma $ & $7.028989 $ & $ 7 $ & $ \frac{17}{3}+\frac{596}{231 n}+\ldots $
 \\ 
$17$ & $ \phi^8 $ & $ 8 - 4\varepsilon + \varepsilon\gamma $ & $8.579668 $ & $ 8 $ & $ \frac{28}{3}-\frac{280}{33 n}+\ldots $
 \\ \hline
\end{tabular}
}
\end{table}

\begin{table}
\centering
\caption{Spinning singlet ($S$) operators. If the ``$n$ gen.'' term contains no ellipses ($\ldots$), it is exact.}\label{tab:singletspinning13}
{\small
\renewcommand{\arraystretch}{1.25}
\begin{tabular}{|c|c|l|lll|}
\hline
\tableheading1
\\\hline
$1$ & $ \partial \phi^6 $ & $ 7 - 3\varepsilon + \varepsilon\gamma $ & $\frac{22}{9} $ & $ \frac{5}{2} $ & $ \frac{2 (4 n-1)}{3 n} $
 \\ 
\hline
\tableheading2
\\\hline
1 & $ T^{\mu\nu}$ & $4-\varepsilon$  & & &
 \\ 
$2$ & $ \partial^2 \phi^4 $ & $ 6 - 2\varepsilon + \varepsilon\gamma $ & $0.5856300 $ & $ \frac{13}{18} $ & $ \frac{2}{3}+\frac{74}{189 n}+\ldots $
 \\ 
$3$ & $ \partial^2 \phi^4 $ & $ 6 - 2\varepsilon + \varepsilon\gamma $ & $0.7262861 $ & $ 0.7343150 $ & $ \frac{4}{3}-\frac{176}{27 n}+\ldots $
 \\ 
$4$ & $ \partial^2 \phi^4 $ & $ 6 - 2\varepsilon + \varepsilon\gamma $ & $1.465862 $ & $ 1.487907 $ & $ \frac{13}{9}+\frac{260}{63 n}+\ldots $
 \\ 
$5$ & $ \partial^2 \square \phi^4 $ & $ 8 - 2\varepsilon + \varepsilon\gamma $ & $0.1403627 $ & $ 0.1634918 $ & $ 0+\frac{80}{81 n}+\ldots $
 \\ 
$6$ & $ \partial^2 \phi^6 $ & $ 8 - 3\varepsilon + \varepsilon\gamma $ & $1.682013 $ & $ 1.604572 $ & $ \frac{4}{3}+\frac{148}{189 n}+\ldots $
 \\ 
$7$ & $ \partial^2 \square \phi^4 $ & $ 8 - 2\varepsilon + \varepsilon\gamma $ & $0.4060371 $ & $ \frac{4}{9} $ & $ \frac{2}{3}-\frac{79}{81 n}+\ldots $
 \\ 
$8$ & $ \partial^2 \square \phi^4 $ & $ 8 - 2\varepsilon + \varepsilon\gamma $ & $0.6769060 $ & $ 0.6959950 $ & $ \frac{8}{9}-\frac{10}{9 n}+\ldots $
 \\ 
$9$ & $ \partial^2 \phi^6 $ & $ 8 - 3\varepsilon + \varepsilon\gamma $ & $2.015478 $ & $ 2.065250 $ & $ 2-\frac{250}{27 n}+\ldots $
 \\ 
$10$ & $ \partial^2 \phi^6 $ & $ 8 - 3\varepsilon + \varepsilon\gamma $ & $2.333382 $ & $ \frac{19}{9} $ & $ 2+\frac{80}{27 n}+\ldots $
 \\ 
$11$ & $ \partial^2 \phi^6 $ & $ 8 - 3\varepsilon + \varepsilon\gamma $ & $2.549260 $ & $ \frac{23}{9} $ & $ \frac{19}{9}+\frac{6318}{665 n}+\ldots $
 \\ 
$12$ & $ \partial^2 \square \phi^4 $ & $ 8 - 2\varepsilon + \varepsilon\gamma $ & $0.9989164 $ & $ 1.084958 $ & $ \frac{4}{3}-\frac{73}{81 n}+\ldots $
 \\ 
$13$ & $ \partial^2 \phi^6 $ & $ 8 - 3\varepsilon + \varepsilon\gamma $ & $2.875551 $ & $ 2.996845 $ & $ \frac{8}{3}+\frac{2666}{945 n}+\ldots $
 \\ 
$14$ & $ \partial^2 \phi^6 $ & $ 8 - 3\varepsilon + \varepsilon\gamma $ & $3.988759 $ & $ \frac{23}{6} $ & $ \frac{38}{9}-\frac{4160}{1197 n}+\ldots $
 \\ \hline
\tableheading3
\\\hline$1$ & $ \partial^3 \phi^4 $ & $ 7 - 2\varepsilon + \varepsilon\gamma $ & $\frac{4}{9} $ & $ \frac{1}{2} $ & $ \frac{2 (n-1)}{3 n} $
 \\ 
$2$ & $ \partial^3 \phi^6 $ & $ 9 - 3\varepsilon + \varepsilon\gamma $ & $1.338380 $ & $ 1.382419 $ & $ \frac{4}{3}-\frac{34}{63 n}+\ldots $
 \\ 
$3$ & $ \partial^3 \square \phi^4 $ & $ 9 - 2\varepsilon + \varepsilon\gamma $ & $\frac{4}{9} $ & $ \frac{1}{2} $ & $ \frac{2 (n-1)}{3 n} $
 \\ 
$4$ & $ \partial^3 \phi^6 $ & $ 9 - 3\varepsilon + \varepsilon\gamma $ & $1.691138 $ & $ 1.723616 $ & $ \frac{5}{3}+\frac{187}{72 n}+\ldots $
 \\ 
$5$ & $ \partial^3 \phi^6 $ & $ 9 - 3\varepsilon + \varepsilon\gamma $ & $2.028790 $ & $ \frac{11}{6} $ & $ 2-\frac{130}{9 n}+\ldots $
 \\ 
$6$ & $ \partial^3 \phi^6 $ & $ 9 - 3\varepsilon + \varepsilon\gamma $ & $2.172610 $ & $ 2.253790 $ & $ 2+\frac{5}{2 n}+\ldots $
 \\ 
$7$ & $ \partial^3 \phi^6 $ & $ 9 - 3\varepsilon + \varepsilon\gamma $ & $2.330307 $ & $ 2.365897 $ & $ \frac{19}{9}+\frac{11593}{840 n}+\ldots $
 \\ 
$8$ & $ \partial^3 \phi^6 $ & $ 9 - 3\varepsilon + \varepsilon\gamma $ & $2.567865 $ & $ 2.667953 $ & $ \frac{8}{3}-\frac{8}{15 n}+\ldots $
 \\ 
$9$ & $ \partial^3 \phi^6 $ & $ 9 - 3\varepsilon + \varepsilon\gamma $ & $3.463503 $ & $ 3.328548 $ & $ \frac{11}{3}-\frac{185}{63 n}+\ldots $
\\\hline
\end{tabular}
}
\end{table}
\begin{table}
\centering
\caption{Spinning singlet ($S$) operators. If the ``$n$ gen.'' term contains no ellipses ($\ldots$), it is exact.}\label{tab:singletspinning46}
{\small
\renewcommand{\arraystretch}{1.25}
\begin{tabular}{|c|c|l|lll|}
\hline
\tableheading4
\\\hline
1 &  $\partial^4\phi^2$ & $6-\varepsilon+\gamma^{(2)}\varepsilon^2$ & $\gamma^{(2)}=\frac{7}{468}$ & $\gamma^{(2)}=\frac7{480}$ & $\gamma^{(2)}=\frac{7(n+2)(n-1)}{540n^2}$
\\
$2$ & $ \partial^4 \phi^4 $ & $ 8 - 2\varepsilon + \varepsilon\gamma $ & $0.1534506 $ & $ \frac{2}{9} $ & $ 0+\frac{44}{45 n}+\ldots $
 \\ 
$3$ & $ \partial^4 \phi^4 $ & $ 8 - 2\varepsilon + \varepsilon\gamma $ & $0.2908402 $ & $ 0.2309175 $ & $ \frac{4}{9}-\frac{665}{666 n}+\ldots $
 \\ 
$4$ & $ \partial^4 \phi^4 $ & $ 8 - 2\varepsilon + \varepsilon\gamma $ & $0.4767270 $ & $ 0.5248214 $ & $ \frac{2}{3}+\frac{-179-7 \sqrt{6609}}{1350 n}+\ldots $
 \\ 
$5$ & $ \partial^4 \phi^4 $ & $ 8 - 2\varepsilon + \varepsilon\gamma $ & $0.5450220 $ & $ \frac{19}{30} $ & $ \frac{2}{3}+\frac{-179+7 \sqrt{6609}}{1350 n}+\ldots $
 \\ 
$6$ & $ \partial^4 \phi^4 $ & $ 8 - 2\varepsilon + \varepsilon\gamma $ & $0.7276645 $ & $ 0.7742878 $ & $ \frac{19}{15}-\frac{34732}{4995 n}+\ldots $
 \\ 
$7$ & $ \partial^4 \phi^4 $ & $ 8 - 2\varepsilon + \varepsilon\gamma $ & $1.295185 $ & $ 1.325529 $ & $ \frac{4}{3}+\frac{6173}{1350 n}+\ldots $
 \\ 
\hline
\tableheading5
\\\hline
$1$ & $ \partial^5 \phi^4 $ & $ 9 - 2\varepsilon + \varepsilon\gamma $ & $\frac{4}{9} $ & $ \frac{1}{2} $ & $ \frac{2 (n-1)}{3 n} $
 \\ 
$2$ & $ \partial^5 \phi^4 $ & $ 9 - 2\varepsilon + \varepsilon\gamma $ & $\frac{4}{9} $ & $ \frac{1}{2} $ & $ \frac{2 (n-1)}{3 n} $
 \\ 
\hline
\tableheading6
\\\hline
1 &  $\partial^6\phi^2$ & $8-\varepsilon+\gamma^{(2)}\varepsilon^2$ & $\gamma^{(2)}=\frac{10}{567}$ & $\gamma^{(2)}=\frac1{56}$ & $\gamma^{(2)}=\frac{(n+2)(n-1)}{63n^2}$
\\\hline
\end{tabular}
}
\end{table}

\item Remaining irreps which exist for $n=3$: $\Delta\leqslant 7$.\footnote{For $B_3$, also the first scalar at $\Delta=9$.} This list includess the vector $V$ (tables~\ref{tab:irrepVscalar} and \ref{tab:irrepVspinning}) and the rank-3 irreps $Z_3,\,XV,\,VB,\,B_3$ (tables~\ref{tab:irrepZ3}, \ref{tab:irrepXVscalar}, \ref{tab:irrepXVspinning}, \ref{tab:irrepVB}, \ref{tab:irrepB3}) as well as the irreps $X,\,B,\,Z$ (tables~\ref{tab:irrepX}, \ref{tab:irrepZscalar}, \ref{tab:irrepZspinning}, \ref{tab:irrepB}) and the rank-$4$ irrep $\XXbar$ (table~\ref{tab:irrepXXbar}).
\item Irreps which exist for $n=4$, but not for $n=3$: For $\ell=0$, $\Delta\leqslant6$. Moreover, the lowest operator at spins $\ell=0$, $\ell=1$ and $\ell=2$. This list includes the rank-$4$ irreps $Z_4,XX,\,XB,\,BZ,\,BB,\,VB_3,\,B_4$ (tables~\ref{tab:irrepZ4}, \ref{tab:irrepXX}, \ref{tab:irrepXZ}, \ref{tab:irrepXB}, \ref{tab:irrepXB}, \ref{tab:irrepBZ}, \ref{tab:irrepBB}, \ref{tab:irrepVB3}, \ref{tab:irrepB4}), the rank-$5$ irrep $\XXbar V$ (table~\ref{tab:irrepXXbarV}), and the rank-$6$ irrep $\XXXbar$ (table~\ref{tab:irrepXXXbar}). These tables are found in appendix~\ref{app:ResultsNgtr4}.
\end{itemize}
In the tables we give a list of the operators for a given spin together with their field content and an expression for their dimension in terms of the anomalous dimension with higher orders ommitted. The anomalous dimension is then given for $n=3$ and $n=4$ separately, and for generic $n$. If the expressions at generic $n$ are too long, we display the large-$n$ expansion. Complete expressions for the anomalous dimensions are given in the data file, as detailed in appendix~\ref{app:datafile}. For the weakly broken higher-spin currents, we give the dimensions as computed in \cite{Dey:2016mcs}.

\begin{table}
\centering
\caption{Scalar irrep $V$ operators. If the ``$n$ gen.'' term contains no ellipses ($\ldots$), it is exact.}\label{tab:irrepVscalar}
{\small
\renewcommand{\arraystretch}{1.25}
\begin{tabular}{|c|c|l|lll|}
\hline
\tableheading0
\\\hline
$1$ & $\phi$ &$1-\frac12\varepsilon+\gamma^{(2)}\varepsilon^2$ & $\gamma^{(2)}=\frac5{486}$&  $\gamma^{(2)}=\frac1{96}$ &  $\gamma^{(2)}=\frac{(n+1)(n-1)}{108n^2}$
 \\ 
$2$ & $ \phi^3 $ & $ 3 - \frac{3}{2}\varepsilon + \varepsilon\gamma $ & $\frac{4}{9} $ & $ \frac{1}{2} $ & $ \frac{2 (n-1)}{3 n} $
 \\ 
$3$ & $ \phi^5 $ & $ 5 - \frac{5}{2}\varepsilon + \varepsilon\gamma $ & $1.590334 $ & $ \frac{5}{3} $ & $ \frac{4}{3}-\frac{4}{3 n}+\ldots $
 \\ 
$4$ & $ \phi^5 $ & $ 5 - \frac{5}{2}\varepsilon + \varepsilon\gamma $ & $1.869760 $ & $ \frac{5}{3} $ & $ \frac{5}{3}+\frac{10}{3 n}+\ldots $
 \\ 
$5$ & $ \phi^5 $ & $ 5 - \frac{5}{2}\varepsilon + \varepsilon\gamma $ & $2.401489 $ & $ \frac{5}{2} $ & $ 2+\frac{8}{3 n}+\ldots $
 \\ 
$6$ & $ \phi^5 $ & $ 5 - \frac{5}{2}\varepsilon + \varepsilon\gamma $ & $3.138417 $ & $ 3 $ & $ \frac{10}{3}-\frac{8}{3 n}+\ldots $
 \\ 
$7$ & $ \phi^7 $ & $ 7 - \frac{7}{2}\varepsilon + \varepsilon\gamma $ & $3.423678 $ & $ \frac{7}{2} $ & $ 2-\frac{2}{n}+\ldots $
 \\ 
$8$ & $ \square \phi^5 $ & $ 7 - \frac{5}{2}\varepsilon + \varepsilon\gamma $ & $1.159827 $ & $ \frac{7}{6} $ & $\frac{9 n-2- \sqrt{n^2+12 n-28}}{6 n}$
 \\ 
$9$ & $ \phi^7 $ & $ 7 - \frac{7}{2}\varepsilon + \varepsilon\gamma $ & $3.852916 $ & $ \frac{7}{2} $ & $ \frac{7}{3}+\frac{20}{3 n}+\ldots $
 \\ 
$10$ & $ \square \phi^5 $ & $ 7 - \frac{5}{2}\varepsilon + \varepsilon\gamma $ & $1.617950 $ & $ \frac{5}{3} $ & $\frac{9 n-2+ \sqrt{n^2+12 n-28}}{6 n}$
 \\ 
$11$ & $ \phi^7 $ & $ 7 - \frac{7}{2}\varepsilon + \varepsilon\gamma $ & --- 
& $ \frac{7}{2} $ & $ \frac{8}{3}+\frac{86}{7 n}+\ldots $
 \\ 
$12$ & $ \phi^7 $ & $ 7 - \frac{7}{2}\varepsilon + \varepsilon\gamma $ & $4.549741 $ & $ \frac{14}{3} $ & $ 3+\frac{8}{n}+\ldots $
 \\ 
$13$ & $ \phi^7 $ & $ 7 - \frac{7}{2}\varepsilon + \varepsilon\gamma $ & $4.980430 $ & $ \frac{14}{3} $ & $ 4+\frac{26}{9 n}+\ldots $
 \\ 
$14$ & $ \phi^7 $ & $ 7 - \frac{7}{2}\varepsilon + \varepsilon\gamma $ & $5.531706 $ & $ \frac{11}{2} $ & $ 5-\frac{2}{7 n}+\ldots $
 \\ 
$15$ & $ \phi^7 $ & $ 7 - \frac{7}{2}\varepsilon + \varepsilon\gamma $ & $6.439307 $ & $ 6 $ & $ 7-\frac{56}{9 n}+\ldots $
 \\
\hline
\end{tabular}
}
\end{table}

\begin{table}
\centering
\caption{Spinning irrep $V$ operators. If the ``$n$ gen.'' term contains no ellipses ($\ldots$), it is exact.}\label{tab:irrepVspinning}
{\small
\renewcommand{\arraystretch}{1.25}
\begin{tabular}{|c|c|l|lll|}
\hline
\tableheading1
\\\hline
$1$ & $ \partial \phi^3 $ & $ 4 - \frac{3}{2}\varepsilon + \varepsilon\gamma $ & $\frac{4}{9} $ & $ \frac{1}{2} $ & $ \frac{2 (n-1)}{3 n} $
 \\ 
$2$ & $ \partial \phi^5 $ & $ 6 - \frac{5}{2}\varepsilon + \varepsilon\gamma $ & $1.274319 $ & $ \frac{5}{4} $ & $ \frac{4}{3}-\frac{4}{3 n}+\ldots $
 \\ 
$3$ & $ \partial \phi^5 $ & $ 6 - \frac{5}{2}\varepsilon + \varepsilon\gamma $ & $\frac{5}{3} $ & $ \frac{5}{3} $ & $ \frac{5}{3}+\frac{2}{3 n}+\ldots $
 \\ 
$4$ & $ \partial \phi^5 $ & $ 6 - \frac{5}{2}\varepsilon + \varepsilon\gamma $ & $2.170125 $ & $ \frac{13}{6} $ & $ 2+\frac{1}{n}+\ldots $
 \\ 
\hline
\tableheading2
\\\hline
$1$ & $ \partial^2 \phi^3 $ & $ 5 - \frac{3}{2}\varepsilon + \varepsilon\gamma $ & $0.1209395 $ & $ 0.1467278 $ & $ 0+\frac{20}{27 n}+\ldots $
 \\ 
$2$ & $ \partial^2 \phi^3 $ & $ 5 - \frac{3}{2}\varepsilon + \varepsilon\gamma $ & $0.2809378 $ & $ \frac{5}{18} $ & $ \frac{5}{9}-\frac{20}{9 n}+\ldots $
 \\ 
$3$ & $ \partial^2 \phi^3 $ & $ 5 - \frac{3}{2}\varepsilon + \varepsilon\gamma $ & $0.5981227 $ & $ 0.6310499 $ & $ \frac{2}{3}+\frac{22}{27 n}+\ldots $
 \\ 
$4$ & $ \partial^2 \phi^5 $ & $ 7 - \frac{5}{2}\varepsilon + \varepsilon\gamma $ & $0.9569331 $ & $ 0.9435369 $ & $ \frac{2}{3}+\frac{214}{189 n}+\ldots $
 \\ 
$5$ & $ \partial^2 \phi^5 $ & $ 7 - \frac{5}{2}\varepsilon + \varepsilon\gamma $ & $1.051044 $ & $ \frac{19}{18} $ & $ 1+\frac{20}{9 n}+\ldots $
 \\ 
$6$ & $ \partial^2 \phi^5 $ & $ 7 - \frac{5}{2}\varepsilon + \varepsilon\gamma $ & $1.258745 $ & $ 1.287113 $ & $ \frac{11}{9}-\frac{46}{39 n}+\ldots $
 \\ 
$7$ & $ \partial^2 \phi^5 $ & $ 7 - \frac{5}{2}\varepsilon + \varepsilon\gamma $ & $1.285195 $ & $ \frac{4}{3} $ & $ \frac{4}{3}-\frac{80+4 \sqrt{562}}{27 n}+\ldots $
 \\ 
$8$ & $ \partial^2 \phi^5 $ & $ 7 - \frac{5}{2}\varepsilon + \varepsilon\gamma $ & $1.422623 $ & $ \frac{4}{3} $ & $ \frac{4}{3}+\frac{4 \sqrt{562}-80}{27 n}+\ldots $ 
 \\ 
$9$ & $ \partial^2 \phi^5 $ & $ 7 - \frac{5}{2}\varepsilon + \varepsilon\gamma $ & $1.514392 $ & $ 1.401469 $ & $ \frac{13}{9}+\frac{1972}{315 n}+\ldots $
 \\ 
$10$ & $ \partial^2 \phi^5 $ & $ 7 - \frac{5}{2}\varepsilon + \varepsilon\gamma $ & $1.875425 $ & $ 1.930585 $ & $ \frac{5}{3}+\frac{2}{3 n}+\ldots $
 \\ 
$11$ & $ \partial^2 \phi^5 $ & $ 7 - \frac{5}{2}\varepsilon + \varepsilon\gamma $ & $2.029382 $ & $ 2.047128 $ & $ 2+\frac{8}{15 n}+\ldots $
 \\ 
$12$ & $ \partial^2 \phi^5 $ & $ 7 - \frac{5}{2}\varepsilon + \varepsilon\gamma $ & $2.532187 $ & $ 2.445724 $ & $ \frac{8}{3}-\frac{226}{117 n}+\ldots $
 \\ 
\hline
\tableheading3
\\\hline
$1$ & $ \partial^3 \phi^3 $ & $ 6 - \frac{3}{2}\varepsilon + \varepsilon\gamma $ & $0.06163627 $ & $ \frac{1}{12} $ & $ 0+\frac{5}{12 n}+\ldots $
 \\ 
$2$ & $ \partial^3 \phi^3 $ & $ 6 - \frac{3}{2}\varepsilon + \varepsilon\gamma $ & $0.1338756 $ & $ 0.1100459 $ & $ \frac{1}{6}-\frac{5}{18 n}+\ldots $
 \\ 
$3$ & $ \partial^3 \phi^3 $ & $ 6 - \frac{3}{2}\varepsilon + \varepsilon\gamma $ & $0.4155993 $ & $ 0.4732875 $ & $ \frac{2}{3}-\frac{29}{36 n}+\ldots $
 \\ 
\hline
\tableheading4
\\\hline
$1$ & $ \partial^4 \phi^3 $ & $ 7 - \frac{3}{2}\varepsilon + \varepsilon\gamma $ & $0 $ & $ 0 $ & $ 0 $
 \\ 
$2$ & $ \partial^4 \phi^3 $ & $ 7 - \frac{3}{2}\varepsilon + \varepsilon\gamma $ & $0.1270235 $ & $ 0.1639079 $ & $ 0+\frac{56}{75 n}+\ldots $
 \\ 
$3$ & $ \partial^4 \phi^3 $ & $ 7 - \frac{3}{2}\varepsilon + \varepsilon\gamma $ & $0.2570099 $ & $ \frac{7}{30} $ & $ \frac{7}{15}-\frac{56}{45 n}+\ldots $
 \\ 
$4$ & $ \partial^4 \phi^3 $ & $ 7 - \frac{3}{2}\varepsilon + \varepsilon\gamma $ & $0.5270778 $ & $ 0.5694254 $ & $ \frac{2}{3}-\frac{38}{225 n}+\ldots $
 \\
\hline
\end{tabular}
}
\end{table}

\begin{table}
\centering
\caption{Irrep $Z_3$ operators. If the ``$n$ gen.'' term contains no ellipses ($\ldots$), it is exact.}\label{tab:irrepZ3}
{\small
\renewcommand{\arraystretch}{1.25}
\begin{tabular}{|c|c|l|lll|}
\hline
\tableheading0
\\\hline
$1$ & $ \phi^3 $ & $ 3 - \frac{3}{2}\varepsilon + \varepsilon\gamma $ & $\frac{2}{3} $ & $ \frac{1}{2} $ & $ \frac{2}{n} $
 \\ 
$2$ & $ \phi^5 $ & $ 5 - \frac{5}{2}\varepsilon + \varepsilon\gamma $ & --- & $ \frac{5}{3} $ & $ \frac{2}{3}+\frac{4}{3 n}+\ldots $
 \\ 
$3$ & $ \phi^5 $ & $ 5 - \frac{5}{2}\varepsilon + \varepsilon\gamma $ & $\frac{26}{9} $ & $ \frac{5}{2} $ & $ 1+\frac{26}{3 n}+\ldots $
 \\ 
$4$ & $ \phi^7 $ & $ 7 - \frac{7}{2}\varepsilon + \varepsilon\gamma $ & $3.738020 $ & $ \frac{7}{2} $ & $ \frac{4}{3}+\frac{2}{3 n}+\ldots $
 \\ 
$5$ & $ \square \phi^5 $ & $ 7 - \frac{5}{2}\varepsilon + \varepsilon\gamma $ & $\frac{10}{9} $ & $ 1 $ & $ \frac{2 (n+2)}{3 n} $
 \\ 
$6$ & $ \phi^7 $ & $ 7 - \frac{7}{2}\varepsilon + \varepsilon\gamma $ & --- & $ \frac{7}{2} $ & $ \frac{5}{3}+\frac{12}{n}+\ldots $
 \\ 
$7$ & $ \square \phi^5 $ & $ 7 - \frac{5}{2}\varepsilon + \varepsilon\gamma $ & --- & $ \frac{7}{6} $ & $ \frac{3 n+2}{3 n} $
 \\ 
$8$ & $ \phi^7 $ & $ 7 - \frac{7}{2}\varepsilon + \varepsilon\gamma $ & --- & --- & $ 2+\frac{10}{n}+\ldots $
 \\ 
$9$ & $ \phi^7 $ & $ 7 - \frac{7}{2}\varepsilon + \varepsilon\gamma $ & --- & $ \frac{14}{3} $ & $ 2+\frac{14}{n}+\ldots $
 \\ 
$10$ & $ \phi^7 $ & $ 7 - \frac{7}{2}\varepsilon + \varepsilon\gamma $ & $6.261980 $ & $ \frac{11}{2} $ & $ \frac{10}{3}+\frac{14}{3 n}+\ldots $
 \\ 
\hline
\tableheading1
\\\hline
$1$ & $ \partial \phi^5 $ & $ 6 - \frac{5}{2}\varepsilon + \varepsilon\gamma $ & --- & $ \frac{5}{4} $ & $\frac{5 n+15-\sqrt{n^2+14 n-47}}{6 n}$
 \\ 
$2$ & $ \partial \phi^5 $ & $ 6 - \frac{5}{2}\varepsilon + \varepsilon\gamma $ & $\frac{16}{9} $ & $ \frac{5}{3} $ & $\frac{5 n+15+\sqrt{n^2+14 n-47}}{6 n}$
 \\ 
\hline
\tableheading2
\\\hline
$1$ & $ \partial^2 \phi^3 $ & $ 5 - \frac{3}{2}\varepsilon + \varepsilon\gamma $ & $\frac{10}{27} $ & $ \frac{5}{18} $ & $ \frac{10}{9 n} $
 \\ 
$2$ & $ \partial^2 \phi^5 $ & $ 7 - \frac{5}{2}\varepsilon + \varepsilon\gamma $ & $0.9265755 $ & $ \frac{8}{9} $ & $ 0+\frac{38}{9 n}+\ldots $
 \\ 
$3$ & $ \partial^2 \phi^5 $ & $ 7 - \frac{5}{2}\varepsilon + \varepsilon\gamma $ & --- & $ 0.9435369 $ & $ \frac{5}{9}+\frac{10}{9 n}+\ldots $
 \\ 
$4$ & $ \partial^2 \phi^5 $ & $ 7 - \frac{5}{2}\varepsilon + \varepsilon\gamma $ & --- & $ \frac{19}{18} $ & $ \frac{2}{3}+\frac{8 (2-\sqrt2)}{9 n}+\ldots $
 \\ 
$5$ & $ \partial^2 \phi^5 $ & $ 7 - \frac{5}{2}\varepsilon + \varepsilon\gamma $ & $1.370846 $ & $ 1.287113 $ & $ \frac{2}{3}+\frac{8 (2+\sqrt2)}{9 n}+\ldots $
 \\ 
$6$ & $ \partial^2 \phi^5 $ & $ 7 - \frac{5}{2}\varepsilon + \varepsilon\gamma $ & --- & $ \frac{4}{3} $ & $ 1+\frac{26-8 \sqrt7}{9 n}+\ldots $
 \\ 
$7$ & $ \partial^2 \phi^5 $ & $ 7 - \frac{5}{2}\varepsilon + \varepsilon\gamma $ & $2.332208 $ & $ 2.047128 $ & $ 1+\frac{26+8 \sqrt7}{9 n}+\ldots $
 \\ 
\hline
\tableheading3
\\\hline
$1$ & $ \partial^3 \phi^3 $ & $ 6 - \frac{3}{2}\varepsilon + \varepsilon\gamma $ & $\frac{1}{9} $ & $ \frac{1}{12} $ & $ \frac{1}{3 n} $
 \\ 
\hline
\tableheading4
\\\hline
$1$ & $ \partial^4 \phi^3 $ & $ 7 - \frac{3}{2}\varepsilon + \varepsilon\gamma $ & $\frac{14}{45} $ & $ \frac{7}{30} $ & $ \frac{14}{15 n} $
 \\ 
\hline
\end{tabular}
}
\end{table}

\begin{table}
\centering
\caption{Scalar irrep $XV$ operators. If the ``$n$ gen.'' term contains no ellipses ($\ldots$), it is exact. The two operators $6$ and $7$ annihilate precisely at $n=3$, and there is one primary with the corresponding anomalous dimension $\frac{10}9$.}\label{tab:irrepXVscalar}
{\small
\renewcommand{\arraystretch}{1.25}
\begin{tabular}{|c|c|l|lll|}
\hline
\tableheading0
\\\hline$1$ & $ \phi^3 $ & $ 3 - \frac{3}{2}\varepsilon + \varepsilon\gamma $ & $\frac{5}{9} $ & $ \frac{1}{2} $ & $ \frac{n+2}{3 n} $
 \\ 
$2$ & $ \phi^5 $ & $ 5 - \frac{5}{2}\varepsilon + \varepsilon\gamma $ & $1.692000 $ & $ \frac{5}{3} $ & $ 1+\frac{4}{3 n}+\ldots $
 \\ 
$3$ & $ \phi^5 $ & $ 5 - \frac{5}{2}\varepsilon + \varepsilon\gamma $ & --- & $ \frac{5}{3} $ & $ \frac{4}{3}+\frac{10}{3 n}+\ldots $
 \\ 
$4$ & $ \phi^5 $ & $ 5 - \frac{5}{2}\varepsilon + \varepsilon\gamma $ & $2.641334 $ & $ \frac{5}{2} $ & $ 2+\frac{4}{3 n}+\ldots $
 \\ 
$5$ & $ \phi^7 $ & $ 7 - \frac{7}{2}\varepsilon + \varepsilon\gamma $ & $3.485992 $ & $ \frac{7}{2} $ & $ \frac{5}{3}+\frac{2}{n}+\ldots $
 \\ 
$6$ & $ \square \phi^5 $ & $ 7 - \frac{5}{2}\varepsilon + \varepsilon\gamma $ & $\frac{10}{9} $ & $ 1 $ & $ 1+\frac{1-\sqrt{17}}{6 n}+\ldots $ 
 \\ 
$7$ & $ \square \phi^5 $ & $ 7 - \frac{5}{2}\varepsilon + \varepsilon\gamma $ & ---
& $ \frac{7}{6} $ & $ 1+\frac{1+\sqrt{17}}{6 n}+\ldots $ 
 \\ 
$8$ & $ \phi^7 $ & $ 7 - \frac{7}{2}\varepsilon + \varepsilon\gamma $ & $3.823138 $ & $ \frac{7}{2} $ & $ 2+\frac{8}{n}+\ldots $
 \\ 
$9$ & $ \square \phi^5 $ & $ 7 - \frac{5}{2}\varepsilon + \varepsilon\gamma $ & $\frac{13}{9} $ & $ \frac{17}{12} $ & $ \frac{4}{3}+\frac{2}{3 n}+\ldots $
 \\ 
$10$ & $ \phi^7 $ & $ 7 - \frac{7}{2}\varepsilon + \varepsilon\gamma $ & --- & $ \frac{7}{2} $ & $ \frac{7}{3}+\frac{26}{3 n}+\ldots $
 \\ 
$11$ & $ \phi^7 $ & $ 7 - \frac{7}{2}\varepsilon + \varepsilon\gamma $ & --- & --- & $ \frac{8}{3}+\frac{118}{21 n}+\ldots $
 \\ 
$12$ & $ \phi^7 $ & $ 7 - \frac{7}{2}\varepsilon + \varepsilon\gamma $ & $4.736930 $ & $ \frac{14}{3} $ & $ 3+\frac{20}{3 n}+\ldots $
 \\ 
$13$ & $ \phi^7 $ & $ 7 - \frac{7}{2}\varepsilon + \varepsilon\gamma $ & --- & $ \frac{14}{3} $ & $ \frac{11}{3}+\frac{10}{3 n}+\ldots $
 \\ 
$14$ & $ \phi^7 $ & $ 7 - \frac{7}{2}\varepsilon + \varepsilon\gamma $ & $5.842829 $ & $ \frac{11}{2} $ & $ 5-\frac{2}{7 n}+\ldots $
 \\ 
\hline
\end{tabular}
}
\end{table}

\begin{table}
\centering
\caption{Spinning irrep $XV$ operators. If the ``$n$ gen.'' term contains no ellipses ($\ldots$), it is exact.}\label{tab:irrepXVspinning}
{\small
\renewcommand{\arraystretch}{1.25}
\begin{tabular}{|c|c|l|lll|}
\hline
\tableheading1
\\\hline
$1$ & $ \partial \phi^3 $ & $ 4 - \frac{3}{2}\varepsilon + \varepsilon\gamma $ & $\frac{2}{9} $ & $ \frac{1}{4} $ & $ \frac{n-1}{3 n} $
 \\ 
$2$ & $ \partial \phi^5 $ & $ 6 - \frac{5}{2}\varepsilon + \varepsilon\gamma $ & $1.291969 $ & $ \frac{5}{4} $ & $ 1+\frac{1}{3 n}+\ldots $
 \\ 
$3$ & $ \partial \phi^5 $ & $ 6 - \frac{5}{2}\varepsilon + \varepsilon\gamma $ & --- & $ \frac{5}{4} $ & $ 1+\frac{4}{3 n}+\ldots $
 \\ 
$4$ & $ \partial \phi^5 $ & $ 6 - \frac{5}{2}\varepsilon + \varepsilon\gamma $ & $\frac{5}{3} $ & $ \frac{5}{3} $ & $ \frac{4}{3}+\frac{5}{3 n}+\ldots $
 \\ 
$5$ & $ \partial \phi^5 $ & $ 6 - \frac{5}{2}\varepsilon + \varepsilon\gamma $ & $1.930253 $ & $ \frac{23}{12} $ & $ 2-\frac{1}{3 n}+\ldots $
 \\ 
\hline
\tableheading2
\\\hline
$1$ & $ \partial^2 \phi^3 $ & $ 5 - \frac{3}{2}\varepsilon + \varepsilon\gamma $ & $0.08512733 $ & $ \frac{1}{9} $ & $ \frac{3 n+2-\sqrt{9 n^2-68 n+164}}{18 n}$
 \\ 
$2$ & $ \partial^2 \phi^3 $ & $ 5 - \frac{3}{2}\varepsilon + \varepsilon\gamma $ & $0.3222801 $ & $ \frac{5}{18} $ & $ \frac{3 n+2+\sqrt{9 n^2-68 n+164}}{18 n}$
 \\ 
$3$ & $ \partial^2 \phi^5 $ & $ 7 - \frac{5}{2}\varepsilon + \varepsilon\gamma $ & $0.9304356 $ & $ \frac{8}{9} $ & $ \frac{1}{3}+\frac{26}{9 n}+\ldots $
 \\ 
$4$ & $ \partial^2 \phi^5 $ & $ 7 - \frac{5}{2}\varepsilon + \varepsilon\gamma $ & --- & $ 0.9435369 $ & $ \frac{2}{3}+\frac{62}{63 n}+\ldots $
 \\ 
$5$ & $ \partial^2 \phi^5 $ & $ 7 - \frac{5}{2}\varepsilon + \varepsilon\gamma $ & $1.118006 $ & $ \frac{19}{18} $ & $ \frac{8}{9}+0+\ldots $
 \\ 
$6$ & $ \partial^2 \phi^5 $ & $ 7 - \frac{5}{2}\varepsilon + \varepsilon\gamma $ & --- & $ \frac{19}{18} $ & $ 1+\frac{20}{9 n}+\ldots $
 \\ 
$7$ & $ \partial^2 \phi^5 $ & $ 7 - \frac{5}{2}\varepsilon + \varepsilon\gamma $ & $1.307693 $ & $ 1.287113 $ & $ 1+\frac{20}{9 n}+\ldots $
 \\ 
$8$ & $ \partial^2 \phi^5 $ & $ 7 - \frac{5}{2}\varepsilon + \varepsilon\gamma $ & $1.326318 $ & $ \frac{47}{36} $ & $ 1+\frac{20}{9 n}+\ldots $
 \\ 
$9$ & $ \partial^2 \phi^5 $ & $ 7 - \frac{5}{2}\varepsilon + \varepsilon\gamma $ & $1.372973 $ & $ \frac{4}{3} $ & $ 1+\frac{20}{9 n}+\ldots $
 \\ 
$10$ & $ \partial^2 \phi^5 $ & $ 7 - \frac{5}{2}\varepsilon + \varepsilon\gamma $ & --- & $ \frac{4}{3} $ & $ \frac{4}{3}+\frac{8}{9 n}+\ldots $
 \\ 
$11$ & $ \partial^2 \phi^5 $ & $ 7 - \frac{5}{2}\varepsilon + \varepsilon\gamma $ & $1.710018 $ & $ \frac{31}{18} $ & $ \frac{13}{9}+\frac{502}{315 n}+\ldots $
 \\ 
$12$ & $ \partial^2 \phi^5 $ & $ 7 - \frac{5}{2}\varepsilon + \varepsilon\gamma $ & $2.160482 $ & $ 2.047128 $ & $ 2-\frac{4}{5 n}+\ldots $
 \\ 
\hline
\tableheading3
\\\hline
$1$ & $ \partial^3 \phi^3 $ & $ 6 - \frac{3}{2}\varepsilon + \varepsilon\gamma $ & $0.07677589 $ & $ \frac{1}{12} $ & $ \frac{2 n-1-\sqrt{4 n^2-24 n+41}}{12 n}$
 \\ 
$2$ & $ \partial^3 \phi^3 $ & $ 6 - \frac{3}{2}\varepsilon + \varepsilon\gamma $ & $0.2010019 $ & $ \frac{5}{24} $ & $ \frac{2 n-1+\sqrt{4 n^2-24 n+41}}{12 n}$
 \\ 
\hline
\tableheading4
\\\hline
$1$ & $ \partial^4 \phi^3 $ & $ 7 - \frac{3}{2}\varepsilon + \varepsilon\gamma $ & $0 $ & $ 0 $ & $ 0 $
 \\ 
$2$ & $ \partial^4 \phi^3 $ & $ 7 - \frac{3}{2}\varepsilon + \varepsilon\gamma $ & $0.09930825 $ & $ \frac{2}{15} $ & $ \frac{5 n+2-\sqrt{25 n^2-204 n+452}}{30 n}$
 \\ 
$3$ & $ \partial^4 \phi^3 $ & $ 7 - \frac{3}{2}\varepsilon + \varepsilon\gamma $ & $0.2784695 $ & $ \frac{7}{30} $ &  $ \frac{5 n+2+\sqrt{25 n^2-204 n+452}}{30 n}$
 \\ 
\hline
\end{tabular}
}
\end{table}

\begin{table}
\centering
\caption{Irrep $VB$ operators. If the ``$n$ gen.'' term contains no ellipses ($\ldots$), it is exact.}\label{tab:irrepVB}
{\small
\renewcommand{\arraystretch}{1.25}
\begin{tabular}{|c|c|l|lll|}
\hline
\tableheading0
\\\hline
$1$ & $ \phi^5 $ & $ 5 - \frac{5}{2}\varepsilon + \varepsilon\gamma $ & $\frac{17}{9} $ & $ \frac{5}{3} $ & $ \frac{3 n+8}{3 n} $
 \\ 
$2$ & $ \square \phi^5 $ & $ 7 - \frac{5}{2}\varepsilon + \varepsilon\gamma $ & --- & $ 1 $ &
$\frac{5 n+9-\sqrt{n^2+14 n-47}}{6 n}$
 \\ 
$3$ & $ \phi^7 $ & $ 7 - \frac{7}{2}\varepsilon + \varepsilon\gamma $ & $3.869010 $ & $ \frac{7}{2} $ & $ \frac{5}{3}+\frac{6}{n}+\ldots $
 \\ 
$4$ & $ \square \phi^5 $ & $ 7 - \frac{5}{2}\varepsilon + \varepsilon\gamma $ & $\frac{13}{9} $ & $ \frac{17}{12} $ & $\frac{5 n+9+\sqrt{n^2+14 n-47}}{6 n}$ 
 \\ 
$5$ & $ \phi^7 $ & $ 7 - \frac{7}{2}\varepsilon + \varepsilon\gamma $ & --- & $ \frac{7}{2} $ & $ 2+\frac{8}{n}+\ldots $
 \\ 
$6$ & $ \phi^7 $ & $ 7 - \frac{7}{2}\varepsilon + \varepsilon\gamma $ & $5.130990 $ & $ \frac{14}{3} $ & $ \frac{10}{3}+\frac{14}{3 n}+\ldots $
 \\ 
\hline
\tableheading1
\\\hline
$1$ & $ \partial \phi^3 $ & $ 4 - \frac{3}{2}\varepsilon + \varepsilon\gamma $ & $\frac{1}{3} $ & $ \frac{1}{4} $ & $ \frac{1}{n} $
 \\ 
$2$ & $ \partial \phi^5 $ & $ 6 - \frac{5}{2}\varepsilon + \varepsilon\gamma $ &--- & $ \frac{5}{4} $ & $\frac{5 n+18-\sqrt{n^2+32 n-80}}{6 n}$
 \\ 
$3$ & $ \partial \phi^5 $ & $ 6 - \frac{5}{2}\varepsilon + \varepsilon\gamma $ & $\frac{4}{3} $ & $ \frac{5}{4} $ & $ \frac{n+1}{n} $ 
 \\ 
$4$ & $ \partial \phi^5 $ & $ 6 - \frac{5}{2}\varepsilon + \varepsilon\gamma $ & $\frac{19}{9} $ & $ \frac{23}{12} $ & $\frac{5 n+18+\sqrt{n^2+32 n-80}}{6 n}$
 \\ 
\hline
\tableheading2
\\\hline
$1$ & $ \partial^2 \phi^3 $ & $ 5 - \frac{3}{2}\varepsilon + \varepsilon\gamma $ & $\frac{4}{27} $ & $ \frac{1}{9} $ & $ \frac{4}{9 n} $
 \\ 
$2$ & $ \partial^2 \phi^5 $ & $ 7 - \frac{5}{2}\varepsilon + \varepsilon\gamma $ & --- & $ \frac{8}{9} $ & $ \frac{5}{9}+\frac{13}{9 n}+\ldots $
 \\ 
$3$ & $ \partial^2 \phi^5 $ & $ 7 - \frac{5}{2}\varepsilon + \varepsilon\gamma $ & $1.108335 $ & $ \frac{19}{18} $ & $ \frac{2}{3}+\frac{13-\sqrt{273}}{18 n}+\ldots $
 \\ 
$4$ & $ \partial^2 \phi^5 $ & $ 7 - \frac{5}{2}\varepsilon + \varepsilon\gamma $ & --- & $ \frac{19}{18} $ & $ \frac{2}{3}+\frac{13+\sqrt{273}}{18 n}+\ldots $
 \\ 
$5$ & $ \partial^2 \phi^5 $ & $ 7 - \frac{5}{2}\varepsilon + \varepsilon\gamma $ & $1.305556 $ & $ \frac{47}{36} $ & $ 1+\frac{0.631604}n+\ldots$
 \\ 
$6$ & $ \partial^2 \phi^5 $ & $ 7 - \frac{5}{2}\varepsilon + \varepsilon\gamma $ & $1.517103 $ & $ \frac{4}{3} $ & $ 1+\frac{2.100793}{ n}+\ldots $
 \\ 
$7$ & $ \partial^2 \phi^5 $ & $ 7 - \frac{5}{2}\varepsilon + \varepsilon\gamma $ & $1.883820 $ & $ \frac{31}{18} $ & $ 1+\frac{4.267603}{x§n}+\ldots $
 \\ 
\hline
\tableheading3
\\\hline
$1$ & $ \partial^3 \phi^3 $ & $ 6 - \frac{3}{2}\varepsilon + \varepsilon\gamma $ & $\frac{5}{18} $ & $ \frac{5}{24} $ & $ \frac{5}{6 n} $
 \\ 
\hline
\tableheading4
\\\hline
$1$ & $ \partial^4 \phi^3 $ & $ 7 - \frac{3}{2}\varepsilon + \varepsilon\gamma $ & $0 $ & $ 0 $ & $ 0 $
 \\ 
$2$ & $ \partial^4 \phi^3 $ & $ 7 - \frac{3}{2}\varepsilon + \varepsilon\gamma $ & $\frac{8}{45} $ & $ \frac{2}{15} $ & $ \frac{8}{15 n} $
 \\ 
\hline
\end{tabular}
}
\end{table}

\begin{table}
\centering
\caption{Irrep $B_3$ operators. If the ``$n$ gen.'' term contains no ellipses ($\ldots$), it is exact. There are no $B_3$ operators of spin $\ell=4$ with $\Delta\leqslant7$.}\label{tab:irrepB3}
{\small
\renewcommand{\arraystretch}{1.25}
\begin{tabular}{|c|c|l|lll|}
\hline
\tableheading0
\\\hline
$1$ & $ \square^2 \phi^5 $ & $ 9 - \frac{5}{2}\varepsilon + \varepsilon\gamma $ & $\frac{25}{27} $ & $ \frac{5}{6} $ & $ \frac{5 (n+2)}{9 n} $
 \\ 
$2$ & $ \square \phi^7 $ & $ 9 - \frac{7}{2}\varepsilon + \varepsilon\gamma $ & --- & $ \frac{5}{2} $ & $\frac{11 n+23-\sqrt{n^2+26 n-71}}{6 n}$
 \\ 
$3$ & $ \square \phi^7 $ & $ 9 - \frac{7}{2}\varepsilon + \varepsilon\gamma $ & $\frac{10}{3} $ & $ \frac{37}{12} $ & $\frac{11 n+23+\sqrt{n^2+26 n-71}}{6 n}$ 
 \\ 
$4$ & $ \phi^9 $ & $ 9 - \frac{9}{2}\varepsilon + \varepsilon\gamma $ & $\frac{59}{9} $ & $ 6 $ & $ \frac{13 n+20}{3 n} $
 \\ 
 \hline
\tableheading1
\\\hline
$1$ & $ \partial \phi^5 $ & $ 6 - \frac{5}{2}\varepsilon + \varepsilon\gamma $ & $\frac{4}{3} $ & $ \frac{5}{4} $ & $ \frac{n+1}{n} $
 \\ 
\hline
\tableheading2
\\\hline
$1$ & $ \partial^2 \phi^5 $ & $ 7 - \frac{5}{2}\varepsilon + \varepsilon\gamma $ & $\frac{29}{27} $ & $ \frac{19}{18} $ & $ \frac{9 n+2}{9 n} $
 \\ 
\hline
\tableheading3
\\\hline
$1$ & $ \partial^3 \phi^3 $ & $ 6 - \frac{3}{2}\varepsilon + \varepsilon\gamma $ & $0 $ & $ 0 $ & $ 0 $
 \\ 
\hline
\end{tabular}
}
\end{table}

\begin{table}
\centering
\caption{Irrep $X$ operators. If the ``$n$ gen.'' term contains no ellipses ($\ldots$), it is exact.}\label{tab:irrepX}
{\small
\renewcommand{\arraystretch}{1.25}
\begin{tabular}{|c|c|l|lll|}
\hline
\tableheading0
\\\hline
$1$ & $ \phi^2 $ & $ 2 - \varepsilon + \varepsilon\gamma $ & $\frac{1}{9} $ & $ \frac{1}{6} $ & $ \frac{n-2}{3 n} $
 \\ 
$2$ & $ \phi^4 $ & $ 4 - 2\varepsilon + \varepsilon\gamma $ & $1 $ & $ 1 $ & $ 1 $
 \\ 
$3$ & $ \phi^4 $ & $ 4 - 2\varepsilon + \varepsilon\gamma $ & $\frac{14}{9} $ & $ \frac{5}{3} $ & $ \frac{2 (3 n-2)}{3 n} $
 \\ 
$4$ & $ \phi^6 $ & $ 6 - 3\varepsilon + \varepsilon\gamma $ & $2.415324 $ & $ \frac{5}{2} $ & $ \frac{5}{3}+\frac{2}{3 n}+\ldots $
 \\ 
$5$ & $ \square \phi^4 $ & $ 6 - 2\varepsilon + \varepsilon\gamma $ & $\frac{5}{9} $ & $ \frac{2}{3} $ & $ \frac{3 n-4}{3 n} $
 \\ 
$6$ & $ \phi^6 $ & $ 6 - 3\varepsilon + \varepsilon\gamma $ & --- & $ \frac{5}{2} $ & $ \frac{7}{3}+\frac{14}{3 n}+\ldots $
 \\ 
$7$ & $ \phi^6 $ & $ 6 - 3\varepsilon + \varepsilon\gamma $ & $\frac{31}{9} $ & $ \frac{7}{2} $ & $ \frac{8}{3}+\frac{62}{21 n}+\ldots $
 \\ 
$8$ & $ \phi^6 $ & $ 6 - 3\varepsilon + \varepsilon\gamma $ & $4.140232 $ & $ \frac{25}{6} $ & $ 5-\frac{30}{7 n}+\ldots $
 \\ 
\hline
\tableheading1
\\\hline
$1$ & $ \partial \phi^6 $ & $ 7 - 3\varepsilon + \varepsilon\gamma $ & $2.083233 $ & $ 2 $ & $ \frac{5}{3}+\frac{2}{3 n}+\ldots $
 \\ 
$2$ & $ \partial \square \phi^4 $ & $ 7 - 2\varepsilon + \varepsilon\gamma $ & $\frac{19}{27} $ & $ \frac{7}{9} $ & $ \frac{9 n-8}{9 n} $
 \\ 
$3$ & $ \partial \phi^6 $ & $ 7 - 3\varepsilon + \varepsilon\gamma $ & $2.601425 $ & $ \frac{5}{2} $ & $ \frac{7}{3}+\frac{8}{3 n}+\ldots $
 \\ 
$4$ & $ \partial \phi^6 $ & $ 7 - 3\varepsilon + \varepsilon\gamma $ & $3.315342 $ & $ \frac{19}{6} $ & $ \frac{8}{3}+\frac{2}{3 n}+\ldots $
 \\ 
\hline
\tableheading2
\\\hline
$1$ & $ \partial^2 \phi^2 $ & $ 4 - \varepsilon + \varepsilon^2\gamma^{(2)} $ & $\gamma^{(2)}=\frac1{81} $ & $ \gamma^{(2)}=\frac1{108}$ & $ \gamma^{(2)}=\frac1{27n}$
 \\ 
$2$ & $ \partial^2 \phi^4 $ & $ 6 - 2\varepsilon + \varepsilon\gamma $ & $0.3925419 $ & $ 0.4535657 $ & $ \frac{1}{3}+\frac{22}{27 n}+\ldots $
 \\ 
$3$ & $ \partial^2 \phi^4 $ & $ 6 - 2\varepsilon + \varepsilon\gamma $ & $0.6620413 $ & $ 0.6747962 $ & $ \frac{2}{3}+\frac{46}{189 n}+\ldots $
 \\ 
$4$ & $ \partial^2 \phi^4 $ & $ 6 - 2\varepsilon + \varepsilon\gamma $ & $0.7510337 $ & $ \frac{13}{18} $ & $ 1-\frac{50}{27 n}+\ldots $
 \\ 
$5$ & $ \partial^2 \phi^4 $ & $ 6 - 2\varepsilon + \varepsilon\gamma $ & $1.194383 $ & $ 1.260527 $ & $ \frac{13}{9}-\frac{34}{63 n}+\ldots $
 \\ 
\hline
\tableheading3
\\\hline
$1$ & $ \partial^3 \phi^4 $ & $ 7 - 2\varepsilon + \varepsilon\gamma $ & $0.3214762 $ & $ 0.3816087 $ & $ \frac{1}{3}+\frac{4}{9 n}+\ldots $
 \\ 
$2$ & $ \partial^3 \phi^4 $ & $ 7 - 2\varepsilon + \varepsilon\gamma $ & $0.5163359 $ & $ \frac{1}{2} $ & $ \frac{2}{3}-\frac{8}{9 n}+\ldots $
 \\ 
$3$ & $ \partial^3 \phi^4 $ & $ 7 - 2\varepsilon + \varepsilon\gamma $ & $0.7547806 $ & $ 0.8128358 $ & $ 1-\frac{7}{9 n}+\ldots $
 \\ 
\hline
\tableheading4
\\\hline
$1$ & $ \partial^4 \phi^2 $ & $ 6 - \varepsilon + \varepsilon^2\gamma^{(2)} $ & $\gamma^{(2)}= \frac{22}{1215}$ & $ \gamma^{(2)}= \frac{5}{288}$ & $ \gamma^{(2)}=\frac{7 n^2+13 n-14}{540 n^2} $
\\\hline
\end{tabular}
}
\end{table}

\begin{table}
\centering
\caption{Scalar irrep $Z$ operators. If the ``$n$ gen.'' term contains no ellipses ($\ldots$), it is exact.}\label{tab:irrepZscalar}
{\small
\renewcommand{\arraystretch}{1.25}
\begin{tabular}{|c|c|l|lll|}
\hline
\tableheading0
\\\hline
$1$ & $ \phi^2 $ & $ 2 - \varepsilon + \varepsilon\gamma $ & $\frac{2}{9} $ & $ \frac{1}{6} $ & $ \frac{2}{3 n} $
 \\ 
$2$ & $ \phi^4 $ & $ 4 - 2\varepsilon + \varepsilon\gamma $ & $1.180858 $ & $ 1 $ & $ \frac{5 n+12-\sqrt{n^2+24 n-48}}{6 n}$
 \\ 
$3$ & $ \phi^4 $ & $ 4 - 2\varepsilon + \varepsilon\gamma $ & $1.819142 $ & $ \frac{5}{3} $ & $ \frac{5 n+12+\sqrt{n^2+24 n-48}}{6 n}$
\\
$4$ & $ \phi^6 $ & $ 6 - 3\varepsilon + \varepsilon\gamma $ & $2.467204 $ & $ \frac{5}{2} $ & $ \frac{4}{3}-\frac{2}{3 n}+\ldots $
 \\ 
$5$ & $ \square \phi^4 $ & $ 6 - 2\varepsilon + \varepsilon\gamma $ & $\frac{2}{3} $ & $ \frac{2}{3} $ & $ \frac{2}{3} $
 \\ 
$6$ & $ \phi^6 $ & $ 6 - 3\varepsilon + \varepsilon\gamma $ & $2.762741 $ & $ \frac{5}{2} $ & $ \frac{5}{3}+\frac{22}{3 n}+\ldots $
 \\ 
$7$ & $ \phi^6 $ & $ 6 - 3\varepsilon + \varepsilon\gamma $ & --- & $ \frac{5}{2} $ & $ 2+\frac{6}{n}-\frac{\sqrt{96}}{n^{3/2}}+\ldots $
 \\ 
$8$ & $ \phi^6 $ & $ 6 - 3\varepsilon + \varepsilon\gamma $ & $3.876133 $ & $ \frac{7}{2} $ & $ 2+\frac{6}{n}+\frac{\sqrt{96}}{n^{3/2}}+\ldots $
 \\ 
$9$ & $ \phi^6 $ & $ 6 - 3\varepsilon + \varepsilon\gamma $ & $4.560589 $ & $ \frac{25}{6} $ & $ \frac{10}{3}+\frac{2}{3 n}+\ldots $
 \\ 
\hline
\end{tabular}
}
\end{table}

\begin{table}
\centering
\caption{Spinning irrep $Z$ operators. If the ``$n$ gen.'' term contains no ellipses ($\ldots$), it is exact.}\label{tab:irrepZspinning}
{\small
\renewcommand{\arraystretch}{1.25}
\begin{tabular}{|c|c|l|lll|}
\hline
\tableheading1
\\\hline
$1$ & $ \partial \phi^4 $ & $ 5 - 2\varepsilon + \varepsilon\gamma $ & $\frac{2}{3} $ & $ \frac{2}{3} $ & $ \frac{2}{3} $
 \\ 
$2$ & $ \partial \phi^6 $ & $ 7 - 3\varepsilon + \varepsilon\gamma $ & $2.083233 $ & $ 2 $ & $ \frac{4}{3}-\frac{2}{3 n}+\ldots $
 \\ 
$3$ & $ \partial \square \phi^4 $ & $ 7 - 2\varepsilon + \varepsilon\gamma $ & $0.4528888 $ & $ \frac{4}{9} $ & $ \frac{5}{9}-\frac{16}{9 n}+\ldots $
 \\ 
$4$ & $ \partial \square \phi^4 $ & $ 7 - 2\varepsilon + \varepsilon\gamma $ & $0.7693334 $ & $ \frac{7}{9} $ & $ \frac{2}{3}+\frac{16}{9 n}+\ldots $
 \\ 
$5$ & $ \partial \phi^6 $ & $ 7 - 3\varepsilon + \varepsilon\gamma $ & --- & $ 2 $ & $ \frac{5}{3}+\frac{60-2\sqrt{85}}{15n}+\ldots $ 
 \\ 
$6$ & $ \partial \phi^6 $ & $ 7 - 3\varepsilon + \varepsilon\gamma $ & $2.601425 $ & $ \frac{5}{2} $ & $ \frac{5}{3}+\frac{60+2\sqrt{85}}{15n}+\ldots $
 \\ 
$7$ & $ \partial \phi^6 $ & $ 7 - 3\varepsilon + \varepsilon\gamma $ & $\frac{26}{9} $ & $ \frac{17}{6} $ & $ 2+\frac{4}{n}+\ldots $
 \\ 
$8$ & $ \partial \phi^6 $ & $ 7 - 3\varepsilon + \varepsilon\gamma $ & $3.315342 $ & $ \frac{19}{6} $ & $ \frac{10}{3}-\frac{4}{3 n}+\ldots $
 \\ 
\hline
\tableheading2
\\\hline
$1$ & $ \partial^2 \phi^2 $ & $ 4 - \varepsilon + \varepsilon^2\gamma^{(2)} $ & $\gamma^{(2)}=\frac{1}{162} $ & $ \gamma^{(2)}=\frac{1}{108}$ & $ \gamma^{(2)}=\frac{n-2}{54 n} $
 \\ 
$2$ & $ \partial^2 \phi^4 $ & $ 6 - 2\varepsilon + \varepsilon\gamma $ & $0.4307964 $ & $ 0.4535657 $ & $ 0+\frac{58}{27 n}+\ldots $
 \\ 
$3$ & $ \partial^2 \phi^4 $ & $ 6 - 2\varepsilon + \varepsilon\gamma $ & $0.5365007 $ & $ \frac{1}{2} $ & $ \frac{5}{9}-\frac{8}{9 n}+\ldots $
 \\ 
$4$ & $ \partial^2 \phi^4 $ & $ 6 - 2\varepsilon + \varepsilon\gamma $ & $0.6763453 $ & $ 0.6747962 $ & $ \frac{2}{3}+\frac{13-\sqrt{889}}{27 n}+\ldots $
 \\ 
$5$ & $ \partial^2 \phi^4 $ & $ 6 - 2\varepsilon + \varepsilon\gamma $ & $0.8669070 $ & $ \frac{13}{18} $ & $ \frac{2}{3}+\frac{13+\sqrt{889}}{27 n}+\ldots $ 
 \\ 
$6$ & $ \partial^2 \phi^4 $ & $ 6 - 2\varepsilon + \varepsilon\gamma $ & $1.341302 $ & $ 1.260527 $ & $ 1+\frac{2}{3 n}+\ldots $
 \\ 
\hline
\tableheading3
\\\hline
$1$ & $ \partial^3 \phi^4 $ & $ 7 - 2\varepsilon + \varepsilon\gamma $ & $0.2720625 $ & $ \frac{5}{18} $ & $ 0+\frac{16}{9 n}+\ldots $
 \\ 
$2$ & $ \partial^3 \phi^4 $ & $ 7 - 2\varepsilon + \varepsilon\gamma $ & $0.4145656 $ & $ 0.3816087 $ & $ \frac{1}{6}+\frac{17}{21 n}+\ldots $
 \\ 
$3$ & $ \partial^3 \phi^4 $ & $ 7 - 2\varepsilon + \varepsilon\gamma $ & $0.5596054 $ & $ \frac{1}{2} $ & $ \frac{5}{9}-\frac{142}{63 n}+\ldots $
 \\ 
$4$ & $ \partial^3 \phi^4 $ & $ 7 - 2\varepsilon + \varepsilon\gamma $ & $0.6378218 $ & $ \frac{7}{12} $ & $ \frac{2}{3}+\frac{15-\sqrt{505}}{18 n}+\ldots $
 \\ 
$5$ & $ \partial^3 \phi^4 $ & $ 7 - 2\varepsilon + \varepsilon\gamma $ & $0.8381670 $ & $ 0.8128358 $ & $ \frac{2}{3}+\frac{15+\sqrt{505}}{18 n}+\ldots $
 \\ 
\hline
\tableheading4
\\\hline
$1$ & $ \partial^4 \phi^2 $ & $ 6 - \varepsilon + \varepsilon^2\gamma^{(2)} $ & $\gamma^{(2)}=\frac{79}{4860}  $ & $ \gamma^{(2)}=\frac{5}{288}$ & $ \gamma^{(2)}=\frac{10 n^2+n-14}{540 n^2}$
\\
\hline
\end{tabular}
}
\end{table}

\begin{table}
\centering
\caption{Irrep $B$ operators. If the ``$n$ gen.'' term contains no ellipses ($\ldots$), it is exact. There are no $B$ operators of spin $\ell=4$ and $\Delta\leqslant7$.}\label{tab:irrepB}
{\small
\renewcommand{\arraystretch}{1.25}
\begin{tabular}{|c|c|l|lll|}
\hline
\tableheading0
\\\hline
$1$ & $ \phi^6 $ & $ 6 - 3\varepsilon + \varepsilon\gamma $ & $\frac{25}{9} $ & $ \frac{5}{2} $ & $ \frac{5 (n+2)}{3 n} $
 \\ 
$2$ & $ \phi^6 $ & $ 6 - 3\varepsilon + \varepsilon\gamma $ & $\frac{32}{9} $ & $ \frac{7}{2} $ & $ \frac{2 (5 n+1)}{3 n} $
 \\ 
\hline
\tableheading1
\\\hline
$1$ & $ \partial \phi^2 $ & $ 3 - \varepsilon + \varepsilon^2\gamma^{(2)} $ & $\gamma^{(2)}= \frac{1}{486}$ & $ \gamma^{(2)}=0$ & $ \gamma^{(2)}= \frac{(n-4)^2}{54 n^2}  $
 \\ 
$2$ & $ \partial \phi^4 $ & $ 5 - 2\varepsilon + \varepsilon\gamma $ & $0.7153830 $ & $ \frac{2}{3} $ & $ \frac{2}{3}-\frac{2}{3 n}+\ldots $
 \\ 
$3$ & $ \partial \phi^4 $ & $ 5 - 2\varepsilon + \varepsilon\gamma $ & $1.173506 $ & $ \frac{7}{6} $ & $ 1+\frac{4}{3 n}+\ldots $
 \\ 
$4$ & $ \partial \square \phi^4 $ & $ 7 - 2\varepsilon + \varepsilon\gamma $ & $0.2573581 $ & $ \frac{5}{18} $ & $ 0+\frac{40}{27 n}+\ldots $
 \\ 
$5$ & $ \partial \phi^6 $ & $ 7 - 3\varepsilon + \varepsilon\gamma $ & $2.051077 $ & $ 2 $ & $ \frac{4}{3}-\frac{4}{3 n}+\ldots $
 \\ 
$6$ & $ \partial \square \phi^4 $ & $ 7 - 2\varepsilon + \varepsilon\gamma $ & $0.5238973 $ & $ \frac{4}{9} $ & $ \frac{5}{9}-\frac{4}{3 n}+\ldots $
 \\ 
$7$ & $ \partial \square \phi^4 $ & $ 7 - 2\varepsilon + \varepsilon\gamma $ & $0.6631890 $ & $ \frac{2}{3} $ & $ \frac{2}{3}+\frac{14}{27 n}+\ldots $
 \\ 
$8$ & $ \partial \phi^6 $ & $ 7 - 3\varepsilon + \varepsilon\gamma $ & $2.133197 $ & $ 2 $ & $ \frac{5}{3}+\frac{40-2\sqrt{205}}{15 n}+\ldots $
 \\ 
$9$ & $ \partial \phi^6 $ & $ 7 - 3\varepsilon + \varepsilon\gamma $ & --- & $ 2 $ & $ \frac{5}{3}+\frac{40+2\sqrt{205}}{15 n}+\ldots $
 \\ 
$10$ & $ \partial \phi^6 $ & $ 7 - 3\varepsilon + \varepsilon\gamma $ & $2.524732 $ & $ \frac{5}{2} $ & $ 2+\frac{4}{n}-\frac{4\sqrt2}{n^{3/2}}+\ldots $
 \\ 
$11$ & $ \partial \phi^6 $ & $ 7 - 3\varepsilon + \varepsilon\gamma $ & $2.924843 $ & $ \frac{17}{6} $ & $ 2+\frac{4}{n}+\frac{4\sqrt2}{n^{3/2}}+\ldots $
 \\ 
$12$ & $ \partial \phi^6 $ & $ 7 - 3\varepsilon + \varepsilon\gamma $ & $3.477263 $ & $ \frac{10}{3} $ & $ \frac{10}{3}-\frac{4}{3 n}+\ldots $
 \\ 
\hline
\tableheading2
\\\hline
$1$ & $ \partial^2 \phi^4 $ & $ 6 - 2\varepsilon + \varepsilon\gamma $ & $0.5047303 $ & $ \frac{1}{2} $ & $ \frac{5}{9}-\frac{2}{3 n}+\ldots $
 \\ 
$2$ & $ \partial^2 \phi^4 $ & $ 6 - 2\varepsilon + \varepsilon\gamma $ & $0.7174919 $ & $ \frac{13}{18} $ & $ \frac{2}{3}+\frac{2}{3 n}+\ldots $
 \\ 
\hline
\tableheading3
\\\hline
$1$ & $ \partial^3 \phi^2 $ & $ 5 - \varepsilon + \varepsilon^2\gamma^{(2)} $ & $\gamma^{(2)}=\frac{17}{972} $ & $ \gamma^{(2)}=\frac{5}{288}$ & $ \gamma^{(2)}= \frac{2 n^2-n+2}{108 n^2}$
\\
$2$ & $ \partial^3 \phi^4 $ & $ 7 - 2\varepsilon + \varepsilon\gamma $ & $0.2617865 $ & $ \frac{5}{18} $ & $ 0+\frac{65}{54 n}+\ldots $
 \\ 
$3$ & $ \partial^3 \phi^4 $ & $ 7 - 2\varepsilon + \varepsilon\gamma $ & $0.3672239 $ & $ 0.2948415 $ & $ \frac{1}{6}+\frac{254}{315 n}+\ldots $
 \\ 
$4$ & $ \partial^3 \phi^4 $ & $ 7 - 2\varepsilon + \varepsilon\gamma $ & $0.4277185 $ & $ 0.4966984 $ & $ \frac{5}{9}-\frac{16}{21 n}+\ldots $
 \\ 
$5$ & $ \partial^3 \phi^4 $ & $ 7 - 2\varepsilon + \varepsilon\gamma $ & $0.5134713 $ & $ \frac{1}{2} $ & $ \frac{2}{3}-\frac{53+\sqrt{1033}}{108 n}+\ldots $
 \\ 
$6$ & $ \partial^3 \phi^4 $ & $ 7 - 2\varepsilon + \varepsilon\gamma $ & $0.6353068 $ & $ \frac{7}{12} $ & $ \frac{2}{3}-\frac{53-\sqrt{1033}}{108 n}+\ldots $
 \\ 
$7$ & $ \partial^3 \phi^4 $ & $ 7 - 2\varepsilon + \varepsilon\gamma $ & $1.072271 $ & $ 1.069571 $ & $ 1+\frac{2}{5 n}+\ldots $
 \\ 
\hline
\tableheading5
\\\hline
$1$ & $ \partial^5 \phi^2 $ & $ 7 - \varepsilon + \varepsilon^2\gamma^{(2)} $ & $\gamma^{(2)}=\frac{47}{2430} $ & $ \gamma^{(2)}=\frac{7}{360}$ & $ \gamma^{(2)}= \frac{5 n^2+2 n-4}{270 n^2}$
\\
\hline
\end{tabular}
}
\end{table}

\begin{table}
\centering
\caption{Irrep $\XXbar$ operators.}\label{tab:irrepXXbar}
{\small
\renewcommand{\arraystretch}{1.25}
\begin{tabular}{|c|c|l|lll|}
\hline
\tableheading0
\\\hline
$1$ & $ \phi^6 $ & $ 6 - 3\varepsilon + \varepsilon\gamma $ & $\frac{23}{9} $ & $ \frac{5}{2} $ & $ \frac{7 n+2}{3 n} $
 \\ 
\hline
\tableheading1
\\\hline
$1$ & $ \partial \phi^4 $ & $ 5 - 2\varepsilon + \varepsilon\gamma $ & $\frac{2}{3} $ & $ \frac{2}{3} $ & $ \frac{2}{3} $
 \\ 
$2$ & $ \partial \phi^6 $ & $ 7 - 3\varepsilon + \varepsilon\gamma $ & --- & $ 2 $ & $\frac{11 n+14-\sqrt{9 n^2+12 n-92}}{6 n} $
 \\ 
$3$ & $ \partial \square \phi^4 $ & $ 7 - 2\varepsilon + \varepsilon\gamma $ & $\frac{10}{27} $ & $ \frac{4}{9} $ & $ \frac{2 (3 n-4)}{9 n} $
 \\ 
$4$ & $ \partial \phi^6 $ & $ 7 - 3\varepsilon + \varepsilon\gamma $ & $\frac{26}{9} $ & $ \frac{17}{6} $ & $\frac{11 n+14+\sqrt{9 n^2+12 n-92}}{6 n} $
 \\ 
\hline
\tableheading2
\\\hline
$1$ & $ \partial^2 \phi^4 $ & $ 6 - 2\varepsilon + \varepsilon\gamma $ & $\frac{5}{9} $ & $ \frac{1}{2} $ & $ \frac{n+2}{3 n} $
 \\ 
\hline
\tableheading3
\\\hline
$1$ & $ \partial^3 \phi^4 $ & $ 7 - 2\varepsilon + \varepsilon\gamma $ & $0.2114258 $ & $ \frac{5}{18} $ & $ \frac{9 n-5-\sqrt{9 n^2-54 n+193}}{18 n} $
 \\ 
$2$ & $ \partial^3 \phi^4 $ & $ 7 - 2\varepsilon + \varepsilon\gamma $ & $0.6033890 $ & $ \frac{7}{12} $ &$ \frac{9 n-5+\sqrt{9 n^2-54 n+193}}{18 n} $
 \\ 
\hline
\end{tabular}
}
\end{table}

\subsubsection{Operators with three fields}
In the case of operators with three fields and spin-$\ell$ Lorentz representation, we can give a complete characterization of the spectrum. This is completely analogous to the observations in \cite{Kehrein:1992fn} for the case of the $O(n)$ model. 

The structure is as follows. In each representation, there is a maximum number of operators at each spin that acquire a non-zero anomalous dimension at order $\varepsilon$. Apart from these operators, at sufficiently large spin new primary operators appear with vanishing order-$\varepsilon$ anomalous dimension. These operators have a degeneracy that grows with spin.

For the vector ($V$) irrep we find that the non-vanishing anomalous dimensions are solutions of the polynomial equation
\begin{align}
&\lambda^3-\left(\frac{3n-2}{3n}+\frac{2(-1)^\ell}{3(\ell+1)}\right)\lambda^2+\left(\frac{2n^2-4}{9n^2}+\frac{(4n^2-2n-4)(-1)^\ell}{9n^2(\ell+1)}-\frac{4n-8}{9n^2(\ell+1)^2}\right)\lambda
\nonumber
\\
&\quad +\frac{(n-1)(n-2)}{27n^3}\left(-4-\frac{12(-1)^\ell}{\ell+1}+\frac{16(-1)^\ell}{(\ell+1)^3}\right)=0.
\end{align}
For the $XV$ irrep we find that the non-vanishing anomalous dimensions are solutions of the polynomial equation
\begin{align}
&\lambda^2-\left(\frac13+\frac{2(-1)^\ell}{3n(\ell+1)}\right)\lambda+\frac{n-2}{9n^2}\left(2+\frac{2(-1)^\ell}{\ell+1}-\frac{4}{(\ell+1)^2}\right)=0.
\end{align}
For the $VB$ irrep, there is one non-vanishing anomalous dimension:
\begin{equation}
\gamma_{\ell}=\frac2{3n}-\frac{2(-1)^\ell}{3n(\ell+1)}, \quad \ell=1,2,3,\ldots.
\end{equation}
For the $Z_3$ irrep, there is one non-vanishing anomalous dimension:
\begin{equation}
\gamma_{\ell}=\frac2{3n}+\frac{4(-1)^\ell}{3n(\ell+1)}, \quad \ell=0,2,3,4,\ldots.
\end{equation}
Finally, in the $B_3$ irrep, there are no operators with non-zero leading-order anomalous dimensions.

\subsection{Comparison to large $n$}
As shown in \cite{Binder:2021vep}, the cubic fixed point can be written as a controlled deformation (in $1/n$) of $n$ decoupled Ising models. In this picture all operators are built from powers of the auxiliary field $\sigma$ with dimension $d-\Delta_{\phi^2}$ together with operators at the Ising fixed point, such as $\phi$, $\phi^2$ etc. See also \cite{Komargodski:2016auf}, where the $C_n$ fixed point is written as a deformation of $n$ decoupled Ising models using conformal perturbation theory. The caveat being that this expansion is uncontrolled in generic $d$.\footnote{In the special cases $d=4-\varepsilon$ and $d=2+\varepsilon$, conformal perturbation theory does become controlled (see e.g. \cite{Komargodski:2016auf}). In the first case it should reduce to the $\varepsilon$ expansion, whereas in the second case one must input coefficients numerically.}

To illustrate these facts, in tables \ref{TableQuadraticLargeN}, \ref{TableCubicLargeN} and \ref{TableQuarticLargeN} we show how operators are constructed in the large $n$ limit around decoupled Ising models. For example, for $Z_4$ we label its large $n$ limit as $4\Delta_\phi$. This is shorthand for $\Delta_{Z_4} = 4\Delta_\phi^{\mathrm{Ising}} + O(1/n)$, where $\Delta_\phi^{\mathrm{Ising}}$ is the scaling dimension of $\phi$ evaluated at the Ising fixed point.

\begin{table}
\caption{Quadratic operators at Large $n$.}
\label{TableQuadraticLargeN}
\centering
\begin{tabular}{|c|c|lr|lr|}
\hline
 Irrep & Large $n$ limit
\\\hline
 $S$ & $d-\Delta_{\phi^2}$  
\\
 $X$ & $\Delta_{\phi^2}$ 
\\
 $Z$ & $2\Delta_\phi$
\\\hline
\end{tabular}
\end{table}

\begin{table}
\caption{Cubic (in powers of the field) operators at Large $n$.}
\label{TableCubicLargeN}
\centering
\begin{tabular}{|c|c|lr|lr|}
\hline
 Irrep & Large $n$ limit
\\\hline
 $(V)_1$ & $d-\Delta_\phi$
 \\
 $(V)_2$ & $\Delta_\phi+d-\Delta_{\phi^2}$
\\
$XV$        & $\Delta_\phi+\Delta_{\phi^2}$
\\
$Z_3$ &  $3\Delta_\phi$
\\\hline
\end{tabular}
\end{table}

\begin{table}
\caption{Quartic operators at Large $n$.}
\label{TableQuarticLargeN}
\centering
\begin{tabular}{|c|c|}
\hline
 Irrep & Large $n$ limit
\\\hline
 $(S)_1$ &   $\Delta_{\phi^4}$
\\
 $(S)_2$ & $2(d-\Delta_{\phi^2})$
\\
$(X)_1$  & $d$ 
\\
$(X)_2$ &   $\Delta_{\phi^4}$
\\
$(Z)_1$  & $\Delta_{\phi^2}+2\Delta_\phi$ 
\\
$(Z)_2$ & $2\Delta_\phi +2$  
\\
$B_2$   & $d$  
\\
$Z_4$ & $4\Delta_\phi$  
\\
$XX$ & $2\Delta_{\phi^2}$  
\\
$XZ$  & $\Delta_{\phi^2}+2\Delta_\phi$
\\\hline
\end{tabular}
\end{table}

\subsection{Comparison to conformal perturbation theory}
As mentioned previously, the $C_n$ theory can be seen as a deformation of the $O(n)$ theory. This is the viewpoint used in \cite{Chester:2020iyt}. In this case, the flow from $O(3)$ to the $C_3$ fixed point is triggered by the an operator in the $T_4$ ($4$-index traceless symmetric) irrep of $O(3)$. Hence, if the scaling dimension of this operator is relevant, then the $O(3)$ fixed point is unstable with respect to the $T_4$ deformation and flows to the cubic fixed point. In \cite{Chester:2020iyt}, $\Delta_{T4}$ was found to be $<3$ but also very close to $\Delta_{T4}=3$, whereas in \cite{Hasenbusch2011} it was found to be  $\Delta_{T4}=2.987(4)$. This is indeed also corroborated from our resummation, see e.g.\ table \ref{TableQuartic} where $\Delta_{T4}=2.992$.

This flow, i.e. from $O(3)$ to $C_3$, may be consequently studied using conformal perturbation theory \cite{Cardy:1996xt}, see also the discussion session in \cite{Chester:2020iyt}. Assume that we deform a theory with an operator $\mathcal O$, such that $\delta \mathcal{L} =g \int d^dx \mathcal O$, then from \cite{Cardy:1996xt} we obtain
\begin{equation}
\beta (g) =-(d-\Delta_{\mathcal O})g + C^{\mathcal O}_{\mathcal O\mathcal O}g^2\,,
\label{conformalpert}
\end{equation}
where $C^{\mathcal O}_{\mathcal O\mathcal O}$ is the OPE coefficient with which the operator $O$ is exchanged in the OPE of $\mathcal O$ with itself. In our case $\mathcal O=T_4$ thus $d-\Delta_{\mathcal O} = 0.008$ and hence $g_{F.P.}=0.008/C^{\mathcal O}_{\mathcal O\mathcal O}$. Assuming $C^{\mathcal O}_{\mathcal O\mathcal O}$ is not very small\footnote{Which indeed we don't expect it to be, since in $d=4-\varepsilon$ dimensions it starts at $O(\varepsilon^0)$.} then $g\ll 1$ and hence we expect the fixed points to be ``close'', in the RG sense of the word. This can indeed be confirmed from our results. See for example tables \ref{TableSplittingT2Quadratic}, \ref{TableSplittingT2Quartic}, \ref{TableSplittingT3} and \ref{TableSplittingT4} (see also \cite{Antipin:2019vdg} for more information on the branching rules as $O(n) \rightarrow C_n$). In these we indeed observe that the scaling dimensions that result from breaking the irreps $T_2$, $T_3$ and $T_4$ to their $C_n$ sub-representations are incredibly close to each other. We also observe that as $n$ is increased from $n=3$ to $n=4$ these do move further apart, as expected. Because as $n$ is raised the scaling dimension of $\Delta_{T_4}$ approaches the value $2+ O(1/n)$, i.e. it moves away from the value $3$. Let us point out that the $T_2$ irrep appears in both tables~\ref{TableSplittingT2Quadratic} and \ref{TableSplittingT2Quartic}, since the operator may be built with two and four powers of the order parameter field respectively. 

\begin{table}
\caption{Splitting of the irreducible representation $T_2$ (quadratic) as one breaks $O(n)$ to $C_n$.}
\label{TableSplittingT2Quadratic}
\centering
\begin{tabular}{|c|c|c|c|c|c|c|c|c|}
\hline
 $n$/Irrep & $T_2$& $X$& $Z$
\\\hline
 $n=3$ & 1.211 &1.204 &1.207
\\
 $n=4$ & 1.189 &1.284 &1.119
\\\hline
\end{tabular}
\end{table}

\begin{table}
\caption{Splitting of the irreducible representation $T_2$ (quartic) as one breaks $O(n)$ to $C_n$.}
\label{TableSplittingT2Quartic}
\centering
\begin{tabular}{|c|c|c|c|c|c|c|c|c|}
\hline
 $n$/Irrep & $T_2$& $X$& $Z$
\\\hline
 $n=3$ & 3.550 &3.554 &3.544
\\
 $n=4$ & 3.484 & 3.685 &3.340
\\\hline
\end{tabular}
\end{table}

\begin{table}
\caption{Splitting of the irreducible representation $T_3$ as one breaks $O(n)$ to $C_n$.}
\label{TableSplittingT3}
\centering
\begin{tabular}{|c|c|c|c|c|c|c|c|c|}
\hline
 $n$/Irrep & $T_3$ & $V$ & $XV$& $VB$
\\\hline
 $n=3$ &  2.043 & 2.054 & 2.043 & 2.066
\\
 $n=4$ &  1.985 & 2.090 &1.975 &1.830 
\\\hline
\end{tabular}
\end{table}

\begin{table}
\caption{Splitting of the irreducible representation $T_4$ as one breaks $O(n)$ to $C_n$.}
\label{TableSplittingT4}
\centering
\begin{tabular}{|c|c|c|c|c|c|c|c|c|}
\hline
 $n$/Irrep & $T_4$ & $S$ & $X$& $Z$&$B$ & $XX$ & $XZ$ &$Z_4$ 
\\\hline
 $n=3$ &  2.992 &3.011 & 3.003& 2.988&3 & & &
\\
 $n=4$ &  2.890 &3.081 &2.951 &2.782 & 3&2.873 &2.725 &2.570
\\\hline
\end{tabular}
\end{table}

\section{Discussion and outlook}

In this work we have performed a comprehensive study of the spectrum of anomalous dimensions in multi-scalar theories with hypercubic global symmetry. This was done with an eye towards the conformal bootstrap. 
Our analysis included a comprehensive treatise of the group theory and a computation of a large part of the low lying operator spectrum to high precision. We were able to corroborate expectations from the large-$n$ expansion as well as conformal perturbation theory. Our results are expected to be crucial for future bootstrap studies, e.g. enabling studies similar to \cite{Henriksson:2022gpa}, which were performed thanks to the detailed knowledge of the Ising spectrum. Additionally, we demonstrated how projectors from the conformal bootstrap can be used to make the computation as mechanical as possible, thus streamlining future computations. For example, given our current setup it will be straightforward to obtain results for other $S_n \ltimes G^n$ theories \cite{Kousvos:2021rar}, a class that includes the experimentally relevant $MN$ models, for which $G=O(M)$.\footnote{These theories were originally studied in the numerical bootstrap in \cite{Stergiou:2019dcv}.} Our results could also be of relevance for experiments regarding crossover exponents, for recent examples see \cite{Aharony:2022ajv,Aharony:2022gjg}. We also demonstrated systematic methods for building operators in a given irrep for generic values of $n$, as well as an automated method for computing anomalous dimensions at one loop for discrete global symmetries. Our work constitutes the first time such a big part of the spectrum of the hypercubic theory was computed. To the best of our knowledge this work presents the perturbative state of the art for this theory.\footnote{Anomalous dimensions and OPE coefficients for spinning operarators of leading twist were computed in \cite{Dey:2016mcs}. We have repeated their study using the Lorentzian inversion formula framework of \cite{Henriksson:2020fqi}, but not improved upon it.}

There are several open directions to pursue in the future. Perhaps the most direct are the aforementioned $MN$ theories. More generally, we would like to obtain precise estimates for operators higher up in the spectrum, e.g. operators irrelevant in $d=4-\varepsilon$. Presumably these were not previously studied since they would either determine corrections to scaling in statistical physics, or sub-leading effects in high energy physics, which would be very hard to measure. However, with the modern resurgence of the conformal bootstrap, such operators become crucial since they provide a quantitative picture of the spectrum, which allows gap assumptions necessarily to isolate theories in small islands in parameter space. Stated slightly differently, knowledge of various irrelevant operators can allow for more precise determinations for the experimentally relevant operators. Steps in this direction were taken in \cite{Bednyakov:2022guj}, where six loop results for operators in the $T_i$ representation of $O(n)$ were obtained for arbitrary $n$ and $i$. A number of irrelevant operators were also computed in \cite{Cao:2021cdt}. To this extent, it would be interesting to also extend the knowledge regarding operators with insertions of derivatives.
The results from \cite{Cao:2021cdt}, which contain higher-loop results for operators with contracted derivatives, albeit for scalar singlet operators only, are another useful extension that could be generalised to non-singlet operators.\footnote{We thank the authors of \cite{Cao:2021cdt} for useful discussions on this point.}
One interesting target is the antisymmetric spin-$1$ operator, which in theories with discrete global symmetries is the broken spin-$1$ flavour current (once these models are viewed as deformations of $O(n)$). Other such operators of interest include the first non-conserved operator in the representation of the stress tensor or the symmetry current (if it exists). The bootstrap has been found to be very sensitive to gaps imposed on such operators. 

While the present work focused on scaling dimensions, we would also like to determine the OPE coefficients of a theory in order to have the complete set of CFT data. Some steps in this direction where taken in \cite{Codello:2017hhh}, \cite{Rose:2021zdk} and \cite{Dey:2016mcs}. It should be possible to recover the OPE coefficient from the second derivative of the beta function in the same way that the scaling dimension is recovered from the first derivative. This idea is derived from the form of the beta function found in \cite{Cardy:1996xt}, where the OPE coefficient multiplies the term quadratic in the coupling.

\acknowledgments
We thank Andreas Stergiou for useful discussions on redundant operators and for providing helpful comments on the manuscript. 
AB is grateful to O. Antipin for clarifying  comments on Ref.~\cite{Antipin:2019vdg} and to A. Pikelner for collaboration on early stages of the project.
The research work of JH and SRK received funding from
the European Research Council (ERC) under the European Union's Horizon 2020 research
and innovation programme (grant agreement no. 758903).

\appendix

\section{More group theory}

\subsection{Tensor products}
\label{app:tensorProducts}
Below we give the tensor products of all irreps with rank up to two, and in general $n \geq 4$.\footnote{Nevertheless, some of these remain correct also for $n \leq 4$. One can always check this by seeing if irreps vanish on the RHS for $n=4$.} The subscripts ${}_{\mathrm{sym}}$ and ${}_{\mathrm{antisym}}$ refer to the behaviour of the operators on the RHS under swapping the operators on the LHS. When the operators on the LHS are different there is no such distinction.
\begin{align}
   V     \otimes      V   & =   (S \oplus  X \oplus  Z)_{\mathrm{sym}} \oplus ( B)_{\mathrm{antisym}}  ,\\
   V     \otimes    X & =  V \oplus  XV ,  \\
   V     \otimes    Z & =      V \oplus  XV  \oplus    VB  \oplus   Z_3,   \\
   V     \otimes    B & =  V \oplus XV  \oplus  B_3 \oplus    VB  , \\
 X   \otimes    X & = ( S\oplus  X \oplus XX   )_{\mathrm{sym}} \oplus ( \XXbar )_{\mathrm{antisym}} ,
 \\
 Z   \otimes    Z & = (  S \oplus{BB} \oplus  Z_4 \oplus  X \oplus  XX  \oplus   {XZ} \oplus  Z )_{\mathrm{sym}} \oplus ( B \oplus  {BZ} \oplus   {XB}  )_{\mathrm{antisym}}  ,  \\
 B   \otimes    B & = (S \oplus  {BB} \oplus    X \oplus  XX  \oplus   {XZ} \oplus Z    \oplus  B_4)_{\mathrm{sym}} \oplus (B \oplus   {XB} \oplus      VB_3   )_{\mathrm{antisym}} , \\
 X   \otimes    Z & =  Z\oplus B \oplus   {XZ} , \\
 X   \otimes    B & =  Z\oplus B \oplus   {XB}  , \\
 Z   \otimes    B & = Z\oplus  B \oplus   {BZ} \oplus  \XXbar  \oplus   {XB} \oplus   {XZ} \oplus    VB_3 \oplus  Z .
\end{align}
Note that the symmetry properties of $B_4$ and $VB_3$ under swapping the operators on the LHS can be worked out using a similar decomposition to that of equation~2.14 of \cite{Kousvos:2021rar}.\footnote{For this decomposition we take all four indices different, which is required in order for the desired irreps to appear.} We first define $T^{ab,cd}=B^{ab}B^{cd}+B^{cd}B^{ab}$ and $\bar{T}^{ab,cd}=B^{ab}B^{cd}-B^{cd}B^{ab}$, thus
\begin{equation}
B^{ab} \otimes B^{cd} \sim T^{ab,cd}+ \bar{T}^{ab,cd},
\end{equation}
which can be decomposed as
\begin{equation}
\begin{split}
B^{ab} \otimes B^{cd} \sim ( T^{ab,cd} +  T^{ad,bc}-T^{ac,bd}) + (T^{ac,bd}-T^{ab,cd}) + (T^{ab,cd}-T^{ad,bc})\\
+(\bar{T}^{ab,cd}-\bar{T}^{ad,cb})+(\bar{T}^{ab,cd}-\bar{T}^{ac,bd})+(\bar{T}^{ad,cb}+\bar{T}^{ac,bd})
\end{split}
\end{equation}
and finally we relabel this as
\begin{equation}
\begin{split}
B^{ab} \otimes B^{cd} \sim {B_4}^{abcd}+BB^{ad,bc}+BB^{ac,bd}\\ 
+{VB_3}^{bdc,a}-{VB_3}^{bcd,a}+{VB_3}^{dcb,a},
\end{split}
\end{equation}
where the $VB_3$ operators are antisymmetric in their first three indices and symmetric in the last two.

\subsection{Branching rules from $O(n)$}
\label{app:branchingRules}
Below we summarize a few branching rules for irreps when breaking $O(n) \rightarrow C_n $. These are valid for $n \geq 4$, as mentioned in the previous subsection. The branching rules are
\begin{align}
\nonumber
 V  &\longrightarrow   V, & 
 T  &\longrightarrow   X \oplus Z, \\\nonumber
 A  &\longrightarrow   B, & 
 T_3  &\longrightarrow   XV \oplus Z_3 \oplus V, \\\nonumber
 T_4  &\longrightarrow  S\oplus Z  \oplus B  \oplus Z_4 \oplus X \oplus XX \oplus XZ, \hspace{-24pt} \\\nonumber
 A_3  &\longrightarrow   B_3, &
 A_4  &\longrightarrow   B_4, \\\nonumber
 H_3  &\longrightarrow    XV  \oplus VB, &
 H_4  &\longrightarrow   Z\oplus  B \oplus BZ \oplus \XXbar  \oplus XB \oplus XZ,  \\
 B_4  &\longrightarrow   BB \oplus XX  \oplus XZ, &
 Y_{2,1,1}  &\longrightarrow   XB \oplus VB_3,
\end{align}
where we have used the naming conventions from \cite{Henriksson:2022rnm} for the $O(n)$ irreps.

\section{Results for irreps that vanish in $n=3$}
\label{app:ResultsNgtr4}
In this section we collect some results for one-loop anomalous dimensions of operators whose irreps do not exist for $n=3$,\footnote{I.e. their group theoretic dimension contains a factor of $n-a$, with $a$ an integer satisfying $a\geq 3$.} see tables~\ref{tab:irrepZ4}, \ref{tab:irrepXX}, \ref{tab:irrepXZ}, \ref{tab:irrepXB}, \ref{tab:irrepBZ}, \ref{tab:irrepBB}, \ref{tab:irrepVB3}, \ref{tab:irrepB4}, \ref{tab:irrepXXbarV} and \ref{tab:irrepXXXbar}. We limit to operators with spins $0$, $1$ and $2$.

\begin{table}
\centering
\caption{Irrep $Z_4$ operators. }\label{tab:irrepZ4}
{\small
\renewcommand{\arraystretch}{1.25}
\begin{tabular}{|c|c|l|ll|}
\hline
\tableheadingF0
\\\hline
$1$ & $ \phi^4 $ & $ 4 - 2\varepsilon + \varepsilon\gamma $ & $1 $ & $ \frac{4}{n} $
 \\ 
$2$ & $ \phi^6 $ & $ 6 - 3\varepsilon + \varepsilon\gamma $ & $\frac{5}{2} $ & $ \frac{5 n+52-\sqrt{n^2+64 n-128}}{6 n}$
 \\ 
$3$ & $ \phi^6 $ & $ 6 - 3\varepsilon + \varepsilon\gamma $ & --- & $ \frac{5 n+52+\sqrt{n^2+64 n-128}}{6 n}$
 \\ 
\hline
\tableheadingF1
\\\hline
$1$ & $ \partial \phi^6 $ & $ 7 - 3\varepsilon + \varepsilon\gamma $ & $2 $ & $ \frac{5 n+34-\sqrt{n^2+28 n-92}}{6 n}$
 \\ 
$2$ & $ \partial \phi^6 $ & $ 7 - 3\varepsilon + \varepsilon\gamma $ & --- & $ \frac{5 n+34+\sqrt{n^2+28 n-92}}{6 n}$
 \\ 
\hline
\tableheadingF2
\\\hline
$1$ & $ \partial^2 \phi^4 $ & $ 6 - 2\varepsilon + \varepsilon\gamma $ & $\frac{13}{18} $ & $ \frac{26}{9 n} $
 \\ 
\hline
\end{tabular}
}
\end{table}

\begin{table}
\centering
\caption{Irrep $XX$ operators.}\label{tab:irrepXX}
{\small
\renewcommand{\arraystretch}{1.25}
\begin{tabular}{|c|c|l|ll|}
\hline
\tableheadingF0
\\\hline
$1$ & $ \phi^4 $ & $ 4 - 2\varepsilon + \varepsilon\gamma $ & $1 $ & $ \frac{2 (n+2)}{3 n} $
 \\ 
$2$ & $ \phi^6 $ & $ 6 - 3\varepsilon + \varepsilon\gamma $ & --- & $ \frac{11 n+28-\sqrt{9 n^2+48 n-192}}{6 n}$
 \\ 
$3$ & $ \square \phi^4 $ & $ 6 - 2\varepsilon + \varepsilon\gamma $ & $\frac{1}{2} $ & $ \frac{2 (n-1)}{3 n} $
 \\ 
$4$ & $ \phi^6 $ & $ 6 - 3\varepsilon + \varepsilon\gamma $ & $\frac{7}{2} $ & $ \frac{11 n+28+\sqrt{9 n^2+48 n-192}}{6 n}$
 \\ 
\hline
\tableheadingF1
\\\hline
$1$ & $ \partial \phi^6 $ & $ 7 - 3\varepsilon + \varepsilon\gamma $ & $2$ & $ \frac{11 n+10-\sqrt{9 n^2-60 n+132}}{6 n} $
 \\ 
$2$ & $ \partial \phi^6 $ & $ 7 - 3\varepsilon + \varepsilon\gamma $ & $\frac52$ & $ \frac{11 n+10+\sqrt{9 n^2-60 n+132}}{6 n} $
 \\ 
\hline
\tableheadingF2
\\\hline
$1$ & $ \partial^2 \phi^4 $ & $ 6 - 2\varepsilon + \varepsilon\gamma $ & $\frac{7}{18} $ & $ \frac{9 n+4-\sqrt{9 n^2-80 n+320}}{18 n}$
 \\ 
$2$ & $ \partial^2 \phi^4 $ & $ 6 - 2\varepsilon + \varepsilon\gamma $ & $\frac{13}{18} $ & $ \frac{9 n+4+\sqrt{9 n^2-80 n+320}}{18 n}$
 \\ 
\hline
\end{tabular}
}
\end{table}

\begin{table}
\centering
\caption{Irrep $XZ$ operators. If the ``$n$ gen.'' term contains no ellipses ($\ldots$), it is exact.}\label{tab:irrepXZ}
{\small
\renewcommand{\arraystretch}{1.25}
\begin{tabular}{|c|c|l|ll|}
\hline
\tableheadingF0
\\\hline
$1$ & $ \phi^4 $ & $ 4 - 2\varepsilon + \varepsilon\gamma $ & $1 $ & $ \frac{n+8}{3 n} $
 \\ 
$2$ & $ \phi^6 $ & $ 6 - 3\varepsilon + \varepsilon\gamma $ & $\frac{5}{2} $ & $ 1+\frac{10}{3 n}+\ldots $
 \\ 
$3$ & $ \square \phi^4 $ & $ 6 - 2\varepsilon + \varepsilon\gamma $ & $\frac{1}{2} $ & $ \frac{n+2}{3 n} $
 \\ 
$4$ & $ \phi^6 $ & $ 6 - 3\varepsilon + \varepsilon\gamma $ & --- & $ \frac{4}{3}+\frac{26}{3 n}+\ldots $
 \\ 
$5$ & $ \phi^6 $ & $ 6 - 3\varepsilon + \varepsilon\gamma $ & $\frac{7}{2} $ & $ 2+\frac{14}{3 n}+\ldots $
 \\ 
\hline
\tableheadingF1
\\\hline
$1$ & $ \partial \phi^4 $ & $ 5 - 2\varepsilon + \varepsilon\gamma $ & $\frac{2}{3} $ & $ \frac{n+4}{3 n} $
\\ 
\hline
\tableheadingF2
\\\hline
$1$ & $ \partial^2 \phi^4 $ & $ 6 - 2\varepsilon + \varepsilon\gamma $ & $\frac{7}{18} $ & $ 0+\frac{58}{27 n}+\ldots $
 \\ 
$2$ & $ \partial^2 \phi^4 $ & $ 6 - 2\varepsilon + \varepsilon\gamma $ & $\frac{1}{2} $ & $ \frac{1}{3}+\frac{22-4 \sqrt{10}}{27 n}+\ldots $
 \\ 
$3$ & $ \partial^2 \phi^4 $ & $ 6 - 2\varepsilon + \varepsilon\gamma $ & $\frac{13}{18} $ & $ \frac{1}{3}+\frac{22+4 \sqrt{10}}{27 n}+\ldots $
 \\ 
\hline
\end{tabular}
}
\end{table}

\begin{table}
\centering
\caption{Irrep $XB$ operators.}\label{tab:irrepXB}
{\small
\renewcommand{\arraystretch}{1.25}
\begin{tabular}{|c|c|l|ll|}
\hline
\tableheadingF0
\\\hline
$1$ & $ \phi^6 $ & $ 6 - 3\varepsilon + \varepsilon\gamma $ & $\frac{5}{2} $ & $ \frac{2 (2 n+7)}{3 n} $
 \\ 
\hline
\tableheadingF1
\\\hline
$1$ & $ \partial \phi^4 $ & $ 5 - 2\varepsilon + \varepsilon\gamma $ & $\frac{2}{3} $ & $ \frac{n+4}{3 n} $
\\ 
\hline
\tableheadingF2
\\\hline
$1$ & $ \partial^2 \phi^4 $ & $ 6 - 2\varepsilon + \varepsilon\gamma $ & $\frac12 $ & $ \frac{n+2}{3 n} $
 \\ 
\hline
\end{tabular}
}
\end{table}

\begin{table}
\centering
\caption{Irrep $BZ$ operators.}\label{tab:irrepBZ}
{\small
\renewcommand{\arraystretch}{1.25}
\begin{tabular}{|c|c|l|ll|}
\hline
\tableheadingF0
\\\hline
$1$ & $ \phi^6 $ & $ 6 - 3\varepsilon + \varepsilon\gamma $ & $\frac{5}{2} $ & $ \frac{n+6}{n} $
 \\ 
\hline
\tableheadingF1
\\\hline
$1$ & $ \partial \phi^4 $ & $ 5 - 2\varepsilon + \varepsilon\gamma $ & $\frac{2}{3} $ & $ \frac{8}{3 n} $
 \\ 
\hline
\tableheadingF2
\\\hline
$1$ & $ \partial^2 \phi^4 $ & $ 6 - 2\varepsilon + \varepsilon\gamma $ & $\frac{1}{2} $ & $ \frac{2}{n} $
 \\ 
\hline
\end{tabular}
}
\end{table}

\begin{table}
\centering
\caption{Irrep $BB$ operators. }\label{tab:irrepBB}
{\small
\renewcommand{\arraystretch}{1.25}
\begin{tabular}{|c|c|l|ll|}
\hline
\tableheadingF0
\\\hline
$1$ & $ \square \phi^4 $ & $ 6 - 2\varepsilon + \varepsilon\gamma $ & $\frac{1}{2} $ & $ \frac{2}{n} $
 \\ 
\hline
\tableheadingF1
\\\hline
$1$ & $ \partial \phi^6 $ & $ 7 - 3\varepsilon + \varepsilon\gamma $ & $2 $ & $ \frac{n+4}{n} $
 \\ 
\hline
\tableheadingF2
\\\hline
$1$ & $ \partial^2 \phi^4 $ & $ 6 - 2\varepsilon + \varepsilon\gamma $ & $\frac{7}{18} $ & $ \frac{14}{9 n} $
 \\ 
\hline
\end{tabular}
}
\end{table}

\begin{table}
\centering
\caption{Irrep $VB_3$ operators.}\label{tab:irrepVB3}
{\small
\renewcommand{\arraystretch}{1.25}
\begin{tabular}{|c|c|l|ll|}
\hline
\tableheadingF0
\\\hline
$1$ & $ \square \phi^6 $ & $ 8 - 3\varepsilon + \varepsilon\gamma $ & $\frac{5}{3} $ & $ \frac{3 n+8}{3 n} $
 \\ 
\hline
\tableheadingF1
\\\hline
$1$ & $ \partial \square \phi^4 $ & $ 7 - 2\varepsilon + \varepsilon\gamma $ & $\frac{2}{9} $ & $ \frac{8}{9 n} $
 \\ 
$2$ & $ \partial \phi^6 $ & $ 7 - 3\varepsilon + \varepsilon\gamma $ & $2 $ & $ \frac{n+4}{n} $
 \\ 
\hline
\tableheadingF2
\\\hline
$1$ & $ \partial^2 \square \phi^4 $ & $ 8 - 2\varepsilon + \varepsilon\gamma $ & $\frac{5}{18} $ & $ \frac{10}{9 n} $
 \\ 
$2$ & $ \partial^2 \phi^6 $ & $ 8 - 3\varepsilon + \varepsilon\gamma $ & $\frac{14}{9} $ & $ \frac{9 n+20}{9 n} $
 \\ 
$3$ & $ \partial^2 \phi^6 $ & $ 8 - 3\varepsilon + \varepsilon\gamma $ & $\frac{16}{9} $ & $ \frac{9 n+28}{9 n} $
 \\ 
\hline
\end{tabular}
}
\end{table}

\begin{table}
\centering
\caption{Irrep $B_4$ operators.}\label{tab:irrepB4}
{\small
\renewcommand{\arraystretch}{1.25}
\begin{tabular}{|c|c|l|ll|}
\hline
\tableheadingF0
\\\hline
$1$ & $ \square^3 \phi^4 $ & $ 10 - 2\varepsilon + \varepsilon\gamma $ & $0 $ & $ 0 $
 \\ 
$2$ & $ \square \phi^8 $ & $ 10 - 4\varepsilon + \varepsilon\gamma $ & $\frac{7}{2} $ & $ \frac{2 (n+3)}{n} $
 \\ 
\hline
\tableheadingF1
\\\hline
$1$ & $ \partial \square \phi^6 $ & $ 9 - 3\varepsilon + \varepsilon\gamma $ & $\frac{4}{3} $ & $ \frac{3 n+4}{3 n} $
 \\ 
\hline
\tableheadingF2
\\\hline
$1$ & $ \partial^2 \square \phi^4 $ & $ 8 - 2\varepsilon + \varepsilon\gamma $ & $0 $ & $ 0 $
 \\ 
\hline
\end{tabular}
}
\end{table}

\begin{table}
\centering
\caption{Irrep $\XXbar V$ operators.}\label{tab:irrepXXbarV}
{\small
\renewcommand{\arraystretch}{1.25}
\begin{tabular}{|c|c|l|ll|}
\hline
\tableheadingF0
\\\hline
$1$ & $ \square \phi^5 $ & $ 7 - \frac{5}{2}\varepsilon + \varepsilon\gamma $ & $1 $ & $ \frac{2 (n+2)}{3 n} $
 \\ 
$2$ & $ \phi^7 $ & $ 7 - \frac{7}{2}\varepsilon + \varepsilon\gamma $ & $\frac{7}{2} $ & $ \frac{7 (n+2)}{3 n} $
 \\ 
\hline
\tableheadingF1
\\\hline
$1$ & $ \partial \phi^5 $ & $ 6 - \frac{5}{2}\varepsilon + \varepsilon\gamma $ & $\frac{5}{4} $ & $ \frac{2 n+7}{3 n} $
 \\ 
\hline
\tableheadingF2
\\\hline
$1$ & $ \partial^2 \phi^5 $ & $ 7 - \frac{5}{2}\varepsilon + \varepsilon\gamma $ & $\frac{8}{9} $ & $ \frac{9 n+34-\sqrt{9 n^2-92 n+260}}{18 n} $
 \\ 
$2$ & $ \partial^2 \phi^5 $ & $ 7 - \frac{5}{2}\varepsilon + \varepsilon\gamma $ & $\frac{19}{18} $ & $ \frac{9 n+34+\sqrt{9 n^2-92 n+260}}{18 n} $
 \\ 
\hline
\end{tabular}
}
\end{table}

\begin{table}
\centering
\caption{Irrep $\XXXbar$ operators.}\label{tab:irrepXXXbar}
{\small
\renewcommand{\arraystretch}{1.25}
\begin{tabular}{|c|c|l|ll|}
\hline
	\tableheadingF0
\\\hline
$1$ & $ \square \phi^8 $ & $ 10 - 4\varepsilon + \varepsilon\gamma $ & $\frac{7}{2} $ & $ \frac{2 (4 n+5)}{3 n} $
 \\ 
\hline
\tableheadingF1
\\\hline
$1$ & $ \partial \square \phi^6 $ & $ 9 - 3\varepsilon + \varepsilon\gamma $ & $\frac{4}{3} $ & $ \frac{2 (n+4)}{3 n} $
 \\ 
$2$ & $ \partial \square \phi^6 $ & $ 9 - 3\varepsilon + \varepsilon\gamma $ & $\frac{4}{3} $ & $ \frac{3 n+4}{3 n} $
 \\ 
$3$ & $ \partial \phi^8 $ & $ 9 - 4\varepsilon + \varepsilon\gamma $ & $4 $ & $ \frac{8 (n+2)}{3 n} $
 \\ 
\hline
\tableheadingF2
\\\hline
$1$ & $ \partial^2 \square \phi^6 $ & $ 10 - 3\varepsilon + \varepsilon\gamma $ & $\frac{19}{18} $ & $ \frac{2 (3 n+7)}{9 n} $
 \\ 
$2$ & $ \partial^2 \square \phi^6 $ & $ 10 - 3\varepsilon + \varepsilon\gamma $ & $\frac{19}{18} $ & $\frac{15 n+27-\sqrt{9 n^2-58 n+209}}{18 n} $
 \\ 
$3$ & $ \partial^2 \square \phi^6 $ & $ 10 - 3\varepsilon + \varepsilon\gamma $ & $\frac{5}{4} $ & $ \frac{2 n+7}{3 n} $
 \\ 
$4$ & $ \partial^2 \square \phi^6 $ & $ 10 - 3\varepsilon + \varepsilon\gamma $ & $\frac{49}{36} $ & $\frac{15 n+27+\sqrt{9 n^2-58 n+209}}{18 n} $
 \\ 
$5$ & $ \partial^2 \phi^8 $ & $ 10 - 4\varepsilon + \varepsilon\gamma $ & $\frac{61}{18} $ & $ \frac{45 n+76-\sqrt{9 n^2-80 n+320}}{18 n}$
 \\ 
$6$ & $ \partial^2 \phi^8 $ & $ 10 - 4\varepsilon + \varepsilon\gamma $ & $\frac{67}{18} $ & $ \frac{45 n+76+\sqrt{9 n^2-80 n+320}}{18 n}$
 \\ 
\hline
\end{tabular}
}
\end{table}

\section{$\phi_a \otimes Z_{bc}$ tensor structures}
\label{ApppendixPhiZ}
We will write the tensor structures in terms of ``blocks''.\footnote{Not to be confused with conformal blocks.}
We have four irreps, namely $V$, $XV$, $VB$ and $Z_3$. Let us start with $V$, and call the block $B_V$.
\begin{equation}
{B_V}_{abcdef} = \delta_{ab}\delta_{de}\delta_{cf}-\delta_{ab}\delta_{cdef}-\delta_{de}\delta_{cabf}+\delta_{abcdef},
\end{equation}
which leads to 
\begin{equation}
P^V_{abcdef}=\frac{1}{2(n-1)}({B_V}_{abcdef}+{B_V}_{abcdfe}+{B_V}_{acbdef}+{B_V}_{acbdfe}).
\end{equation}
For $Z_3$ we have (this time we need two blocks)
\begin{equation}
{B_{Z_3}}_{abcdef}=(\delta_{af} \delta_{be} \delta_{cd} - \delta_{cd} \delta_{a b e f} - 
  \delta_{be} \delta_{a c d f} - \delta_{af} \delta_{c b e d}+   2 \delta_{a b c d ef})
\end{equation}
and
\begin{equation}
{BB_{Z_3}}_{abcdef}={B_{Z_3}}_{a b c d e f} + {B_{Z_3}}_{a b c e d f},
\end{equation}
which lead to 
\begin{equation}
{P^{Z3}}_{a b c d e f} = \frac{1}{6} ({BB_{Z3}}_{a b c d e f} + 
    {BB_{Z3}}_{a b c d f e} + {BB_{Z3}}_{a b c e f d}).
\end{equation}
Lastly, for $VB$ and $XV$ we have
\begin{align}
\begin{split}
{B_{VB}}_{a b c d e  f} &= (\delta_{cf} \delta_{ad}\delta_{be}  -\delta_{adbe}\delta_{cf} -\delta_{adcf}\delta_{be}  -\delta_{cfbe} \delta_{ad} + 2\delta_{abcdef})         \\&\quad         - (\delta_{cf}\delta_{ae}\delta_{bd} -\delta_{adbe}\delta_{cf}  -\delta_{aecf}\delta_{bd} -\delta_{cfbd}\delta_{ae}  + 2\delta_{abcdef}) 
\end{split}
\end{align}
and
\begin{equation}
{B_{XV}}_{a b c d e  f} = \left((\delta_{cf}\delta_{abed} -\delta_{abcdef}) - \frac{1}{n - 1}(\delta_{abcdef} - 
\delta_{ab}\delta_{decf} -\delta_{de}\delta_{abcf} +
\delta_{ab}\delta_{de}\delta_{cf} )\right),
\end{equation}
which lead to 
\begin{equation}
{P^{VB}_{abcdef}}=\frac16 (({B_{VB}}_{a b c d e f} + 
     {B_{VB}}_{a b c d f e}) + ({B_{VB}}_{a c b d e f} + 
     {B_{VB}}_{a c b d f e}))
\end{equation}
and
\begin{equation}
{P^{XV}_{abcdef}}=\frac12 (({B_{XV}}_{a b c d e  f} + 
    {B_{XV}}_{a b c d f e}) + ({B_{XV}}_{a c b d e f} + 
     {B_{XV}}_{a c b d f e})).
\end{equation}

\section{Products of four $\phi$ fields}
\label{fourfields}
To calculate all possible operators that can be built with four powers of the order parameter field we will again use the associativity of tensor products
\begin{equation}
\phi_a \phi_b \phi_c \phi_d = (\phi_a \otimes \phi_b) \otimes (\phi_c \otimes \phi_d) =(\delta_{ab} S +X_{ab}+Z_{ab}+B_{ab})\otimes (\delta_{cd}S +X_{cd}+Z_{cd}+B_{cd}),
\end{equation}
where we have plugged in the decomposition for the product of two fields. The products involving $S$ are very simple, we just use the tensor structures obtained in \eqref{twofields}. For example $S \otimes X_{cd} \sim \delta_{ab}(\delta_{cdef}-\frac{1}{n}\delta_{cd}\delta_{ef}) \phi_a \phi_b \phi_e \phi_f$. Thus, the non-trivial products to compute are $X_{ab}\otimes X_{cd}$, $X_{ab}\otimes Z_{cd}$ and $Z_{ab} \otimes Z_{cd}$.

\paragraph{The product $X_{ab} \otimes X_{cd}$}
This product decomposes into four irreps. These are $S$, $X$ and two new ones which we call $XX$ and $\XXbar$. They have dimensions 
\begin{equation}
\begin{split}
& \dim S = 1,\\
& \dim X = n-1,\\
& \dim XX = \frac{n(n-3)}{2},\\
& \dim \XXbar = \frac{(n-1)(n-2)}{2},\\
\end{split}
\label{4ptdimensions}
\end{equation}
which add up to $(n-1)^2$ as expected. The reader may have noticed that this decomposition is precisely that of two fundamental operators of $S_n$ (or $S_n \otimes \mathbb Z_2$\footnote{The only effect of the extra $\mathbb Z_2$ is that the $X$ on the LHS and RHS of $X \otimes X \sim S+X +XX+\XXbar$ have different $\mathbb Z_2$ parities. However the actual tensor structures are identical.}) which are the global symmetries of the Potts and hyper-tetrahedral theories respectively. This is expected, since $X$ is a singlet of all the $\mathbb Z_2$s appearing in $S_n \ltimes (\mathbb Z_2)^n$ and is also traceless. Hence, for all intents and purposes it is a fundamental of $S_n$.

The projectors corresponding to these irreps can be found to be
\begin{align}
\nonumber
 P^X_{i j k l m n o p}& = \delta_{i j k l m n o p} -\frac{1}{n} (\delta_{ij}\delta_{klmnop} + \delta_{kl}\delta_{ijmnop}
      +\delta_{mn}\delta_{klijop} 
      + \delta_{op}\delta_{klmnij})
  - \frac{1}{n} (\delta_{ijkl} \delta_{mnop})\\\nonumber
&   + \frac{2}{n^2} (\delta_{ij}\delta_{kl}\delta_{mnop}  + \delta_{mn}\delta_{op} \delta_{ijkl} )      - \frac{4}{n^3} (\delta_{ij}\delta_{kl}\delta_{mn}\delta_{op})\\\nonumber
& + \frac{1}{n^2} (\delta_{ij}\delta_{op}\delta_{mnkl}  +
     \delta_{ij}\delta_{mn} \delta_{opkl}  +
     \delta_{kl}\delta_{op}\delta_{mnij}  +
     \delta_{kl}\delta_{mn}\delta_{opij} ),\\\nonumber
 P^{\XXbar}_{i j k l m n o p}& = P^X_{ijmn}P^X_{klop}-P^X_{ijop}P^X_{klmn},\\\nonumber
 P^S_{ijklmnop} &= P^X_{ijkl}P^X_{mnop},\\\nonumber
 {P^{XX}}_{ij k l m n o p} &= -\delta_{ijklmnop} + \frac{1}{n} 
(\delta_{ij}\delta_{klmnop}   + \delta_{kl}\delta_{ijmnop} +  \delta_{mn} \delta_{klijop} + \delta_{op}\delta_{ijmnkl}  ) \\\nonumber
&      + \frac{1}{n(n-1)} (\delta_{ijkl}\delta_{mnop} ) +\frac{n-2}{2n} (\delta_{ijmn}\delta_{klop}  + \delta_{ijop}\delta_{klmn} )\\\nonumber
&       - \frac{1}{n(n-1)} (\delta_{ijkl}\delta_{mn}\delta_{op} +\delta_{mnop}\delta_{ij}\delta_{kl} )\\\nonumber
&  - \frac{1}{2n} (\delta_{ijmn}\delta_{kl}\delta_{op}  +\delta_{klop}\delta_{ij}\delta_{mn}  +\delta_{ijop}\delta_{mn}\delta_{kl}  + \delta_{klmn}\delta_{ij}\delta_{op} )\\
& + \frac{1}{n(n-1)} (\delta_{ij}\delta_{kl}\delta_{mn}\delta_{op} )  ,
\label{XXprojectors}
\end{align}
where the four-index projectors are the ones found when studying the product of two fields.

\paragraph{The product $Z_{ab} \otimes X_{cd}$}
This product decomposes on to three irreps. Let us call them $Z$,\footnote{This is indeed the same irrep as the one that appears in the product of two $\phi$ fields, hence we give it the same name.} $B$ and $XZ$, we have
\begin{equation}
\begin{split}
& \dim Z=\frac{n(n-1)}{2} ,\\
& \dim B=\frac{n(n-1)}{2},\\
&  \dim XZ=\frac{n(n-1)(n-3)}{2}
\end{split}
\end{equation}
and
\begin{equation}
\dim Z +\dim B +\dim XZ =\frac{n(n-1)^2}{2}.
\end{equation}
The first two irreps appear if $a$ or $b$ are equal to $c$ (which we remind is equal to $d$) for example
\begin{equation}
Z_{12}\otimes X_{22} \sim Z_{12}(X_{22}+X_{11}) + Z_{12}(X_{22}-X_{11}),
\end{equation}
where the first term on the RHS is now in the $Z$ irrep and the second in the $B$ irrep. For the $XZ$ irrep we need to take $a \neq b \neq c$. If for example we take $a=1$, $b=2$ and $c=3$ we have
\begin{equation}
Z_{12}\otimes X_{33} \sim Z_{12}\bigg(X_{33}-\frac{1}{n-2}\sum_{i\neq 1, i \neq 2}^n X_{ii}\bigg)+Z_{12}\frac{1}{n-2}\sum_{i\neq 1, i \neq 2}^n X_{ii}.
\end{equation}
The first term on the RHS is now in the $XZ$ irrep. The second is actually in the $Z$ irrep; to see this remember that
\begin{equation}
\sum_i^n X_{ii}=0 , 
\end{equation}
thus
\begin{equation}
\sum_{i\neq 1, i \neq 2}^n X_{ii} = -(X_{11}+X_{22}).
\end{equation}
We now conclude by writing their projectors. For these it is convenient to define the following intermediate tensors
\begin{equation}
\begin{split}
& P^1_{abcdefgh}=P^X_{aecd}P^X_{bfgh},\\
& P^2_{abcdefgh}=\delta_{abcdefgh}-\frac{1}{n}\delta_{aecdbf}\delta_{gh}-\frac{1}{n}\delta_{aeghbf}\delta_{cd} + \frac{1}{n^2} \delta_{aebf}\delta_{cd}\delta_{gh},\\
& P^3_{abcdefgh}=\delta_{bf}\left(\delta_{acdegh}  -\frac{1}{n}(\delta_{cd}\delta_{aegh}+\delta_{gh}\delta_{aecd})+\frac{1}{n^2}\delta_{cd}\delta_{gh}\delta_{ae}\right).
\end{split}
\end{equation}
Next, define
\begin{equation}
\begin{split}
 P^v_{abcdefgh}&=P^1_{abcdefgh}-P^2_{abcdefgh}, \\
 P^u_{abcdefgh}&=P^3_{abcdefgh}-P^2_{abcdefgh}.
\end{split}
\end{equation}
From these we obtain the projectors as
\begin{equation}
\begin{split}
& P^Z_{abcdefgh} =((P^u_{abcdefgh}+P^v_{abcdefgh})+ (a \leftrightarrow b))+(d \leftrightarrow e ),\\
& P^{B}_{abcdefgh}= ((P^u_{abcdefgh}-P^v_{abcdefgh})+ (a \leftrightarrow b))+(d \leftrightarrow e ).
\end{split}
\end{equation}
A compact way to then define the projector of $XZ$ is 
\begin{equation}
P^{XZ}_{abcdefgh}=P^Z_{abef}  P^X_{cdgh} -\frac{n}{4(n-2)}P^Z_{abcdefgh} -\frac{1}{4} P^{B}_{abcdefgh}.
\end{equation}

\paragraph{The product $Z_{ab} \otimes Z_{cd}$}
The last product we need to analyse is also the most complicated. For this reason we have collected the explicit projectors in Appendix \ref{AppendixZZ}. However, we will give a qualitative description of the irreps that appear. These can be separated into three groups, those that have two pairs of indices equal (e.g. $a=c$ and $b=d$), those that have one pair of indices equal (e.g. $a=c$ but $b\neq d$), and those with no pairs of indices equal (i.e. $a\neq b \neq c \neq d$). The first group consists of $S$, $X$ and $XX$, the first two we have seen already at the level of the product of two fields, whereas the last one also appears in $X \otimes X$, it has four indices and it is traceless. The second group consists of $XB$, $XZ$, $SB=B$ and $SZ=Z$. The first one of these is new, the second already appeared in the product $Z \otimes X$, and the last two we have already seen before in the product of two $\phi$ fields. Lastly, the third group consists of $BB$, $BZ$ and $Z_4$ which are all new. More specifically, $Z_4$ is a four index totally symmetric irrep, $BZ$ is symmetric in some indices but antisymmetric in others and satisfies antisymmetry under $(ab)\leftrightarrow (cd)$, whereas $BB$ is antisymmetric under exchanges of various indices but symmetric under $(ab)\leftrightarrow (cd)$. Let us report the dimensions of these representations 
\begin{equation}
\begin{split}
& \dim S=1, \\
& \dim X=n-1 ,\\
& \dim XX=\frac{n(n-3)}{2}, \\
& \dim XB=\frac{n(n-1)(n-3)}{2}, \\
& \dim XZ=\frac{n(n-1)(n-3)}{2}, \\
& \dim SB=\frac{n(n-1)}{2}, \\
& \dim SZ=\frac{n(n-1)}{2} ,\\
& \dim Z_4=\frac{n(n-1)(n-2)(n-3)}{24} ,\\
& \dim BB=2  \dim Z_4, \\
& \dim BZ=3  \dim Z_4.
\end{split}
\end{equation}
These add up to $\frac{n^2(n-1)^2}{4}$ and one can also check that they agree with the irreps reported in \cite{Antipin:2019vdg}.

We would now like to write a general formula for obtaining scaling dimensions from \cite{Bednyakov:2021ojn}. For example, let us say we want to calculate the scaling dimension of the $Z_4$ field that appears in the $Z \otimes Z$ product. Firstly, we write 
\begin{equation}
{Z_4}_{abcd} \sim P^{Z_4}_{abcdefgh}Z_{ef}Z_{gh},
\end{equation} 
then remembering that
\begin{equation}
Z_{ab} \sim P^Z_{abcd} \phi_c \phi_d
\end{equation}
we have
\begin{equation}
{Z_4}_{abcd} \sim P^{Z_4}_{abcdefgh}P^Z_{efkl}P^Z_{ghmn}\phi_k \phi_l \phi_m \phi_n.
\end{equation}
Recalling the simplification encountered in \eqref{threepointprojectorssimplified}, this is simplified to 
\begin{equation}
{Z_4}_{abcd} \sim P^{Z_4}_{abcdefgh}\phi_e \phi_f \phi_g \phi_h.
\end{equation}

We may now deform the theory with $\delta \mathcal{L} = gP^{Z_4}_{1234efgh}\phi_e \phi_f \phi_g \phi_h$, which corresponds to making the substitution $\lambda_{abcd} \rightarrow \lambda^S_{abcd}+ g P^{Z_4}_{1234efgh}$ in \eqref{lagr}.

\section{$Z_{ab} \otimes Z_{cd}$ tensor structures}
\label{AppendixZZ}
The tensor structures in this appendix are just those of \cite{Kousvos:2021rar} and the corresponding appendix there, but with the $i,j,k,l,m,n,o,p$ indices in that reference removed (since these correspond to $G$ indices in $S_n \ltimes G^n$, and for $G=\mathbb Z_2$ we can drop them).

We start by defining the following convenient tensor quantities
\begin{align}
\begin{split}
  {R_1}_{abcdefgh}&=(\delta_{ac}\delta_{bd}-\delta_{acbd})(\delta_{eg}\delta_{fh}-\delta_{egfh})\,,\\
  {R_2}_{abcdefgh}&=\delta_{aceg}\delta_{bd}\delta_{fh}-\delta_{abcdeg}\delta_{fh}-\delta_{afcheg}\delta_{bd}+\delta_{abcdefgh}\,,\\
  {R_3}_{abcdefgh}&=\delta_{aceg}\delta_{bdfh}-\delta_{abcdefgh}\,.
\end{split}
\label{threetensors}
\end{align}
Using these tensors we can write the blocks for representations where we have two pairs of indices equal (such as e.g. $a=c$ and $b=d$)
\begin{align}
\begin{split}
{B_S}_{abcdefgh} &=  {R_1}_{abcdefgh}\,,\\
{B_{X}}_{abcdefgh} &=  \Big({R_2}_{a b c d e f gh} - \frac{1}{n} {R_1}_{abcdefgh}\Big)\,,\\
{B_{XX}}_{abcdefgh}&= \Big({R_3}_{b a d c f e h  g} -\frac{1}{n-2} ({R_2}_{a b c defg h} +
{R_2}_{b a d c f e h g})+ \frac{1}{(n-1)(n-2)} {R_1}_{a b c d e f g h}\Big)\,.
\end{split}
\label{blocks}
\end{align}
We then get the projectors by performing symmetrizations. Letting $g$ stand for either $S$, $X$ or $XX$, we have
\begin{align}
\begin{split}
  {{{B_g}'} \lsp}^{abcdefgh} &= {{{B_g}} \lsp}^{a b c d e f g
h} + {{{B_g}} \lsp}^{a b c d e f h g} \,,\\
{{{B_g}''} \lsp}^{abcdefgh} &= {{{B_g}'}
\lsp}^{abcdefgh} + {{{B_g}'} \lsp}^{a b c d f e g h }\,,\\
{{B_g}'''\lsp}^{abcdefgh} &=
{{B_g}'' \lsp}^{abcdefgh}+
{{B_g}''\lsp}^{a b d c e f, g h }\,,\\
{{P_g}\lsp}^{abcdefgh} &= {{{B_g}'''
}\lsp}^{abcdefgh} +{{B_g}'''}{\lsp}^{b a c
d e f g h }\,,
\end{split}
\label{symmetrizations}
\end{align}
The last line in the above is what corresponds to the projector. 

Next, we move to representations where one pair of indices is equal, e.g. $a=c$ but $b\neq d$. Let us define some new convenient tensors
\begin{equation}
\begin{aligned}
{RR_1}_{a b c d e f g  h} &= \delta_{a c eg} (\delta_{bf} \delta_{dh}   -\delta_{bfdh}) - (\delta_{acegbf}\delta_{dh} -\delta_{acegbfdh}) - (\delta_{acegdh}\delta_{bf} - \delta_{acegbfdh})\\
{RR_2}_{a b c d e f g  h}   &=\delta_{ac}\delta_{eg} ( \delta_{bf}\delta_{dh} - \delta_{bfdh})
 - \delta_{eg} (\delta_{acbf}\delta_{dh} +\delta_{acdh}\delta_{bf} -
 2\delta_{acbfdh})\,,\\
 &\quad-\delta_{ac} (\delta_{egbf}\delta_{dh} +\delta_{egdh}\delta_{bf} -
 2\delta_{egbfdh})
 + (\delta_{acbfeg}\delta_{dh} +\delta_{acdheg}\delta_{bf} -
    2\delta_{acegbfdh})\\
    &\quad+ (\delta_{acbf}\delta_{egdh} +\delta_{acdh}\delta_{bfeg} -
    2\delta_{acegbfdh})\,.
\end{aligned}
\end{equation}
The blocks in terms of these will then be
\begin{equation}
\begin{aligned}
{B_{XB}}_{a b c d e f g h}&=\left({RR_1}_{a b c d e f g h}-\frac{1}{n-2}{RR_2}_{a b c d e f g
h}\right)-\left({RR_1}_{a b c d e h g f}-\frac{1}{n-2}{RR_2}_{a b c d e h g f}\right)
,\\
{B_{XZ}}_{a b c d e f g h}&=\left({RR_1}_{a b c d e f g h}-\frac{1}{n-2}{RR_2}_{a b c d e f g
h}\right)+\left({RR_1}_{a b c d e h g f}-\frac{1}{n-2}{RR_2}_{a b c d e h g f}\right)
,\\
{B_{SB}}_{a b c d e f g h}&={RR_2}_{a b c d e f g h} 
-  {RR_2}_{a b c d e h g f} \,,\\
{B_{SZ}}_{a b c d e f g h}&={RR_2}_{a b c d e f g h} 
+  {RR_2}_{a b c d e h g f} \,,\\
\end{aligned}
\end{equation}
the corresponding projectors are then obtained by applying \eqref{symmetrizations} to the above blocks. 

The last of group of representations we wish to study are those where all indices are different. For these we define the following quantity
\begin{equation}
\begin{aligned}
{R_4}^{abcdefgh}&=(\delta_{ae}\delta_{bf}\delta_{cg}\delta_{dh}-
   \delta_{cg}\delta_{dh}\delta_{abef} -
   \delta_{bf}\delta_{dh}\delta_{aceg} -
   \delta_{bf}\delta_{cg}\delta_{adeh} -
   \delta_{ae}\delta_{dh}\delta_{bcfg} \\
   &\quad-\delta_{ae}\delta_{cg}\delta_{bdfh} -
   \delta_{ae}\delta_{bf}\delta_{cdgh} +  \delta_{aedh}\delta_{bcfg} +
   \delta_{aecg} \delta_{bdfh} + \delta_{aebf} \delta_{cdgh}\\
   &\quad+
   2 (\delta_{dh}\delta_{abcefg} +
      \delta_{cg}\delta_{abdefh} +
      \delta_{bf}\delta_{adcehg} +
      \delta_{ae}\delta_{dbchfg} ) -
    6  \delta_{aecgbfdh}) \,,
 \end{aligned}
\end{equation}
which can then be used to define the blocks
\begin{equation}
\begin{aligned}
{B_{BB}\lsp} ^{abcdefgh}&= {R_4}^{abcdefgh} - {R_4}^{a
b c d g f e h} - {R_4}^{a b c d e h g f} + {R_4}^{a b c d g h e f}\,,\\
{B_{BZ}\lsp} ^{abcdefgh}&={R_4}^{abcdefgh} - {R_4}^{a b
c d g f e h} + {R_4}^{a b c d e h g f} - {R_4}^{a b c d g h e f}\,,\\
{B_{\TotS}\lsp} ^{abcdefgh}& = {R_4}^{abcdefgh} +
{R_4}^{a b c d e f h g }\,.
\end{aligned}
\end{equation}
To get projectors we may again use \eqref{symmetrizations}, except for the $Z_4$ irrep, for which we use the following relations  
\begin{equation}
\begin{aligned}
{{{B_{\TotS}}^\prime }\lsp}^{abcdefgh}&= {B_{\TotS}\lsp}
^{abcdefgh} + {B_{\TotS}\lsp} ^{a b c d e g f h}
+ {B_{\TotS}\lsp} ^{a b c d e h g f}\,,\\
{{{P_{\TotS}} }\lsp}^{abcdefgh} &= {{{B_{\TotS}}^\prime
}\lsp}^{abcdefgh} +  {{{B_{\TotS}}^\prime }\lsp}^{a b c d f e g
h} +  {{{B_{\TotS}}^\prime }\lsp}^{a b c d g f e h}  +  {{{B_{\TotS}}^\prime }\lsp}^{a b c d h f g e}\,.
\end{aligned}
\end{equation}

\section{Representation theory for $n=4$}
\label{app:n4reps}

For $n=4$, the hypercubic group has $384$ elements, organized in $20$ conjugacy classes. There are $20$ irreducible representations.
The character table is given in table~\ref{tab:charTab4}. Here we have chosen to label the representations according to \cite{Baake:1981qe}; the relation to our labels is as follows:
\begin{align}
 r_1 &= \chi ^1_1 = S ,&
 r_2 &= \chi ^1_2 = Z_4 ,&
 r_3 &= \chi ^1_3 = \XXXbar ,&
 r_4 &= \chi ^1_4 = B_4,&
 r_5 &= \chi ^2_1 = XX, \nonumber\\
 r_6 &= \chi ^2_2 = BB ,&
 r_7 &= \chi ^3_1 = X ,&
 r_8 &= \chi ^3_2 = BZ, &
 r_9 &= \chi ^3_3 = \XXbar . &
 r_{10} &= \chi ^3_{4} = VB_3,\nonumber \\
 r_{11} &= \chi ^4_{1} = V, &
 r_{12} &= \chi ^4_{2} = Z_3, &
 r_{13} &= \chi ^4_{3} = \XXbar V, &
 r_{14} &= \chi ^4_{4} = B_3 ,&
 r_{15} &= \chi ^6_{1} = B, \nonumber\\
 r_{16} &= \chi ^6_{2} = {XZ}, &
 r_{17} &= \chi ^6_{3} = Z ,&
 r_{18} &= \chi ^6_{4} = {XB}, &
 r_{19} &= \chi ^8_{1} =  XV , &
 r_{20} &= \chi ^8_{2} = VB .&
\end{align}

Some functions transforming in the different irreps are $(\phi_1,\phi_2,\phi_3,\phi_4)$ in $r_{11}=V$, $(\phi_1^2-\phi_2^2,\phi_1^2-\phi_3^2,\phi_1^2-\phi_4^2)$ in $r_7=X$ and $(\phi_1\phi_2,\phi_1\phi_3,\phi_1\phi_4,\phi_2\phi_3,\phi_2\phi_4,\phi_3\phi_4)$ in $r_{17}=Z$.\footnote{See the caption of table~4 in \cite{Veysseyre1984}, however the irreps there are given in a different order compared to here.}

\begin{table}
\centering
\caption{Character table for the hypercubic group $C_4$ at $n=4$, taken from \cite{Baake:1981qe}.}\label{tab:charTab4}
{\scriptsize
\renewcommand{\arraystretch}{1.25}
\begin{tabular}{|c|cccccccccccccccccccc|}
\hline
    &  $\! 1 \! $ & $ \! 12 \! $ & $ \! 48 \! $ & $ \! 4 \! $ & $ \! 32 \! $ & $ \! 12 \! $ & $ \! 12 \! $ & $ \! 48 \! $ & $ \! 32 \! $ & $ \! 32 \! $ & $ \! 24 \! $ & $ \! 6 \! $ & $ \! 32 \! $ & $ \! 12 \! $ & $ \! 24 \! $ & $ \! 24 \! $ & $ \! 12 \! $ & $ \! 4 \! $ & $ \!
   12 \! $ & $ \! 1 $  \\\hline
$ r_1 \! $ & $ \! 1 \! $ & $ \! 1 \! $ & $ \! 1 \! $ & $ \! 1 \! $ & $ \! 1 \! $ & $ \! 1 \! $ & $ \! 1 \! $ & $ \! 1 \! $ & $ \! 1 \! $ & $ \! 1 \! $ & $ \! 1 \! $ & $ \! 1 \! $ & $ \! 1 \! $ & $ \! 1 \! $ & $ \! 1 \! $ & $ \! 1 \! $ & $ \! 1 \! $ & $ \! 1 \! $ & $ \! 1 \! $ & $ \! 1 $  \\
$ r_2 \! $ & $ \! 1 \! $ & $ \! 1 \! $ & $ \! 1 \! $ & $ \! -1 \! $ & $ \! 1 \! $ & $ \! -1 \! $ & $ \! 1 \! $ & $ \! -1 \! $ & $ \! -1 \! $ & $ \! -1 \! $ & $ \! -1 \! $ & $ \! 1 \! $ & $ \! 1 \! $ & $ \! 1 \! $ & $ \! -1 \! $ & $ \! 1 \! $ & $ \! 1 \! $ & $ \! -1 \! $ & $ \! -1 \! $ & $ \! 1 $  \\
$ r_3 \! $ & $ \! 1 \! $ & $ \! -1 \! $ & $ \! -1 \! $ & $ \! 1 \! $ & $ \! 1 \! $ & $ \! -1 \! $ & $ \! 1 \! $ & $ \! -1 \! $ & $ \! 1 \! $ & $ \! 1 \! $ & $ \! 1 \! $ & $ \! 1 \! $ & $ \! 1 \! $ & $ \! 1 \! $ & $ \! -1 \! $ & $ \! -1 \! $ & $ \! -1 \! $ & $ \! 1 \! $ & $ \! -1 \! $ & $ \! 1 $  \\
$ r_4 \! $ & $ \! 1 \! $ & $ \! -1 \! $ & $ \! -1 \! $ & $ \! -1 \! $ & $ \! 1 \! $ & $ \! 1 \! $ & $ \! 1 \! $ & $ \! 1 \! $ & $ \! -1 \! $ & $ \! -1 \! $ & $ \! -1 \! $ & $ \! 1 \! $ & $ \! 1 \! $ & $ \! 1 \! $ & $ \! 1 \! $ & $ \! -1 \! $ & $ \! -1 \! $ & $ \! -1 \! $ & $ \! 1 \! $ & $ \! 1 $  \\
$ r_5 \! $ & $ \! 2 \! $ & $ \! 0 \! $ & $ \! 0 \! $ & $ \! 2 \! $ & $ \! -1 \! $ & $ \! 0 \! $ & $ \! 2 \! $ & $ \! 0 \! $ & $ \! -1 \! $ & $ \! -1 \! $ & $ \! 2 \! $ & $ \! 2 \! $ & $ \! -1 \! $ & $ \! 2 \! $ & $ \! 0 \! $ & $ \! 0 \! $ & $ \! 0 \! $ & $ \! 2 \! $ & $ \! 0 \! $ & $ \! 2 $  \\
$ r_6 \! $ & $ \! 2 \! $ & $ \! 0 \! $ & $ \! 0 \! $ & $ \! -2 \! $ & $ \! -1 \! $ & $ \! 0 \! $ & $ \! 2 \! $ & $ \! 0 \! $ & $ \! 1 \! $ & $ \! 1 \! $ & $ \! -2 \! $ & $ \! 2 \! $ & $ \! -1 \! $ & $ \! 2 \! $ & $ \! 0 \! $ & $ \! 0 \! $ & $ \! 0 \! $ & $ \! -2 \! $ & $ \! 0 \! $ & $ \! 2 $  \\
$ r_7 \! $ & $ \! 3 \! $ & $ \! 1 \! $ & $ \! -1 \! $ & $ \! 3 \! $ & $ \! 0 \! $ & $ \! 1 \! $ & $ \! -1 \! $ & $ \! -1 \! $ & $ \! 0 \! $ & $ \! 0 \! $ & $ \! -1 \! $ & $ \! 3 \! $ & $ \! 0 \! $ & $ \! -1 \! $ & $ \! 1 \! $ & $ \! 1 \! $ & $ \! 1 \! $ & $ \! 3 \! $ & $ \! 1 \! $ & $ \! 3 $  \\
$ r_8 \! $ & $ \! 3 \! $ & $ \! 1 \! $ & $ \! -1 \! $ & $ \! -3 \! $ & $ \! 0 \! $ & $ \! -1 \! $ & $ \! -1 \! $ & $ \! 1 \! $ & $ \! 0 \! $ & $ \! 0 \! $ & $ \! 1 \! $ & $ \! 3 \! $ & $ \! 0 \! $ & $ \! -1 \! $ & $ \! -1 \! $ & $ \! 1 \! $ & $ \! 1 \! $ & $ \! -3 \! $ & $ \! -1 \! $ & $ \! 3 $  \\
$ r_9 \! $ & $ \! 3 \! $ & $ \! -1 \! $ & $ \! 1 \! $ & $ \! 3 \! $ & $ \! 0 \! $ & $ \! -1 \! $ & $ \! -1 \! $ & $ \! 1 \! $ & $ \! 0 \! $ & $ \! 0 \! $ & $ \! -1 \! $ & $ \! 3 \! $ & $ \! 0 \! $ & $ \! -1 \! $ & $ \! -1 \! $ & $ \! -1 \! $ & $ \! -1 \! $ & $ \! 3 \! $ & $ \! -1 \! $ & $ \! 3 $  \\
$ r_{10} \! $ & $ \! 3 \! $ & $ \! -1 \! $ & $ \! 1 \! $ & $ \! -3 \! $ & $ \! 0 \! $ & $ \! 1 \! $ & $ \! -1 \! $ & $ \! -1 \! $ & $ \! 0 \! $ & $ \! 0 \! $ & $ \! 1 \! $ & $ \! 3 \! $ & $ \! 0 \! $ & $ \! -1 \! $ & $ \! 1 \! $ & $ \! -1 \! $ & $ \! -1 \! $ & $ \! -3 \! $ & $ \! 1 \! $ & $ \! 3
   $  \\\hline
$ r_{11} \! $ & $ \! 4 \! $ & $ \! 2 \! $ & $ \! 0 \! $ & $ \! 2 \! $ & $ \! 1 \! $ & $ \! 2 \! $ & $ \! 0 \! $ & $ \! 0 \! $ & $ \! -1 \! $ & $ \! 1 \! $ & $ \! 0 \! $ & $ \! 0 \! $ & $ \! -1 \! $ & $ \! 0 \! $ & $ \! 0 \! $ & $ \! 0 \! $ & $ \! -2 \! $ & $ \! -2 \! $ & $ \! -2 \! $ & $ \! -4 $  \\
$ r_{12}\! $ & $ \! 4 \! $ & $ \! 2 \! $ & $ \! 0 \! $ & $ \! -2 \! $ & $ \! 1 \! $ & $ \! -2 \! $ & $ \! 0 \! $ & $ \! 0 \! $ & $ \! 1 \! $ & $ \! -1 \! $ & $ \! 0 \! $ & $ \! 0 \! $ & $ \! -1 \! $ & $ \! 0 \! $ & $ \! 0 \! $ & $ \! 0 \! $ & $ \! -2 \! $ & $ \! 2 \! $ & $ \! 2 \! $ & $ \! -4 $  \\
$ r_{13} \! $ & $ \! 4 \! $ & $ \! -2 \! $ & $ \! 0 \! $ & $ \! 2 \! $ & $ \! 1 \! $ & $ \! -2 \! $ & $ \! 0 \! $ & $ \! 0 \! $ & $ \! -1 \! $ & $ \! 1 \! $ & $ \! 0 \! $ & $ \! 0 \! $ & $ \! -1 \! $ & $ \! 0 \! $ & $ \! 0 \! $ & $ \! 0 \! $ & $ \! 2 \! $ & $ \! -2 \! $ & $ \! 2 \! $ & $ \! -4 $  \\
$ r_{14} \! $ & $ \! 4 \! $ & $ \! -2 \! $ & $ \! 0 \! $ & $ \! -2 \! $ & $ \! 1 \! $ & $ \! 2 \! $ & $ \! 0 \! $ & $ \! 0 \! $ & $ \! 1 \! $ & $ \! -1 \! $ & $ \! 0 \! $ & $ \! 0 \! $ & $ \! -1 \! $ & $ \! 0 \! $ & $ \! 0 \! $ & $ \! 0 \! $ & $ \! 2 \! $ & $ \! 2 \! $ & $ \! -2 \! $ & $ \! -4 $  \\
$ r_{15} \! $ & $ \! 6 \! $ & $ \! 0 \! $ & $ \! 0 \! $ & $ \! 0 \! $ & $ \! 0 \! $ & $ \! 2 \! $ & $ \! -2 \! $ & $ \! 0 \! $ & $ \! 0 \! $ & $ \! 0 \! $ & $ \! 0 \! $ & $ \! -2 \! $ & $ \! 0 \! $ & $ \! 2 \! $ & $ \! -2 \! $ & $ \! 0 \! $ & $ \! 0 \! $ & $ \! 0 \! $ & $ \! 2 \! $ & $ \! 6 $  \\
$ r_{16} \! $ & $ \! 6 \! $ & $ \! 0 \! $ & $ \! 0 \! $ & $ \! 0 \! $ & $ \! 0 \! $ & $ \! -2 \! $ & $ \! -2 \! $ & $ \! 0 \! $ & $ \! 0 \! $ & $ \! 0 \! $ & $ \! 0 \! $ & $ \! -2 \! $ & $ \! 0 \! $ & $ \! 2 \! $ & $ \! 2 \! $ & $ \! 0 \! $ & $ \! 0 \! $ & $ \! 0 \! $ & $ \! -2 \! $ & $ \! 6 $  \\
$ r_{17} \! $ & $ \! 6 \! $ & $ \! 2 \! $ & $ \! 0 \! $ & $ \! 0 \! $ & $ \! 0 \! $ & $ \! 0 \! $ & $ \! 2 \! $ & $ \! 0 \! $ & $ \! 0 \! $ & $ \! 0 \! $ & $ \! 0 \! $ & $ \! -2 \! $ & $ \! 0 \! $ & $ \! -2 \! $ & $ \! 0 \! $ & $ \! -2 \! $ & $ \! 2 \! $ & $ \! 0 \! $ & $ \! 0 \! $ & $ \! 6 $  \\
$ r_{18} \! $ & $ \! 6 \! $ & $ \! -2 \! $ & $ \! 0 \! $ & $ \! 0 \! $ & $ \! 0 \! $ & $ \! 0 \! $ & $ \! 2 \! $ & $ \! 0 \! $ & $ \! 0 \! $ & $ \! 0 \! $ & $ \! 0 \! $ & $ \! -2 \! $ & $ \! 0 \! $ & $ \! -2 \! $ & $ \! 0 \! $ & $ \! 2 \! $ & $ \! -2 \! $ & $ \! 0 \! $ & $ \! 0 \! $ & $ \! 6 $  \\
$ r_{19} \! $ & $ \! 8 \! $ & $ \! 0 \! $ & $ \! 0 \! $ & $ \! 4 \! $ & $ \! -1 \! $ & $ \! 0 \! $ & $ \! 0 \! $ & $ \! 0 \! $ & $ \! 1 \! $ & $ \! -1 \! $ & $ \! 0 \! $ & $ \! 0 \! $ & $ \! 1 \! $ & $ \! 0 \! $ & $ \! 0 \! $ & $ \! 0 \! $ & $ \! 0 \! $ & $ \! -4 \! $ & $ \! 0 \! $ & $ \! -8 $  \\
$ r_{20} \! $ & $ \! 8 \! $ & $ \! 0 \! $ & $ \! 0 \! $ & $ \! -4 \! $ & $ \! -1 \! $ & $ \! 0 \! $ & $ \! 0 \! $ & $ \! 0 \! $ & $ \! -1 \! $ & $ \! 1 \! $ & $ \! 0 \! $ & $ \! 0 \! $ & $ \! 1 \! $ & $ \! 0 \! $ & $ \! 0 \! $ & $ \! 0 \! $ & $ \! 0 \! $ & $ \! 4 \! $ & $ \! 0 \! $ & $ \! -8 \!$
\\\hline
\end{tabular}
}
\end{table}

\section{More tensor structures for local operators}
\label{app:moreStructuresDil}

In this appendix we give tensor structures used in the one-loop dilatation operator study. We present the irreducible representations not covered in the main text, in order to describe all irreps that exist for $n\leqslant 4$.
\begin{description}
\item[Irrep $\boldsymbol{\XXbar}$] We introduce
\begin{equation}
\mathbf{XXbar}^{(m,p)}_{a_1\cdots a_mb_1\cdots b_p}=\mathbf X^{(m)}_{a_1a_2\cdots a_m}\mathbf X'^{(p)}_{b_1b_2\cdots b_p}-\mathbf X'^{(m)}_{a_1a_2\cdots a_m}\mathbf X^{(p)}_{b_1b_2\cdots b_p},
\end{equation}
where $x^a x'_a=0$.
\item[Irrep $\boldsymbol{B_4}$] We introduce
\begin{equation}
\mathbf{B4}^{(m,n,p,q)}_{a_1\cdots a_mb_1\cdots b_nc_1\cdots c_pd_1\cdots d_q}=(b^{(4)})^{ijkl}\delta^{(m+1)}_{ia_1\cdots a_m}\delta^{(n+1)}_{jb_1\cdots b_n}\delta^{(p+1)}_{kc_1\cdots c_p}\delta^{(q+1)}_{ld_1\cdots d_q},
\end{equation}
where $b^{(4)}$ is completely antisymmetric.
\item[Irrep $\boldsymbol{Z_4}$] We introduce
\begin{equation}
\mathbf{Z4}^{(m,n,p,q)}_{a_1\cdots a_mb_1\cdots b_nc_1\cdots c_pd_1\cdots d_q}=z^iz^jz^kz^l\delta^{(m+1)}_{ia_1\cdots a_m}\delta^{(n+1)}_{jb_1\cdots b_n}\delta^{(p+1)}_{kc_1\cdots c_p}\delta^{(q+1)}_{ld_1\cdots d_q},
\end{equation}
\item[Irrep $\boldsymbol{XX}$] We introduce
\begin{equation}
\mathbf{XX}^{(m,n)}_{a_1\cdots a_mb_1\cdots b_n}=\mathbf X^{(m)}_{a_1a_2\cdots a_m}\mathbf X'^{(p)}_{b_1b_2\cdots b_n}+\mathbf X'^{(m)}_{a_1a_2\cdots a_m}\mathbf X^{(p)}_{b_1b_2\cdots b_n},
\end{equation}
where $x^a x'_a=0$.
\item[Irrep $\boldsymbol {VB_3}$] We introduce
\begin{equation}
\mathbf{VB3}^{(m,n)}_{a_1\cdots a_mb_1\cdots b_n}=\mathbf {B3}^{(m,n,p)}_{a_1a_2\cdots a_mb_1\cdots b_nc_1\cdots c_p}\mathbf {V}^{(q)}_{d_1d_2\cdots d_q},
\end{equation}
where $b^{ijk} z_k=0$.
\item[Irrep $\boldsymbol{\XXXbar}$] We introduce
\begin{equation}
\hspace{-30pt}\mathbf{XXXbar}^{(m_1,m_2,m_3)}_{a_{11}\cdots a_{1m_1}a_{21}\cdots a_{2m_2}a_{31}\cdots a_{3m_3}}=\sum_{\sigma\in S_3}\mathrm{sign}(\sigma)\mathbf X^{(m_{\sigma(1)})}_{a_{\sigma(1)1}\cdots a_{\sigma(1)m_{\sigma(1)}}}\mathbf X'^{(m_{\sigma(2)})}_{a_{\sigma(2)1}\cdots a_{\sigma(2)m_{\sigma(2)}}}\mathbf X''^{(m_{\sigma(3)})}_{a_{\sigma(3)1}\cdots a_{\sigma(3)m_{\sigma(3)}}},
\end{equation}
where all the $x$, $x'$ and $x''$ are different, and the sum is over all six permutations.
\item[Irrep $\boldsymbol{\XXbar V}$] We introduce
\begin{equation}
\mathbf{XXbarV}^{(m,n,p)}_{a_1\cdots a_mb_1\cdots b_n}=\mathbf {XXbar}^{(m,n)}_{a_1a_2\cdots a_mb_1\cdots b_n}\mathbf {V}^{(q)}_{c_1c_2\cdots c_p},
\end{equation}
where $x^{a} v_a=x'^{a} v_a=0$.
\end{description}

\section{Presentation of ancillary file}
\label{app:datafile}

Attached to the Arxiv submission of this paper is an ancillary data file \texttt{cubic-spectrum.nb} in Mathematica format. It contains four sections:

\begin{itemize}
\item Section \texttt{Representation theory} contains names of the irreps (\texttt{irrepNamesN3}, which is a length-$10$ vector), character table (\texttt{charTableN3}, $10\times 10$ matrix), sizes of the conjugacy classes (\texttt{conjugacyClassSizesN3}, length-$10$ vector) and class representatives (\texttt{classRepresentativesN3}, list of $10$ matrices of size $3\times3$) in the case of $n=3$, and correspondingly for $n=4$.
\item Section \texttt{Conformal data: ancillary results} containts more orders for the six-loop results computed in section~\ref{sec:resultsSixloop}. They are given on the format \texttt{moreOrders[$\langle rep\rangle$,$i$]} for the $i$th operator in irrep $\langle rep\rangle$.
\item Section \texttt{Conformal data: representations 1-10} contains results for the 10 irreps that exist for $n=3$, with complete data for Lorentz traceless-symmetric operators up to $\Delta\leqslant 7$ ($\Delta\leqslant 9$ for singlet). The command is
\begin{equation}
\label{eq:operatorlistcommand}
\texttt{operatorList[}\langle rep\rangle\texttt{,}\langle spin\rangle\texttt{]}
\end{equation}
which gives a list of operators in the irrep $\langle rep\rangle$ and spin $\langle spin\rangle$. This in turn contains a list of operators given in the format
\begin{equation}
\texttt{\{}\delta^{l}\,\texttt{box}^m\,\phi^p\texttt{,}\langle dim \rangle\texttt{\}}
\end{equation}
where the first part determines the form of the operator as constructed from $p$ fields, $l$ uncontracted derivatives and $2m$ contracted derivatives, and the second part is the scaling dimension of the operator given as a series in the variable \texttt{e}, with the $O(\varepsilon^k)$ term represented by $\texttt{ord e}^k$. For the six-loop results, higher orders are given in the form \texttt{more[$\langle rep\rangle$,$i$]} and are recovered by the substitution \texttt{/. more -> moreOrders}.

In addition, we give results for twist families in the format
\begin{equation}
\texttt{twistFamilies[}\langle rep\rangle\texttt{]}
\end{equation}
where the data is a function of the free parameter \texttt l (spin), and otherwise are in the same format as above. Here we also included the results from \cite{Dey:2016mcs}.
\item Section \texttt{Conformal data: representations 11-20} contains results for the 10 irreps that exist for $n=4$ but not for $n=3$. It contains complete data for $\Delta\leqslant 6$, and the first operators in spins $0$, $1$ and $2$. The format is the same as for the previous irreps, \eqref{eq:operatorlistcommand}.
\end{itemize}

{
\bibliographystyle{JHEP.bst}
\bibliography{Anom_Dims}
}

\end{document}